\documentclass[11pt,a4paper]{article}
\pdfoutput=1
\usepackage{jstyle}
\usepackage{amsmath, amsfonts, amssymb}
\usepackage{graphicx, hyperref, color, verbatim}
\usepackage[dvipsnames]{xcolor}
\usepackage{float}
\usepackage{caption}
\usepackage{subcaption}

\DeclareGraphicsExtensions{.pdf,.png,.jpg,.mps}
\usepackage{booktabs}
\usepackage{multirow}
\usepackage{tikz}

\usetikzlibrary{calc,matrix,decorations.pathmorphing,decorations.markings,arrows,positioning,intersections,mindmap,backgrounds,patterns,arrows.meta,chains}
\usepackage{cancel}

\usepackage{physics}
\usepackage{diagbox}
\usepackage{pifont}

\newcommand{\zb}{{\bar{z}}}

\newcommand{\bs}{\mathtt{b}_s}
\newcommand{\bt}{\mathtt{b}_t}
\newcommand{\bu}{\mathtt{b}_u}

\newcommand{\cs}{\mathtt{c}_{s}}
\newcommand{\ct}{\mathtt{c}_{t}}
\newcommand{\cu}{\mathtt{c}_{u}}

\newcommand{\dst}{\mathtt{d}_{st}}
\newcommand{\dsu}{\mathtt{d}_{su}}
\newcommand{\dtu}{\mathtt{d}_{tu}}

\title{\LARGE{Simplicity of AdS Super Yang-Mills at One Loop}}

\author[]{Zhongjie Huang$^{a,b}$,}
\author[]{Bo Wang$^{a,b}$,}
\author[]{Ellis Ye Yuan$^{a,b}$,}
\author[]{Xinan Zhou$^{c,d}$}
\affiliation[]{$^{a}$Zhejiang Institute of Modern Physics, School of Physics, Zhejiang University, \\Hangzhou, Zhejiang 310058, China }
\affiliation[]{$^{b}$Joint Center for Quanta-to-Cosmos Physics, Zhejiang University,
	\\Hangzhou, Zhejiang 310058, China}
\affiliation[]{$^{c}$Kavli Institute for Theoretical Sciences,
	University of Chinese Academy of Sciences, \\Beijing 100190, China}
 \affiliation[]{$^{d}$School of Physical Science and Technology, ShanghaiTech University,\\ Shanghai 201210, China }

\abstract{We perform a systematic bootstrap analysis of four-point one-loop Mellin amplitudes for super gluons in $\mathrm{AdS}_5\times\mathrm{S}^3$ with arbitrary Kaluza-Klein weights. The analysis produces the general expressions for these amplitudes at extremalities two and three, as well as analytic results for many other special cases. From these results we observe remarkable simplicity. We find that the Mellin amplitudes always contain only simultaneous poles in two Mellin-Mandelstam variables, extending a previous observation in the simplest case with the lowest Kaluza-Klein weights. Moreover, we discover a substantial extension of the implication of the eight-dimensional hidden conformal symmetry, which goes far beyond the Mellin poles associated with the leading logarithmic singularities. This leaves only a small finite set of poles which can be determined on a case-by-case basis from the contributions of protected operators in the OPE.}

\emailAdd{zjhuang@zju.edu.cn}
\emailAdd{b\_w@zju.edu.cn}
\emailAdd{eyyuan@zju.edu.cn}
\emailAdd{xinan.zhou@ucas.ac.cn}

\begin{document}
\maketitle
\tableofcontents
	
\newpage
		
\section{Introduction and summary}
	
Since its introduction about one and a half decades ago \cite{Mack:2009mi,Mack:2009gy}, the Mellin space representation of conformal correlators (also known as the Mellin amplitude) has been playing an increasingly more significant role in the study of conformal field theories, especially in those theories that admit a perturbative AdS dual description through holography \cite{Penedones:2010ue,Fitzpatrick:2011ia}. The use of the Mellin formalism considerably simplies the analytic structure and made it possible to  bootstrap the holographic correlators of four arbitrary chiral primary operators (CPOs) in $\mathcal{N}=4$ super Yang-Mills at the bulk tree level in the supergravity limit \cite{Rastelli:2016nze,Rastelli:2017udc}. The strategy was later generalized to deal with various other SCFTs with maximal and half-maximal  supersymmetries \cite{Zhou:2017zaw,Rastelli:2017ymc,Zhou:2018ofp,Rastelli:2019gtj,Alday:2020dtb,Alday:2020lbp,Alday:2021odx}.\footnote{See \cite{Bissi:2022mrs} for a review.} The resulting Mellin amplitudes allowed people to extract the anomalous dimensions of the double-particle spectrum in these theories \cite{Aprile:2017xsp,Aprile:2018efk,Drummond:2022dxd}, which later led to the observation of the so-called hidden conformal symmetries \cite{Caron-Huot:2018kta,Rastelli:2019gtj,Alday:2021odx}. More recently, significant progress has also been made to extend the bootstrap techniques to tree-level correlators at higher points \cite{Goncalves:2019znr,Alday:2022lkk,Goncalves:2023oyx,Alday:2023kfm}. These developments have greatly deepened our quantitative understanding of the properties of these holographic CFTs at large central charge, as well as the bulk particle dynamics of the dual theories.
	
Given the above success it is natural to go beyond tree level and study the Mellin amplitudes at one loop. This was initiated in \cite{Alday:2018kkw,Alday:2019nin,Alday:2021ajh}, which studied the simplest correlators in the 4d $\mathcal{N}=4$ and $\mathcal{N}=2$ SCFTs, in correspondence to super gravitons and super gluons with the lowest Kaluza-Klein (KK) weight in $\mathrm{AdS}_5\times\mathrm{S}^5$  and $\mathrm{AdS}_5\times\mathrm{S^3}$ respectively.\footnote{Loop-level correlators have also been studied in position space in \cite{Aprile:2017bgs,Aprile:2017qoy,Aprile:2019rep,Drummond:2019hel,Drummond:2020uni,Huang:2021xws,Drummond:2022dxw,Huang:2023oxf}.} In this paper we carry out a systematic study of the one-loop Mellin amplitudes in the super gluon case to include all KK weights. The specific construction of the $\mathcal{N}=2$ SCFTs hosting these super gluons will be reviewed in the next section. For the time being it suffices to just keep in mind that the CPOs in correspondence to gluons are labeled by a Kaluza-Klein weight $p=2,3,\ldots$, with protected conformal dimension $p$. We schematically denote these operators as $\mathcal{O}_p$ (omitting the color and other internal indices). By the superconformal Ward identities the correlator $\langle\mathcal{O}_{p_1}\mathcal{O}_{p_2}\mathcal{O}_{p_3}\mathcal{O}_{p_4}\rangle\equiv\langle p_1p_2p_3p_4\rangle$ can be equivalently described by a so-called reduced correlator $\mathcal{H}_{p_1p_2p_3p_4}(U,V)$. The reduced correlator depends on the spacetime cross ratios $U$, $V$ (and other cross ratios for internal symmetries which we omit for simplicity), and encodes all the information about interactions in the correlators. We are interested in its expansion with respect to the bulk coupling at the second order $\mathcal{H}_{\{p_i\}}^{(2)}$, which describes the scattering of four super gluons in AdS at one loop. Compared to the supergravity case, the new feature  of AdS gauge theories is that the correlators can be organized in terms color factors
\begin{equation}
\mathcal{H}_{\{p_i\}}^{(2)}=\underbrace{\parbox{2.25cm}{\tikz{\draw [thick] (135:1) node [left] {\scriptsize 1} -- (135:0.5) -- (-135:0.5) -- (-135:1) node [left] {\scriptsize 2};\draw [thick] (45:1) node [right] {\scriptsize 4} -- (45:0.5) -- (-45:0.5) -- (-45:1) node [right] {\scriptsize 3};\draw [thick] (135:0.5) rectangle (-45:0.5);}}}_{\text{color}}\,\mathcal{H}_{\{p_i\},st}^{(2)}+(\text{two other channels})\;.
\end{equation}
The appearance of the distinct color structures allows us to disentangle different parts of the correlator and therefore gives us an extra handle to understand the structure of the loop-level dynamics in AdS. In this decomposition of the reduced correlator the subscript $st$ is a reminder of the corresponding color factor in front. By Bose symmetry it is sufficient to just focus on the color-stripped partial contribution $\mathcal{H}_{\{p_i\},st}^{(2)}$.
	
The Mellin amplitude for this process is defined in the usual way
\begin{equation}\label{eq:Mstdef}
\begin{split}
\mathcal{H}_{\{p_i\},st}^{(2)}(U,V)=&\int\frac{\mathrm{d}s\mathrm{d}t}{(2\pi i)^2}U^{\frac{s}{2}}V^{\frac{t}{2}}\widetilde{\mathcal{M}}_{\{p_i\},st}^{(2)}(s,t)\\
&\times\Gamma(\frac{p_1+p_2-s}{2})\Gamma(\frac{p_3+p_4-s}{2})\Gamma(\frac{p_1+p_4-t}{2})\Gamma(\frac{p_2+p_3-t}{2})\cdots.
\end{split}
\end{equation}
Here we have omitted two other factors of Gamma functions and multiplicative powers of $U$ and $V$ which are not relevant at present.  The simplest case  of $\widetilde{\mathcal{M}}_{2222,st}^{(2)}$ was computed in \cite{Alday:2021ajh}, where it was observed that the amplitude contains \emph{only} simultaneous poles of the Mellin variables $s$ and $t$
\begin{equation}\label{eq:simplestMst}
\widetilde{\mathcal{M}}_{2222,st}^{(2)}=\sum_{m,n=0}^\infty\frac{c_{m,n}}{(s-4-2m)(t-4-2n)}\;.
\end{equation}
The coefficients $c_{m,n}$ are just rational functions of $m$, $n$. Note that the poles of a Mellin amplitude are in correspondence to the twists of operators exchanged in the OPE in the channel specified by the Mellin variable. The above pole structure is very intuitive as it matches precisely the spectrum of double-particle operators in the two channels suggested by the color factor in front of $\widetilde{\mathcal{M}}_{2222,st}^{(2)}$. From this simplest example, it is natural to conjecture that such a pattern continues to hold for Mellin amplitudes of super gluons with arbitrary KK weights
\begin{equation}\label{eq:KKMst}
\widetilde{\mathcal{M}}_{\{p_i\},st}^{(2)}=\sum_{m,n=-\mathcal{E}+2}^\infty\frac{c_{m,n}^{st}}{(s-s_0-2m)(t-t_0-2n)}\;.
\end{equation}
Here $s_0=\min(p_1+p_2,p_3+p_4)$, $t_0=\min(p_1+p_4,p_2+p_3)$ and $\mathcal{E}$ is a parameter called extremality. For simplicity, if we assume $p_1\leq p_2\leq p_3\leq p_4$ then $\mathcal{E}=\min(p_1,\frac{1}{2}(p_1+p_2+p_3-p_4))$ (so the case in \eqref{eq:simplestMst} has $\mathcal{E}=2$). The general definition for $\mathcal{E}$ will be given later. The above minimum values of $m$ and $n$ are constrained by selection rules from superconformal symmetries. 
	
In this paper we will verify this conjecture via explicit computations, and determine the analytic expressions for the coefficients $c_{m,n}^{st}$ in \eqref{eq:KKMst}. Schematically, the strategy can be summarized as follows. We take the ansatz \eqref{eq:KKMst} as the starting point and deform the $s$ and $t$ contours to wrap around the poles in $\widetilde{\mathcal{M}}_{\{p_i\},st}^{(2)}$ and Gamma functions in \eqref{eq:Mstdef}. Taking residues gives an expansion of the reduced correlator $\mathcal{H}_{\{p_i\},st}^{(2)}(U,V)$ in small $U$ and $V$. On the other hand, a part of the expansion is fully determined by the CFT data from the previous two orders, i.e., the disconnected and tree-level scatterings. This allows us to fully determine the $c_{m,n}^{st}$ coefficients. Finally, further consistency checks are performed to show that \eqref{eq:KKMst} is the full answer and no other singularities are allowed in addition to the simultaneous poles. 
	
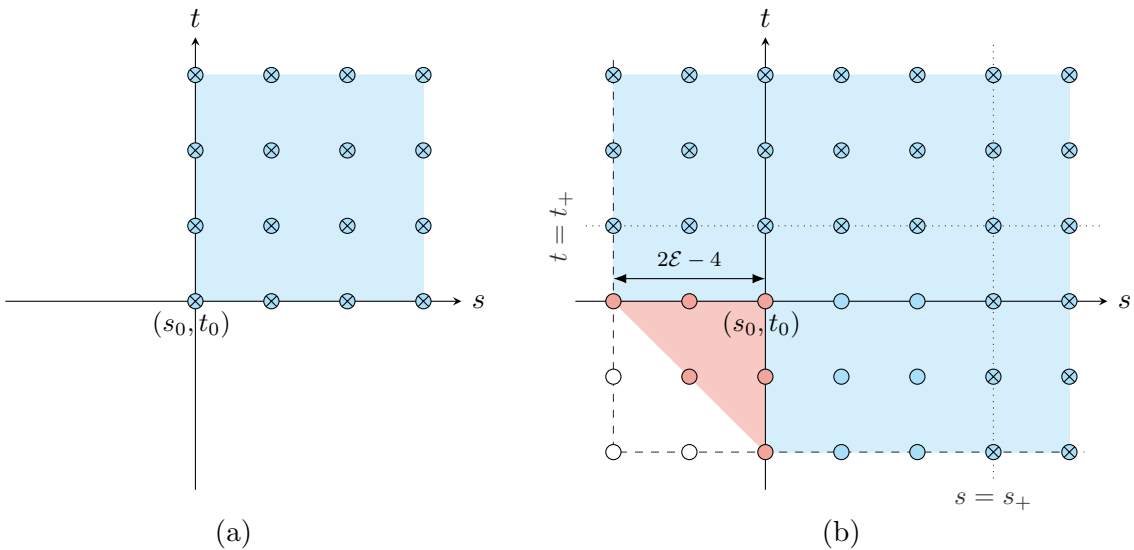
\begin{figure}[ht]
		\centering
		\begin{tikzpicture}
			\definecolor{cr1}{HTML}{fcf6bd}
			\definecolor{cr2}{HTML}{d0f4de}
			\definecolor{cr3}{HTML}{a9def9}
			\definecolor{naturegreen}{HTML}{00A686}
			\definecolor{naturered}{HTML}{E64B35}
			\definecolor{naturecolor3}{HTML}{A9DEF9}
			\begin{scope}
				\draw[draw=cr3, fill=cr3, opacity=0.5] (0,0) rectangle (3,3) ;
				\draw[-stealth] (-2.5,0)--(3.5,0) node [right] {$s$};
				\draw[-stealth] (0,-2.5)--(0,3.5) node [above] {$t$}; 
				\foreach \x in {0,1,2,3} \foreach \y in {0,1,2,3} \draw [fill=cr3] (\x,\y) circle (0.1cm);
				\foreach \x in {0,1,2,3} \foreach \y in {0,1,2,3} \draw ($(\x,\y)+(-135:0.1cm)$) -- +(45:0.2cm) ($(\x,\y)+(-45:0.1cm)$) -- +(135:0.2cm);
				\node [anchor=north] at (-0.05,0) {\small $(s_0,t_0)$};
				\node [anchor=north] at (0.5,-2.75) {(a)};
			\end{scope}
			\begin{scope}[xshift=7.5cm]
				\draw[draw=cr3, fill=cr3, opacity=0.5] (-2,0) rectangle (4,3) ;
				\draw[draw=cr3, fill=cr3, opacity=0.5] (0,-2) rectangle (4,0) ;
				\draw[draw=naturered, fill=naturered, opacity=0.3] (-2,0) -- (0,0) -- (0,-2) -- cycle;
				\draw[-stealth] (-2.5,0)--(4.5,0) node [right] {$s$};
				\draw[-stealth] (0,-2.5)--(0,3.5) node [above] {$t$}; 
				\draw[color=black,thin,opacity=0.8,dashed] (-2,3.2) -- (-2,-2) -- (4.2,-2);
				\draw[color=black,opacity=0.8,dotted] (3,3.4) -- (3,-2.4) node [below] {\small $s=s_+$};
				\draw[color=black,opacity=0.8,dotted] (4.4,1) -- (-2.4,1) node [above,rotate=90] {\small $t=t_+$};
				\foreach \x in {-2,-1} \draw [fill=white] (\x,-2) circle (0.1cm);
				\draw [fill=white] (-2,-1) circle (0.1cm);
				\foreach \x in {-2,-1,...,4} \foreach \y in {1,2,3} \draw [fill=cr3] (\x,\y) circle (0.1cm);
				\foreach \x in {1,2,3,4} \foreach \y in {-2,-1,0} \draw [fill=cr3] (\x,\y) circle (0.1cm);
				\foreach \x in {-2,-1,...,4} \foreach \y in {1,2,3} \draw ($(\x,\y)+(-135:0.1cm)$) -- +(45:0.2cm) ($(\x,\y)+(-45:0.1cm)$) -- +(135:0.2cm);
				\foreach \x in {3,4} \foreach \y in {-2,-1,0} \draw ($(\x,\y)+(-135:0.1cm)$) -- +(45:0.2cm) ($(\x,\y)+(-45:0.1cm)$) -- +(135:0.2cm);
				\foreach \x in {-2,-1,0} \draw [fill=naturered!50] (\x,0) circle (0.1cm);
				\foreach \x in {-1,0} \draw [fill=naturered!50] (\x,-1) circle (0.1cm);
				\draw [fill=naturered!50] (0,-2) circle (0.1cm);
				\draw [{Latex}-{Latex}] (-2,0.3) -- +(2,0);
				\node [anchor=south] at (-1,0.3) {\scriptsize $2\mathcal{E}-4$};
				\node [anchor=north] at (-0.05,0) {\small $(s_0,t_0)$};
				\node [anchor=north] at (1,-2.75) {(b)};
			\end{scope}
		\end{tikzpicture}
\caption{Pole structure of $\widetilde{\mathcal{M}}_{\{p_i\},st}^{(2)}$. Poles responsible for the leading log singularities are marked with a cross. (a) The case of $\langle2222\rangle$, where all poles are responsible for the leading log singularity. (b) The case of generic $\langle p_1p_2p_3p_4\rangle$, where $s_+=\max(p_1+p_2,p_3+p_4)$ and $t_+=\max(p_1+p_4,p_2+p_3)$ are the double-twist thresholds in the s- and t-channel respectively. The explicit graph shown here corresponds to extremality $\mathcal{E}=4$, and may represent $\widetilde{\mathcal{M}}_{st}^{(2)}$ of, e.g., $\langle 5579\rangle$.}
\label{fig:intro}
\end{figure}

This strategy largely follows the one used in \cite{Alday:2021ajh}. In particular, for the $s$ integral for instance, one can easily observe that the poles in the region $s\geq s_+\equiv\max(p_1+p_2,p_3+p_4)$ are triple poles. The presence of the factor $U^{\frac{s}{2}}$ in \eqref{eq:Mstdef} leads to contributions to the residues proportional to $\log^2U$ which are called leading logarithmic singularities. These singularities are associated to unprotected double-particle operators above the double-twist threshold $s_+$, and are dictated by hidden conformal symmetries \cite{Caron-Huot:2018kta}. In fact, the statement of hidden conformal symmetries can be made precise in Mellin space and gives rise to a general expression of $c_{m,n}^{st}$ for all poles in this region. Similar properties hold for the $t$ integral as well. In the simplest case $\widetilde{\mathcal{M}}_{2222,st}^{(2)}$ this region covers the entire set of $m$, $n$ values. The leading logarithmic singularity already determines all the coefficients in the ansatz, as shown in figure \ref{fig:intro} (a). For the general case with arbitrary KK weights, the main challenge is that the constraints from leading log singularities no longer cover all the poles in the ansatz \eqref{eq:KKMst}, see figure \ref{fig:intro} (b). In order to constrain those extra poles we need additional data of the long operators below the double-twist threshold as well as the protected operators. While these extra data can in principle be worked out from disconnected and tree-level correlators as well, their computation has to be performed separately and will be described in detail later. 
	
The necessity of involving a different set of data for these poles might naturally lead to the expectation that the coefficients are more complicated and have very distinct structures compared to the ones responsible for the leading logarithmic singularities.	However, from computing  many non-trivial examples we observe that their structure turn out to be much simpler in two aspects. On the one hand, coefficients of the poles with $m>0$ or $n>0$ turn out to share the same unified formula with those responsible for the leading log singularities (marked blue in figure \ref{fig:intro} (b)).\footnote{This also holds for poles with $m=0$ or $n=0$ in case  when they are also related to the leading logarithmic singularities, as in the simplest case $\widetilde{\mathcal{M}}_{2222,st}$.} This hints at an extension of the original tree-level hidden symmetries to one loop. On the other hand, in the remaining corner region where $m,n\leq 0$, only the poles with $m+n\geq2-\mathcal{E}$ have non-vanishing coefficients (marked red in figure \ref{fig:intro} (b)), while the other poles turn out to be absent (marked white in figure \ref{fig:intro} (b)).
	
As a result, we can organize all the one-loop Mellin amplitudes of CPOs according to the extremality $\mathcal{E}$. The complexity of the correlators grows as $\mathcal{E}$ increases but is insensitive to how the weights $\{p_i\}$ are distributed. For every fixed $\mathcal{E}$ it is possible to work out a unified analytic result for the Mellin amplitudes of all correlators in this class. Although with more computational power we can obtain the results with arbitrary high extremality, in practice we carried out this computation for $\mathcal{E}=2$ and $\mathcal{E}=3$. For instance, the simplest class $\mathcal{E}=2$ has
\begin{align}
c_{m,n}^{st}&=(\alpha\beta)^{-\frac{\kappa_t+\kappa_u}{4}}\frac{(p_1+p_2+p_3+p_4-4)F_{m,n}}{24(m+n)_3\qty(\frac{\kappa_s}{2})!\qty(\frac{\kappa_t}{2})!\qty(\frac{\kappa_u}{2})!},\qquad m>0\text{ or }n>0\;,\\
F_{m,n}&=(\kappa_u  +2) (m+1) (n+1) (m+n)\nonumber \\
&\quad+( \kappa_t  +2) \,n\, (m+1)  (m+n+2)+(\kappa_s +2) \,m\, (n+1) (m+n+2) \;,
\end{align}
where $\kappa_s=|p_1+p_2-p_3-p_4|$, $\kappa_t=|p_1+p_4-p_2-p_3|$, $\kappa_u=|p_1+p_3-p_2-p_4|$, and $\alpha$, $\beta$ are cross ratios for internal symmetries, while
\begin{equation}
c^{st}_{0,0}= (\alpha\beta)^{-\frac{\kappa_t+\kappa_u}{4}}\frac{(\kappa_u+6) (p_1+p_2+p_3+p_4 -4)}{48  \left( \frac{\kappa_s}{2}\right) ! \left( \frac{\kappa_t}{2}\right) !\left( \frac{\kappa_u}{2}\right) !}\;.
\end{equation}
	
The outline of this paper is as follows. In section \ref{sec:preliminaries} we review the necessary background which includes the theory under study as well as the basic kinematic properties of the correlators. In section \ref{sec:mellin} we switch to Mellin space and motivate the ansatz which we will use for our computation. In section \ref{sec:bootstrap} we describe the bootstrap algorithm for computing the one-loop correlators, and discuss how to generate the data needed for the bootstrap by the use of hidden conformal symmetries and by unitarity recursion. This strategy is then put to practice in section \ref{sec:examples}, where we provide several prototypical examples to illustrate in detail several subtleties that one may encounter in this computation. The results and the observations mentioned above are collected and further discussed in section \ref{sec:oneloopgeneral}. We also provide several appendices which contain further technical details of the computation.

\section{Preliminaries}\label{sec:preliminaries}
	
In this paper we focus on super gluons in AdS which exist in the holographic dual of certain SCFTs with 4d $\mathcal{N}=2$ superconformal symmetry. These SCFTs can be engineered by using a stack of $N$ D3-branes probing an F-theory 7-brane singularity \cite{Aharony:1998xz}, or $N$ D3-branes with a small number of probe D7-branes \cite{Karch:2002sh}. In the near-horizon limit the super gluons arise as localized degrees of freedom living on an $\mathrm{AdS}_5\times\mathrm{S}^3$ subspace inside a bigger 10d space which is locally $\mathrm{AdS}_5\times\mathrm{S}^5$.\footnote{In the F-theory 7-brane setup, the internal space has a deficit angle.} Since these degrees of freedom have Lorentz spin at most 1 and are supersymmetric, the spectrum is fixed by the superconformal representation theory \cite{Aharony:1998xz}. Moreover, they decouple from the bulk gravitational modes in the large $N$ limit and therefore the theory in the subspace is the same for the two setups \cite{Karch:2002sh,Alday:2021odx}. Essentially, we  have an 8d $\mathcal{N}=1$ SYM on $\mathrm{AdS}_5\times\mathrm{S}^3$ when considering the leading contribution (tree-level) at large $N$. However, in this paper we will take this decoupling further and turn off gravity to all orders in $1/N$. The purpose is to single out the SYM theory and study the properties of AdS amplitudes in this supersymmetric gauge theory. In particular, we will study loop-level amplitudes which are subleading in $1/N$.
	
The super gluons are scalar fields labelled by a KK weight $k=2,3,\ldots$  corresponding to the angular momentum on $S^3$. In the dual SCFT, they are super primaries of half-BPS multiplets which have protected dimensions $k$.\footnote{More precisely, the super gluons are single-particle states which are dual to linear combinations of the ``single-trace operators'' and ``multi-trace operators'' with $1/N$-suppressed coefficients such that they are orthogonal to the multi-particle states. Such a basis arises naturally from the bulk perspective and is analogous to the supergravity case \cite{Aprile:2019rep,Alday:2019nin,Aprile:2020uxk}. See also appendix \ref{appx:singleparticle} for the explicit construction for such single-particle states for super gluons. \label{footnotesp}} The spinning gluons with Lorenz spin 1 are superconformal descendants of the multiplets and only appear in internal lines in processes with only super gluon external lines. The super gluon operators transform in the adjoint representation of the flavor group $G_F$, which is a gauge group from the AdS perspective.\footnote{Therefore, we will use ``flavor'' and ``color'' interchangeably in this paper depending on the context.}  They also transform in the spin-$\frac{k}{2}$ and spin-$\frac{k-2}{2}$ representation with respect to the $SU(2)_R$ R-symmetry group and $SU(2)_L$ flavor group correspondingly.\footnote{Note that the $SU(2)_L$ flavor group has a geometric origin and has a different nature than $G_F$. The isometry group $SO(4)$ of $\mathrm{S}^3$ can be written as $SU(2)_R\times SU(2)_L$, where one factor becomes the R-symmetry group while the other becomes a flavor symmetry. Therefore, $SU(2)_L$ should just be treated as a global symmetry for the amplitudes rather than a gauge symmetry because the associated gauge field has the same $1/N$ suppression as the gravity (which includes the R-symmetry current) and is decoupled in our setup.} These $SU(2)$ representations can be conveniently kept track of by contracting the indices with two-component polarization spinors $v_a$ and $\bar{v}_{\bar{a}}$. We have
\begin{equation}\label{eq:vvb}
\mathcal{O}^I_k(x;v,\bar v)=\mathcal{O}^{I;a_1,\cdots,a_k;\bar a_1,\dots,\bar a_{k-2}}_k (x) v_{a_1}\cdots v_{a_k} \bar v_{\bar a_1}\cdots \bar v_{\bar a_{k-2}}\;,
\end{equation} 
where $I=1,\ldots,{\rm dim}(G_F)$ is the color group index, and $a,\bar a=1,2$ are the $SU(2)_R$ and $SU(2)_L$ indices respectively.

\subsection{Super gluon four-point functions}\label{sec:gluon4pt}
	
The four-point functions of $\mathcal{O}_{p_i}^{I_i}(x_i;v_i,\bar{v}_i)$, with kinematic factors stripped off, are functions of cross ratios
\begin{equation}\label{eq:4pt}
\begin{split}
\left\langle \mathcal{O}_{p_1}^{I_1}{\mathcal{O}}_{p_2}^{I_2}{\mathcal{O}}_{p_3}^{I_3}{\mathcal{O}}_{p_4}^{I_4}\right\rangle = & \left(\frac{t_{12}}{x_{12}^2}\right)^{\Sigma_{12}/2}\left(\frac{t_{34}}{x_{34}^2}\right)^{\Sigma_{34}/2}\left(\frac{x_{14}^2 t_{24}}{x_{24}^2 t_{14}}\right)^{p_{21}/2}\left(\frac{x_{14}^2 t_{13}}{x_{13}^2 t_{14}}\right)^{p_{34}/2}  \\
& \times (\bar{v}_{12})^{-2} (\bar{v}_{34})^{-2}\, \mathcal{G}^{I_1I_2I_3I_4}_{\left\{p_i\right\}}(U,V; \alpha, \beta) \;,
\end{split}
\end{equation}
where $x_{ij}=x_i-x_j$, $v_{ij} = \epsilon^{a b}v_a v_b$ (similarly for $\bar v$), $t_{ij}={v}_{ij}\bar{v}_{ij}$, and we abbreviate
\begin{equation}
\Sigma_{ij}=p_i+p_j\;, \qquad p_{ij}=p_i-p_j\;.
\end{equation}
The cross ratios are defined as
\begin{align}
U&=\frac{x_{12}^2 x_{34}^2}{x_{13}^2 x_{24}^2} = z \zb\;, \quad V=\frac{x_{14}^2 x_{23}^2}{x_{13}^2 x_{24}^2} = (1-z)(1-\zb) \;,\\
\alpha&=\frac{v_{12} v_{34}}{v_{13} v_{24}}\;, \quad\qquad \hspace{7.7pt}  \beta=\frac{\bar{v}_{12} \bar{v}_{34}}{\bar{v}_{13} \bar{v}_{24}}\;.
\end{align}
For later convenience we will often abbreviate $\left\langle\mathcal{O}_{p_1}^{I_1}{\mathcal{O}}_{p_2}^{I_2}{\mathcal{O}}_{p_3}^{I_3}{\mathcal{O}}_{p_4}^{I_4}\right\rangle=\langle p_1p_2p_3p_4\rangle$. It is also useful to define a quantity called the extremality
\begin{equation}\label{eq:extremality}
\mathcal{E} = \frac{1}{4}\left(\Sigma - |\Sigma_{12}-\Sigma_{34}| - |\Sigma_{13}-\Sigma_{24}| - |\Sigma_{14}-\Sigma_{23}|\right)\;,
\end{equation}
where $\Sigma = p_1+p_2+p_3+p_4$. R-symmetry requires $\mathcal{E}$ to be non-negative and it serves as a measure of complexity of the correlator. More precisely, it follows from the definition that the correlator is a polynomial in $1/\alpha$ and $1/\beta$, where the degrees depend only on the extremality and are $\mathcal{E}$ and $\mathcal{E}-2$ respectively (there is no dependence on $\beta$ for $\mathcal{E}\leq2$). The cases of extremal ($\mathcal{E}=0$) and next-to-extremal ($\mathcal{E}=1$) are special. For these cases, the holographic correlators vanish in the single-particle basis (see footnote \ref{footnotesp}) \cite{Aprile:2019rep,Alday:2019nin}. Therefore, we will restrict our attention to the non-trivial correlators with $\mathcal{E}\geq2$.
	
In writing the correlators as functions of cross ratios we have only used the bosonic part of the superconformal symmetry. The fermionic generators impose additional constraints on the four-point correlator which relate the dependence on the conformal and R-symmetry cross ratios. More precisely, the four-point function obeys the superconformal Ward identities \cite{Nirschl:2004pa}
\begin{equation}\label{eq:wardid}
\partial_{\zb}\, \mathcal{G}^{I_1I_2I_3I_4}_{\left\{p_i\right\}}(z,\zb;\zb,\beta) = 0\;,
\end{equation}
and similarly with $z$ and $\bar{z}$ interchanged. These identities can be solved as follows 
\begin{equation}\label{eq:GHdecp}
\mathcal{G}(z, \zb ; \alpha,\beta)=\mathcal{G}_{\text{protected}}(z, \zb ; \alpha,\beta) + \frac{(z - \alpha)(\zb - \alpha)}{z \zb \alpha^2} \mathcal{H}(z, \zb;\alpha,\beta)\;,
\end{equation}
where we have suppressed the color indices and dependence on $\{p_i\}$ to lighten the notation. The solution splits the correlator into a protected part $\mathcal{G}_{\text{protected}}$ and a {\it reduced correlator} $\mathcal{H}$ which is multiplied by a factor determined by superconformal symmetry.\footnote{When the theory admits a marginal coupling, $\mathcal{G}_{\text{protected}}$ directly relates to the correlator in the free theory. See appendix \ref{appx:Gprotected} for details.} Both $\mathcal{G}_{\text{protected}}$  and $\mathcal{H}$ have Bose symmetry, i.e., they are symmetric under permutations of the operators, by the crossing relations
\begin{align}
\mathcal{G}^{I_1I_2I_3I_4}_{\text{protected}}(z,\zb;\alpha,\beta) =& \left(\frac{\beta}{1-\beta}\right)^2 \qty(\frac{U}{\sigma})^{\frac{\Sigma_{12}}{2}}\qty(\frac{\tau}{V})^{\frac{\Sigma_{23}}{2}} \mathcal{G}^{I_3I_2I_1I_4}_{\text{protected}}\left(1-z,1-\zb\,;\,1-\alpha,1-\beta\right)\nonumber\\
=& \beta^{2}\left(\frac{U}{\sigma}\right)^\frac{\Sigma_{14}}{2} \mathcal{G}^{I_1I_3I_2I_4}_{\text{protected}}\left(\frac{1}{z},\frac{1}{\zb}\,;\,\frac{1}{\alpha},\frac{1}{\beta}\right),\\
\label{eq:Hcrossing}\mathcal{H}^{I_1I_2I_3I_4}(z,\zb;\alpha,\beta) =&  \frac{U^{\frac{\Sigma_{12}}{2}+1}}{V^{\frac{\Sigma_{23}}{2}+1}}\frac{\tau^{\frac{\Sigma_{23}}{2}-2}}{\sigma^{\frac{\Sigma_{12}}{2}-2}}\mathcal{H}^{I_3I_2I_1I_4}\left(1-z,1-\zb\,;\,1-\alpha,1-\beta\right)\nonumber\\
=& U^{\frac{\Sigma_{14}}{2}+1}\sigma^{-\frac{\Sigma_{14}}{2}+2} \mathcal{H}^{I_1I_3I_2I_4}\left(\frac{1}{z},\frac{1}{\zb}\,;\,\frac{1}{\alpha},\frac{1}{\beta}\right),
\end{align}
where we define
	\begin{equation}\label{eq:defst}
		\sigma=\alpha\beta\;,\qquad \tau=(1-\alpha)(1-\beta)\;.
	\end{equation}
Note that the factor multiplying $\mathcal{H}$ is different from that of $\mathcal{G}_{\text{protected}}$. This happens because the superconformal factor $(z-\alpha)(\bar{z}-\alpha)/(z\bar{z}\alpha^2)$ also carries nontrivial conformal and R-symmetry weights which effectively changes the weights of the reduced correlator.
	
Finally, as we mentioned earlier, we want to study the AdS super gluon amplitudes in a $1/N$ expansion. More precisely, the expansion is performed with respect to the large flavor central charge $C_\mathcal{J}$ which is proportional to $N$ 
\begin{equation}\label{eq:aFexpansion}
\mathcal{G}_{\{p_i\}}^{ I_1 I_2 I_3 I_4}= \mathcal{G}^{(0),\,I_1 I_2 I_3 I_4}_{\{p_i\}}+a_F\, \mathcal{G}^{(1),\,I_1 I_2 I_3 I_4}_{\{p_i\}}+a^{2}_F\, \mathcal{G}^{(2),\,I_1 I_2 I_3 I_4}_{\{p_i\}}+\cdots\;,
\end{equation}
where $a_F=6/C_{\mathcal{J}}$.\footnote{The proportionality constant depends on the theory. The flavor central charge $C_{\mathcal{J}}$ is defined by the two-point function of the flavor current as $\langle \mathcal{J}\mu^I(x)\mathcal{J}\nu^J(0)\rangle=\frac{C_{\mathcal{J}}}{2\pi^2}\frac{\delta^{IJ}(\delta_{\mu\nu}-2\frac{x_\mu x_\nu}{x^2})}{x^6}$. Furthermore, through supersymmetry, it is related to the three-point function coefficient $C_{2,2,2}^2$ of $\langle \mathcal{O}_{2}\mathcal{O}_{2}\mathcal{O}_{2}\rangle$ as $C_{2,2,2}^2=\frac{6}{\mathcal{C}_{\mathcal{J}}}$.} Similar expansion also applies to the reduced correlator $\mathcal{H}_{\{p_i\}}^{I_1I_2I_3I_4}$. The leading contribution $\mathcal{G}^{(0)}$ is the disconnected piece of the 2-to-2 scattering. The first non-trivial piece is the tree-level amplitude $\mathcal{G}^{(1)}$ which has been computed for general $\{p_i\}$ in \cite{Alday:2021odx}. In this paper, we will focus on the one-loop amplitude $\mathcal{G}^{(2)}$ (or equivalently $\mathcal{H}^{(2)}$) and explore its dependence on different $\{p_i\}$.

\subsection{Color structures}\label{color}
	
The super gluons carry color indices, which gives rise to different color structures in the four-point function $\mathcal{G}_{\{p_i\}}^{I_1 I_2 I_3 I_4}$ at different perturbation levels. These color structures depend on the way in which gluons are exchanged during the scattering process, as dictated by Feynman rules in AdS. In general these color structures can be written as products of structure constants $f^{IJK}$ of the color group $G_F$. 
	
The presence of these color structures allows us to simplify the analysis of the gluon scattering by just focusing on the ``partial correlator'' accompanying each color structure, as will be described in more detail in the next section. Here we present a list of the color structures that arise in $\mathcal{G}_{\{p_i\}}^{I_1 I_2 I_3 I_4}$ up to one loop, with the purpose of setting up conventions for later discussions. 
	
The simplest cases of the color structures appear at the disconnected level $\mathcal{G}^{(0)}$, where no gluons are exchanged. At this level, the four-point function can be obtained by performing Wick contraction in AdS spacetime, resulting in
\begin{equation}\label{eq:G0pqqp}
\mathcal{G}^{(0),\,I_1 I_2 I_3 I_4}_{pqpq} = \delta_{pq}\,\mathtt{b}_s +  \delta_{pq}\,\beta^2 \left(\frac{U}{\sigma}\right)^{\frac{p+q}{2}} \mathtt{b}_t +\frac{\beta^2}{(1-\beta)^2}\left(\frac{U}{\sigma}\right)^{p}\left(\frac{\tau}{V}\right)^{\frac{p+q}{2}}\mathtt{b}_u \;,
\end{equation}
where the color structures are given by
\begin{equation}
\mathtt{b}_s = \delta^{I_1 I_2} \delta^{I_3 I_4}\;,\quad \mathtt{b}_t= \delta^{I_1 I_4} \delta^{I_2 I_4}\;,\quad \mathtt{b}_u = \delta^{I_1 I_3} \delta^{I_2 I_4}\;.
\end{equation}
To clarify the notation, we will always use a typewriter font for color structures and omit the indices they carry. At the tree level, the gluons interact by exchanging one gluon or via a contact vertex, which results in the familiar color structures
\begin{equation}\label{eq:cs}
\mathtt{c}_s = f^{I_1 I_2 J} f^{J\, I_3 I_4},\quad \mathtt{c}_t= f^{I_1 I_4 J} f^{J\, I_2 I_3},\quad \mathtt{c}_u = f^{I_1 I_3 J} f^{J\, I_4 I_2}\;,
\end{equation}
where $f^{IJK}$ is the structure constant of the color group. These three structures also satisfy the Jacobi identity
\begin{equation}
\cs+\ct+\cu = 0\;.
\end{equation}
At the one-loop level, color structures corresponding to one-loop gluon exchanges may contribute. To obtain all possible color structures at this level, one can draw all one-loop diagrams consisting of three-point vertices and assign each vertex a structure constant $f^{IJK}$. By using the Jacobi identity, all of these structures can be reduced to a linear combination of three box diagrams in color space
\begin{subequations}\label{eq:dst}
\begin{align}
\dst= f^{J\, I_1 K} f^{K\, I_2 L} f^{L\, I_3 M} f^{M\, I_4 J}\;,\\
\dsu= f^{J\, I_1 K} f^{K\, I_2 L} f^{L\, I_4 M} f^{M\, I_3 J}\;,\\
\dtu= f^{J\, I_1 K} f^{K\, I_3 L} f^{L\, I_2 M} f^{M\, I_4 J}\;.
\end{align}
\end{subequations}
Tree-level graviton exchanges also appear in the correlator $\mathcal{G}^{(2)}$. Since the graviton is color-neutral, we may encounter color structures $\bs,\bt,\bu$ at the one-loop level. However, since we focus only on the gluon sector, we will discard contributions from gravitons, i.e., those with color structure $\bs,\bt,\bu$ in the one-loop correlator $\mathcal{G}^{(2)}$.
\begin{figure}
\centering
		\begin{tikzpicture}
			\centering
			\draw [line width=1.5pt] (-4.7,0) -- (-5.3,0.6);
			\draw [line width=1.5pt](-4.7,0) -- (-5.3,-0.6);
			\draw [line width=1.5pt](-4.7,0) -- (-3.5,0);
			\draw [line width=1.5pt](-3.5,0) -- (-2.9,-0.6);
			\draw [line width=1.5pt](-3.5,0) -- (-2.9,0.6);

			\node [anchor=east] at (-5.1,1) {$I_1$};
			\node [anchor=west] at (-3.1,1) {$I_4$};
			\node [anchor=west] at (-3.1,-1) {$I_3$};
			\node [anchor=east] at (-5.1,-1) {$I_2$};
			\node at (-6,0) {${\mathtt{c}}_{s} = $};

			\draw [line width=1.5pt] (0,0.6) -- (-0.6,1.2);
			\draw [line width=1.5pt](0,0.6) -- (0.6,1.2);
			\draw [line width=1.5pt](0,0.6) -- (0,-0.6);
			\draw [line width=1.5pt] (0,-0.6) -- (-0.6,-1.2);
			\draw [line width=1.5pt] (0,-0.6) -- (0.6,-1.2);
			\draw   (-1.2,0) -- (-0.95,0);
			\node at (-1.8,0) {${\mathtt{c}}_{t} =$};
			\node [anchor=east] at (-0.6,1.2) {$I_1$};
			\node [anchor=west] at (0.6,1.2) {$I_4$};
			\node [anchor=west] at (0.6,-1.2) {$I_3$};
			\node [anchor=east] at (-0.6,-1.2) {$I_2$};
		\end{tikzpicture}
		\begin{tikzpicture}
			\draw [line width=1.5pt] (0,0.6) -- (-0.6,1.2);
			\draw [line width=1.5pt](0,0.6) -- (0.6,-1.2);
			\draw [line width=1.5pt](0,0.6) -- (0,-0.6);
			\draw [line width=1.5pt] (0,-0.6) -- (-0.6,-1.2);
			\draw [draw=white,line width=2pt] (0.15,0.15) -- (0.25,-0.15);
			\draw [line width=1.5pt] (0,-0.6) -- (0.6,1.2);
			\node at (-1.5,0) {${\mathtt{c}}_{u} =$};
			\node [anchor=east] at (-0.6,1.2) {$I_1$};
			\node [anchor=west] at (0.6,1.2) {$I_4$};
			\node [anchor=west] at (0.6,-1.2) {$I_3$};
			\node [anchor=east] at (-0.6,-1.2) {$I_2$};
		\end{tikzpicture}\\[0.5em]
		\begin{tikzpicture}
			\draw [line width=1.5pt] (0.5,0.5) -- (0.5,-0.5);
			\draw [line width=1.5pt](0.5,-0.5) -- (-0.5,-0.5);
			\draw [line width=1.5pt](-0.5,-0.5) -- (-0.5,0.5);
			\draw [line width=1.5pt](-0.5,0.5) -- (0.5,0.5);
			\draw [line width=1.5pt](0.5,0.5) -- (1.0,1.0);
			\draw [line width=1.5pt](0.5,-0.5) -- (1.0,-1.0);
			\draw [line width=1.5pt] (-0.5,-0.5) -- (-1.0,-1.0);
			\draw [line width=1.5pt] (-0.5,0.5) -- (-1.0,1.0);
			\node at (-2,0) {${\mathtt{d}}_{st} = $};
			\node [anchor=east] at (-1.0,1.0) {$I_1$};
			\node [anchor=west] at (1.0,1.0) {$I_4$};
			\node [anchor=west] at (1.0,-1.0) {$I_3$};
			\node [anchor=east] at (-1.0,-1.0) {$I_2$};
		\end{tikzpicture}
		\begin{tikzpicture}
			\draw [line width=1.5pt] (0.5,0.5) -- (0.5,-0.5);
			\draw [line width=1.5pt](0.5,-0.5) -- (-0.5,-0.5);
			\draw [line width=1.5pt](-0.5,-0.5) -- (-0.5,0.5);
			\draw [line width=1.5pt](-0.5,0.5) -- (0.5,0.5);
			\draw [line width=1.5pt](0.5,0.5) -- (1.0,1.0);
			\draw [line width=1.5pt](0.5,-0.5) -- (1.0,-1.0);
			\draw [line width=1.5pt] (-0.5,-0.5) -- (-1.0,-1.0);
			\draw [line width=1.5pt] (-0.5,0.5) -- (-1.0,1.0);
			\node at (-2,0) {${\mathtt{d}}_{su} =$};
			\node [anchor=east] at (-1.0,1.0) {$I_1$};
			\node [anchor=west] at (1.0,1.0) {$I_3$};
			\node [anchor=west] at (1.0,-1.0) {$I_4$};
			\node [anchor=east] at (-1.0,-1.0) {$I_2$};
		\end{tikzpicture}
		\begin{tikzpicture}
			\draw [line width=1.5pt] (0.5,0.5) -- (0.5,-0.5);
			\draw [line width=1.5pt](0.5,-0.5) -- (-0.5,-0.5);
			\draw [line width=1.5pt](-0.5,-0.5) -- (-0.5,0.5);
			\draw [line width=1.5pt](-0.5,0.5) -- (0.5,0.5);
			\draw [line width=1.5pt](0.5,0.5) -- (1.0,1.0);
			\draw [line width=1.5pt](0.5,-0.5) -- (1.0,-1.0);
			\draw [line width=1.5pt] (-0.5,-0.5) -- (-1.0,-1.0);
			\draw [line width=1.5pt] (-0.5,0.5) -- (-1.0,1.0);
			\node at (-2,0) {${\mathtt{d}}_{tu} =$};
			\node [anchor=east] at (-1.0,1.0) {$I_1$};
			\node [anchor=west] at (1.0,1.0) {$I_4$};
			\node [anchor=west] at (1.0,-1.0) {$I_2$};
			\node [anchor=east] at (-1.0,-1.0) {$I_3$};
		\end{tikzpicture}
\caption{Color structures at tree and one-loop levels.}
\label{fig:colortreeoneloop}
\end{figure}
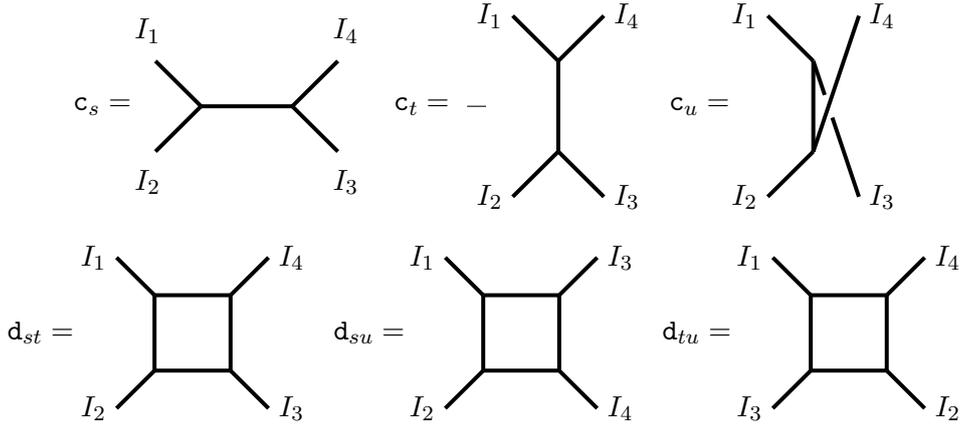
	
Apart from the above, an alternative description of color structures is presented in appendix \ref{appx:colordecp}, which projects them onto irreducible representations in the tensor product of two adjoint representations $\mathbf{adj}\otimes\mathbf{adj}$. This representation is fully equivalent to the one used above and is very useful in performing conformal block decomposition.

\subsection{Superconformal block decomposition}\label{sec:scfblockdecom}
	
Our computation of the one-loop correlators $\mathcal{G}^{(2),\,I_1 I_2 I_3 I_4}_{\{p_i\}}$ heavily relies on the knowledge about CFT data that are extracted from the correlators at lower loop orders. So even though the correlator itself is more naturally presented from the bulk perspective, in the intermediate computations we need to switch to the boundary perspective and study OPE and the associated block decomposition in a chosen channel. As we will see later in section \ref{sec:mellinpoles}, the structure revealed in the (perturbative expansion of) superconformal block decomposition also has a very close connection to the pattern of poles in the Mellin amplitudes.
	
Supersymmetry organizes operators into superconformal multiplets (see \cite{Nirschl:2004pa} for the representation theory of the 4d $\mathcal{N}=2$ superconformal algebra and classification of the superconformal multiplets). In the OPE of two half-BPS operators
\begin{equation}\label{eq:fusion}
\mathcal{B}_{R_1} \times \mathcal{B}_{R_2} \simeq \bigoplus_{R=|R_1-R_2|}^{R_1+R_2} \mathcal{B}_R + \bigoplus_{R=|R_1-R_2|}^{R_1+R_2-1}\mathcal{C}_{R,\ell} + \bigoplus _{R=|R_1-R_2|}^{R_1+R_2-2}\,\mathcal{A}_{\tau,R,\ell}\;,
\end{equation}
we encounter three types of multiplets: the half-BPS (short) multiplets $\mathcal{B}_{R}$, the semi-short multiplets $\mathcal{C}_{R,\ell}$ and the long multiplets $\mathcal{A}_{\tau,R,\ell}$.\footnote{Here we use the conformal twist $\tau=\Delta-\ell$ instead of the dimension $\Delta$ to label the operators. While the symbol $\tau$ was also used in \eqref{eq:defst} to denote one of the cross ratios, we hope the meaning is clear concerning the context.} These multiplets, along with their quantum numbers, are tabulated in table \ref{table:multiplets}. Both the half-BPS and semi-short multiplets are protected and have quantized conformal dimensions, while the long multiplets can develop anomalous dimensions and are only required to satisfy the unitarity bound $\tau\geq 2+2R$. 
\begin{table}[h]
\centering
\begin{tabular}{c|c|c|c}
\toprule
Multiplet type & Label & $SU(2)_R$ charge & Dimension $\Delta$ and spin $\ell$ \\
\midrule
half-BPS &$\mathcal{B}_{R}$ & {$R$} & $\Delta=2 R $, $\ell=0$ \\
Semi-short &$\mathcal{C}_{R,\ell}$ & {$R$} & $\Delta=2+2R+\ell$ \\
Long &  $\mathcal{A}_{\tau,R,\ell}$ & ${ {R} }$ & $\Delta\geq2+2R+\ell$ \\
\bottomrule
\end{tabular}
\caption{Superconformal multiplets appearing in the OPE of two half-BPS operators.}
\label{table:multiplets}
\end{table}
	
Correspondingly, the decomposition of the correlator into conformal blocks is regrouped into superconformal blocks
\begin{equation}\label{eq:superblockdecp}
\mathcal{G}(z, \zb ; \alpha,\beta) = \sum_R b_R\, \mathcal{B}_{R}(z,\zb;\alpha) + \sum_{R,\ell} c_{R,\ell}\, \mathcal{C}_{R,\ell}(z,\zb;\alpha) + \sum_{\tau,R,\ell} a_{\tau,R,\ell}\, \mathcal{A}_{\tau,R,\ell}(z,\zb;\alpha)\;.
\end{equation}
Here we are abusing the notation slightly by using the name of the multiplet to denote the corresponding superconformal block. We have also suppressed the color indices, the dependence of the $SU(2)_L$ cross ratio $\beta$, and the labels of the external weights $\{p_i\}$. To conveniently describe the superconformal blocks, let us write the solution to the superconformal Ward identities (\ref{eq:wardid}) in a slightly different way
\begin{equation}\label{eq:solscfWI2}
\begin{split}
\mathcal{G}(z,\zb;\alpha) = &\ \frac{z-\alpha}{z-\zb}\chi(\zb;\alpha)f(z) + \frac{\zb-\alpha}{\zb-z}\chi(z;\alpha)f(\zb) \\
&+ \frac{(z-\alpha)(\zb-\alpha)}{z\zb\alpha^2} \mathcal{F}(z,\zb;\alpha)\;,
\end{split}
\end{equation}
where we have defined 
\begin{equation}\label{eq:chidef}
\chi(z;\alpha) = \left( \frac{z}{\alpha} \right)^{1+\max(|p_{21}|,|p_{34}|)/2}\left( \frac{1-\alpha}{1-z} \right)^{\max (p_{21}+p_{34},0)/2}.
\end{equation}
The first line in (\ref{eq:solscfWI2}) is the protected part and is related to $\mathcal{G}_{\rm protected}$ in (\ref{eq:GHdecp}) by the property that they give rise to the same meromorphic function $f(z)$ upon setting $\zb=\alpha$. However, these two protected objects are not identical. Constructed in this way, the first line of (\ref{eq:solscfWI2}) is not invariant under Bose symmetry while $\mathcal{G}_{\rm protected}$ transforms as a correlator. Consequently, the transformation of $\mathcal{F}$ under Bose symmetry also receives inhomogeneous contributions from the single-variable function $f$. Nevertheless, the advantage of this alternative representation is that the contribution of a superconformal block can be expressed in a very simple way in terms of $f$ and $\mathcal{F}$. Taking the long superconformal block $\mathcal{A}_{\tau,R,\ell}$ in the s-channel OPE as an example, its decomposition \eqref{eq:solscfWI2} is 
\begin{equation}\label{eq:longfF}
f^{\mathcal{A}}=0\;,\quad \mathcal{F}^{\mathcal{A}}=g^{p_{21},p_{34}}_{\tau+2,\ell}(z,\zb) P^{p_{21},p_{34}}_R(\alpha)\equiv\mathfrak{h}^{p_{21},p_{34}}_{\tau,R,\ell}(z,\zb;\alpha)\;,
\end{equation}
where $g^{r,s}_{\tau,\ell}(z,\zb)$ is the bosonic conformal block in 4d defined by
\begin{align}\label{eq:defblock}
g^{r,s}_{\tau, \ell}(z,\zb)&=\frac{z \bar{z}}{\bar{z}-z}\left(k^{r,s}_{\frac{\tau-2}{2}}(z) k^{r,s}_{\frac{\tau+2 \ell}{2}}(\zb)-k^{r,s}_{\frac{\tau-2}{2}}(\zb) k^{r,s}_{\frac{\tau+2 \ell}{2}}(z)\right), \\ k^{r,s}_h(z)&=z^h\  _2 F_1(h+\frac{r}{2}, h+\frac{s}{2}, 2 h, z)\;,
\end{align}
and $P^{r,s}_R(\alpha)$ is the harmonic function of the R-symmetry group $SU(2)_R$
\begin{equation}\label{eq:PRdef}
P_R^{r,s}(\alpha)=k^{-r,-s}_{-R}(\alpha)\;.
\end{equation}
For convenience we also introduced an abbreviation $\mathfrak{h}^{p_{21},p_{34}}_{\tau,R,\ell}(z,\zb;\alpha)$ in \eqref{eq:longfF}, which will be frequently used later (and we will often omit the superscripts when there is no confusion). Therefore the long superconformal block is simply
\begin{equation}
\mathcal{A}_{\tau,R,\ell}=\frac{(z-\alpha)(\bar{z}-\alpha)}{z\bar{z}\alpha^2}\,\mathfrak{h}_{\tau,R,\ell}(z,\zb;\alpha)\;.
\end{equation}
This means that long multiplets only contribute to the reduced correlator $\mathcal{H}$ in the decomposition \eqref{eq:GHdecp} and to $\mathcal{F}$ in \eqref{eq:solscfWI2}. In view of the OPE the two quantities $\mathcal{H}$ and $\mathcal{F}$ differ by the amount of contribution from the protected multiplets. Detailed expressions for the superconformal blocks of the protected multiplets are collected in appendix \ref{appx:superblocks}.

Now we incorporate the bulk perturbative expansion into the superconformal block decomposition \eqref{eq:superblockdecp}. From the boundary perspective the presence of the small parameter $a_F$ indicates that various CFT data can be perturbatively expanded. Because the twists (or dimensions) of the protected multiplets are fixed by superconformal symmetries, their superconformal blocks $\mathcal{B}_R$ and $\mathcal{C}_{R,\ell}$ are protected and do not receive any correction. In contrast, the twists of long multiplets are expanded into the classical twists plus corrections from anomalous dimensions
\begin{equation}
\tau=\tau_0+a_F\gamma^{(1)}+a_F^2\gamma^{(2)}+\cdots.
\end{equation}
Hence the long superconformal blocks can be expanded as
\begin{equation}
\begin{split}
\mathcal{A}_{\tau,R,\ell}=&\frac{(z-\alpha)(\bar{z}-\alpha)}{z\bar{z}\alpha^2}\underbrace{\qty(\mathfrak{h}_{\tau,R,\ell}+a_F\,\gamma^{(1)}\partial_{\tau_0}\mathfrak{h}_{\tau_0,R,\ell}+\cdots)}_{\text{contribution to }\mathcal{H}}.
\end{split}
\end{equation}
	
The expansion behavior of OPE coefficients is more involved. First let us focus on the long multiplets. Up to one loop in the perturbative expansion the long multiplets allowed to be exchanged are the double-particle operators, which have the schematic form
\begin{equation}
[pq]_{\tau,R,\ell}=\;:\!\mathcal{O}_p\square^{\frac{1}{2}(\tau-p-q)}\partial^\ell\mathcal{O}_q\!:\big|_{R}\;.
\end{equation}
Here we have dropped the color indices. The OPE coefficients of $[pq]_{\tau,R,\ell}$ with a pair of super gluon operators $\mathcal{O}_{p_i}$ and $\mathcal{O}_{p_j}$ has the generic expansion
\begin{equation}\label{eq:Cseries0}
C_{p_ip_j,[pq]} = C^{(0)}_{p_ip_j,[pq]} + a_F\, C^{(1)}_{p_ip_j,[pq]}+\ldots\;,
\end{equation}
When the twist of the long multiplet satisfies $\tau\geq p_i+p_j$, the leading contribution $C^{(0)}_{p_ip_j,[pq]}$ above is non-trivial and is determined by Wick contraction. On the other hand, when $|p_i-p_j|+4<\tau<p_i+p_j$ it vanishes and so the OPE coefficient $C_{p_ip_j,[pq]}$ is suppressed by $a_F$. 

\begin{figure}[ht]
		\centering
        \begin{tikzpicture}
		\draw[thick] (-1,2) -- (-1,-2) -- (1,-2)--(1,2);
		\draw[dashed,thick,BrickRed] (-4,0.48)--(1,0.48);
		\draw[dashed,thick,RoyalBlue] (-1,-0.48)--(4,-0.48);
		\draw[dashed,thick] (-4,-2) -- (4,-2) node [right] {\small $\tau^{\rm min}-2(\mathcal{E}-2)$}; 
		\draw[{stealth}-{stealth},BrickRed,thick] (-3.5,0.48)--(-3.5,2); 
		\node [left] at (-3.5,1.24) {$\small\color{BrickRed}{\mathcal{O}(1)}$};
		\draw[{stealth}-{stealth},BrickRed,thick] (-3.5,0.48)--(-3.5,-2); 
		\node [left] at (-3.5,-0.76) {$\color{BrickRed}{\small \mathcal{O}(a_F)}$};
		\node [left] at (-4.5,0) {\small $\color{BrickRed}{C_{p_1 p_2 ,[pq]}=}$};
		\draw[{stealth}-{stealth},RoyalBlue,thick] (3.5,-0.48)--(3.5,2); 
		\node [right] at (3.5,0.76) {\small $\color{RoyalBlue}{\mathcal{O}(1)}$};
		\draw[{stealth}-{stealth},RoyalBlue,thick] (3.5,-0.48)--(3.5,-2); 
		\node [right] at (3.5,-1.24) {\small $\color{RoyalBlue}{\mathcal{O}(a_F)}$};
		\node [right] at (4.5,0) {\small $\color{RoyalBlue}{=C_{p_3 p_4,[pq]}}$};
		\node[left,align=right] at (-1,-0.48) {\small $\tau^{\rm min}=p_3+p_4$};
		\node[right,align=left] at (1,0.48) {\small $\tau^{\rm max}=p_1+p_2$};
		\node [align=center] at (0,0) {\small Window};
		\node [align=center] at (0,1.24) {\small Above\\[-0.5em]\small Threshold};
		\node [align=center] at (0,-1.24) {\small Below\\[-0.5em]\small Window};
	\end{tikzpicture}
\caption{Large central charge $\mathcal{C_{J}}$ ($a_F=6/\mathcal{C}_\mathcal{J}$) behavior of OPE coefficients $C_{p_i p_j,[pq]}$. In this figure we assume $p_1+p_2 > p_3+p_4$ for illustration.}
\label{fig:OPE}
\end{figure}
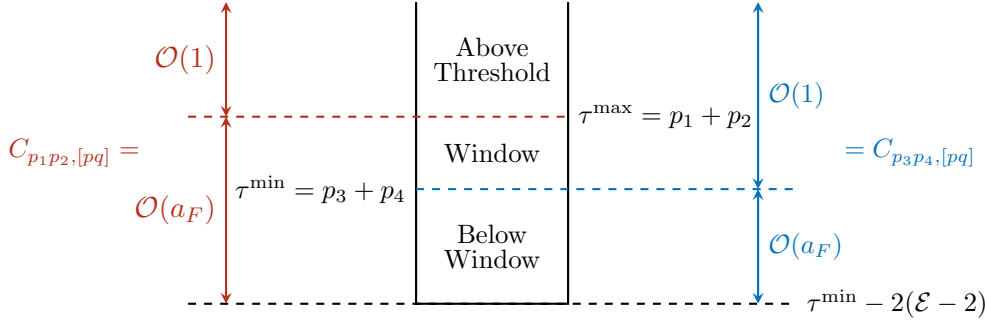

Expressed in terms of the reduced correlator $\mathcal{H}$, the long sector of the superconformal block expansion \eqref{eq:superblockdecp} reads
\begin{equation}\label{eq:Hlongdecp}
\mathcal{H}\big|_{\rm long}=\sum_{\tau,R,\ell}a_{\tau,R,\ell}\,\mathfrak{h}_{\tau,R,\ell}(z,\bar{z};\alpha)\;,
\end{equation}
where the expansion coefficients are related to the above OPE coefficients by $a_{\tau,R,\ell}\sim C_{p_1p_2,[pq]} C_{p_3p_4,[pq]}$ (here we only focus on the dependence on $a_F$, while the explicit relation is postponed to section \ref{sec:recursion}). These coefficients receive the perturbative expansion
\begin{equation}
a = a^{(0)} + a_F\, a^{(1)} + a_F^2\, a^{(2)} + \cdots.
\end{equation}
Taking this into consideration, the $a_F$ counting of $C_{p_ip_j,[pq]}$ then naturally divides the twist spectrum of the exchanged long operators at one loop into three regions \cite{Aprile:2019rep}
\begin{itemize}
\item {\it Above Threshold:} $\tau\geq \tau^{\rm max}=\max(p_1+p_2,p_3+p_4)$. In this region both $ C_{p_1p_2,[pq]}$ and $ C_{p_3p_4,[pq]}$ are of order $\mathcal{O}(1)$, and so $a_{\tau,R,\ell}$ starts at order $\mathcal{O}(1)$ as well. 
\item {\it Window:} $\tau^{\rm max}>\tau\geq \tau^{\rm min}=\min(p_1+p_2,p_3+p_4)$. In this region one of the OPE coefficient is $\mathcal{O}(1)$ while the other is $\mathcal{O}(a_F)$. Hence $a^{(0)}$ vanishes and $a_{\tau,R,\ell}$ starts at order $\mathcal{O}(a_F)$. 
\item {\it Below Window:} $\tau<\tau^{\rm min}$. Both OPE coefficents are $\mathcal{O}(a_F)$. Hence $a^{(0)}$ and $a^{(1)}$ both vanish and $a_{\tau,R,\ell}$ starts at order $\mathcal{O}(a_F^2)$.
\end{itemize}
This division of the exchanged spectrum is schematically illustrated in figure \ref{fig:OPE}.

With the above expansions of the CFT data for long multiplets, we now check each perturbative order (up to one loop) of the long sector of the reduced correlator $\mathcal{H}$. These have the schematic form
\begin{subequations}\label{eq:Hexpansion}
\begin{align}
\mathcal{H}^{(0)}\big|_\text{long} =&\ \langle a^{(0)} \rangle_{\tau_0,R,\ell} \ \mathfrak{h}_{\tau_0, R, \ell}\;,\\
\mathcal{H}^{(1)}\big|_\text{long} =& \ \langle a^{(1)}\rangle_{\tau_0,R,\ell} \ \mathfrak{h}_{\tau_0, R, \ell} \,+\, \langle a^{(0)}\gamma^{(1)} \rangle_{\tau_0,R,\ell} \ \partial_{\tau_0} \mathfrak{h}_{\tau_0, R, \ell}\;,\\
\label{eq:H2expansion}\mathcal{H}^{(2)}\big|_\text{long} =&\ \langle a^{(2)}\rangle_{\tau_0,R,\ell} \ \mathfrak{h}_{\tau_0, R, \ell} \,+\, \langle a^{(1)}\gamma^{(1)} + a^{(0)}\gamma^{(2)} \rangle_{\tau_0,R,\ell} \ \partial_{\tau_0} \mathfrak{h}_{\tau_0, R, \ell}\nonumber\\
&+ \frac{1}{2} \langle a^{(0)}(\gamma^{(1)})^2 \rangle_{\tau_0,R,\ell} \ \partial_{\tau_0}^2 \mathfrak{h}_{\tau_0, R, \ell}\;,
\end{align}
\end{subequations}
where we implicitly sum over all possible $\tau_0,R,\ell$, and the angle brackets $\langle \cdots \rangle$ here are to emphasize the fact that there may be multiple operators with the same classical twist $\tau_0$ and they mix with each other in the decomposition. The distinction between the three regions indicates that some of the coefficients in \eqref{eq:Hexpansion} simplify or even vanish when the twist is below some specific thresholds. Specifically, when $\tau_0 < \tau_{\rm max}$ we have
\begin{equation}\label{eq:a0vanish}
\langle a^{(0)}(\gamma^{(1)})^{\geq 0}\rangle_{\tau_0,R,\ell}=0,\qquad 
\langle a^{(1)}\gamma^{(1)}+a^{(0)}\gamma^{(2)}\rangle_{\tau_0,R,\ell}=\langle a^{(1)}\gamma^{(1)}\rangle_{\tau_0,R,\ell}\;,
\end{equation}
and when $\tau_0 < \tau_{\rm min}$ we further have
\begin{equation}\label{eq:a1vanish}
\langle a^{(1)}\rangle_{\tau_0,R,\ell}=0\;,\qquad
\langle a^{(1)}\gamma^{(1)}+a^{(0)}\gamma^{(2)}\rangle_{\tau_0,R,\ell}=0\;.
\end{equation}
	
A crucial feature of the perturbative expansion \eqref{eq:Hexpansion} is that the derivatives $\partial_{\tau_0}$ acting on $\mathfrak{h}_{\tau_0, R, \ell}$ will generate $\log U$ singularities in the correlator
\begin{equation}
(\partial_{\tau_0})^{k} \mathfrak{h}^{p_{21},p_{34}}_{\tau_0, R, \ell} \sim 2^{-k} \log^{k}(U)\,\mathfrak{h}_{\tau_0, R, \ell} +\cdots,
\end{equation}
where the dots here represent terms with lower power of $\log U$. This indicates that the leading $\log U$ singularity of $\mathcal{H}^{(2)}$ should be $\log^2U$. By \eqref{eq:a0vanish} it is easy to see that these leading singularities receive contributions from long multiplets in the above threshold region only. In contrast, the long multiplets in the window region can contribute at most to the subleading singularity $\log^1U$, while those in the below window region can only have rational contributions. A similar pattern also holds in the other OPE channels. This qualitative pattern will have important implications for the structure of Mellin amplitudes, as well be discussed in detail in the next section.
	
We now consider the contribution from the protected super multiplets in $\mathcal{H}$
\begin{equation}
\mathcal{H}\big|_{\rm protected}=\sum_R b_R\, \mathcal{H}^{\mathcal{B}} + \sum_{R,\ell} c_{R,\ell}\, \mathcal{H}^{\mathcal{C}}\;,
\end{equation}
where $\mathcal{H}^{\mathcal{B}}$ means to extract the reduced part of $\mathcal{B}_R(z,\bar{z};\alpha)$ in \eqref{eq:GHdecp}, and similarly for $\mathcal{H}^{\mathcal{C}}$. Because of the protected nature of these operators, they do not make any contribution to terms with $\log^{>0}U$. On the other hand, by the relations between their R-charges and twists, one can see that the protected multiplets only show up when their twists sit in the below window region or at the lower bound of the window region. In this case their OPE coefficients have the expansions
\begin{align}
b =& b^{(0)} + a_F\, b^{(1)} + a_F^2\, b^{(2)} + \cdots,\\
c =& c^{(0)} + a_F\, c^{(1)} + a_F^2\, c^{(2)} + \cdots.
\end{align}
At one-loop level that we are going to study, a simplification occurs in the contribution of the protected sector. As is discussed in detail at the end of appendix \ref{appx:Gprotected}, at one loop the protected part $\mathcal{G}^{(2)}_\text{protected}$ of the correlator in gluon sector has to vanish. Consequently, even though each individual block $\mathcal{B}$ and $\mathcal{C}$ may in general contribute to $\mathcal{G}^{(2)}_\text{protected}$, these contributions have to cancel out at this level so that the sum over them will entirely belong to the reduced correlator $\mathcal{H}^{(2)}$. Hence at one loop the protected sector can equivalently be computed by
\begin{equation}\label{eq:H2protectedblocks}
\mathcal{H}^{(2)}\big|_{\rm protected}=\frac{z\bar{z}\alpha^2}{(z-\alpha)(\bar{z}-\alpha)}\qty(\sum_R \langle b^{(2)}\rangle_R\, \mathcal{B}_R + \sum_{R,\ell} \langle c^{(2)}\rangle_{R,\ell}\, \mathcal{C}_{R,\ell})\;.
\end{equation}

\section{Mellin space}\label{sec:mellin}
	
Now we move into Mellin space to study the four-point correlators of super gluons.  In this formalism, the analytic structure of these physical observables are greatly simplified and they exhibit features analogous to flat-space scattering in many aspects. As in \cite{Alday:2021ajh}, we focus on the {\it reduced} Mellin amplitude which is directly related to the reduced correlator. Recall that the reduced correlator captures all the dynamical information of scattering in AdS and has already taken superconformal symmetry into account. This makes the reduced Mellin amplitude a simpler object to study.\footnote{The Mellin amplitude is defined with respect to the full correlator and is related to the reduced Mellin amplitude by the action of a difference operator. See \cite{Alday:2021odx} for details.} Hereafter  we will always mean the reduced Mellin amplitude whenever referring to the Mellin ampitude.
	
The reduced Mellin amplitude is defined by 
\begin{align}\label{eq:mellindef}
\mathcal{H}_{\{p_i\}}^{I_1 I_2 I_3 I_4}(U,V;\alpha,\beta)=\int_{-i \infty}^{i \infty} \frac{\dd s \dd t}{(2 \pi i)^2} U^{\frac{s+2}{2}} V^{\frac{t-\Sigma_{23}}{2}} \widetilde{\mathcal{M}}^{I_1 I_2 I_3 I_4}_{\{p_i\}}(s,t;\alpha,\beta) \Gamma_{\{p_i\}}(s,t)\;,
\end{align}
where
\begin{equation}
\Gamma_{\{p_i\}}(s,t)=\Gamma(\frac{\Sigma_{12}-s}{2}) \Gamma(\frac{\Sigma_{34}-s}{2}) \Gamma(\frac{\Sigma_{23}-t}{2})\Gamma(\frac{\Sigma_{14}-t}{2}) \Gamma(\frac{\Sigma_{13}-\tilde{u}}{2})\Gamma(\frac{\Sigma_{24}-\tilde{u}}{2})\;,
\end{equation}
and
\begin{equation}
s + t + \tilde{u}=\Sigma-2\;, \qquad \Sigma=p_1+p_2+p_3+p_4\;,\qquad \Sigma_{ij}=p_i+p_j\;.
\end{equation}
$s$, $t$ and $\tilde{u}$ are the Mellin analogue of the usual Mandelstam variables for four-point amplitudes in flat space.\footnote{Here a tilde is added to $\tilde{u}$ to be in agreement with the notation used in \cite{Alday:2021odx} and $\tilde{u}$ is related to the $u$ variable appearing in the Mellin amplitude by a shift of 2. Under exchanging external operators, $\{s,t,u\}$ and $\{s,t,\tilde{u}\}$ are permuted for the Mellin amplitude and the reduced Mellin amplitude respectively.} In the inverse Mellin transformation (\ref{eq:mellindef}), the integration contours run in parallel to the imaginary axis and separate the sequences of Gamma function poles facing opposite directions.  
	
In the following, we will first review the structure of the one-loop Mellin amplitude for the lowest KK mode obtained in \cite{Alday:2021ajh}. This then motivates an ansatz that we propose for the reduced Mellin amplitude of general KK modes. Along the way we also discuss how information of the exchanged operators is encoded in this ansatz. This will be used in the next section where we present an algorithm to bootstrap the Mellin amplitudes.

\subsection{Ansatz for the Mellin amplitude}
	
The one-loop Mellin amplitude for the case $p_i=2$ was first obtained in \cite{Alday:2021ajh}. Let us review this result and briefly explain how to arrive at it. The schematic discussion only serves to motivate our ansatz for the general one-loop amplitude and a more detailed discussion can be found in the next subsection and later when we discuss our algorithm. The one-loop amplitude with $p_i=2$ can be written as the sum of three channels
\begin{equation}\label{eq:2222oneloop}
\widetilde{\mathcal{M}}^{(2)}_{2222}(s,t) = \dst \widetilde{\mathcal{M}}_{2222,st} + \dsu \widetilde{\mathcal{M}}_{2222,su} +\dtu \widetilde{\mathcal{M}}_{2222,tu}\;.
\end{equation}
In each channel we have a color box diagram multiplying an ``AdS box diagram'', in terms of which the Mellin amplitude is given as
\begin{align}\label{eq:2222Bst}
&\widetilde{\mathcal{M}}_{2222,st} = \sum_{m,n=0}^{\infty} \frac{c_{m,n}}{(s-4-2m)(t-4-2n)}\;,&\\
&\widetilde{\mathcal{M}}_{2222,su} = \widetilde{\mathcal{M}}_{2222,st}\big|_{t\leftrightarrow\tilde{u}}\;,\quad 
\widetilde{\mathcal{M}}_{2222,tu} = \widetilde{\mathcal{M}}_{2222,st}\big|_{s\leftrightarrow t}\;.&
\end{align}
An interesting feature of this one-loop amplitude is that it contains only simultaneous poles with constant numerators which are given by a simple rational function of the pole locations
\begin{equation}
c_{m,n} = \frac{(3 m^2 n+2 m^2+3 m n^2+8 m n+3 m+2 n^2+3 n)}{3 (m+n) (m+n+1) (m+n+2)}\;.
\end{equation}
The appearance of the simultaneous poles is easy to understand. In Mellin space, poles are related to the twists of exchanged operators. The exchanged two-particle long operators in the s-channel with twists $\tau=4,6,8,\ldots$ lead to poles at $s=4,6,8,\ldots$. Meanwhile, these long operators have infinite support on spins, which requires singular $t$-dependence in the Mellin amplitude. Crossing symmetry between $s$ and $t$ then implies the existence of simultaneous poles as in (\ref{eq:2222Bst}). On the other hand, that $c_{m,n}$ are merely constants (in other words, the Mellin amplitude contains no other singularities such as single poles in $s$) is highly nontrivial and {\it a priori} unclear. It follows from the analysis of the leading logarithmic singularity $\log^2U$ which is a function of $V$ and can be computed on the CFT side from the tree-level data. From Mellin space, we can compute the leading logarithmic singularity by closing the $s$-contour to take the residues at the poles $s=4+2m$ and selecting the highest power of $\log U$. Simiarly the $t$ integral leads to terms with $\log V$ dependence up to $\log^2V$. In principle the coefficients $c_{m,n}$ can already be determined by matching the $\log^2U\log^2V$ part. The validity of (\ref{eq:2222Bst}) is confirmed by the fact that the entire $V$-dependence of the $\log^2U$ singularity is fully reproduced. 
	
The $p_i=2$ amplitude (\ref{eq:2222oneloop}) serves as the starting point of our generalization to one-loop amplitudes with general weights $p_i$. We propose the following ansatz which includes (\ref{eq:2222oneloop}) as a special case
\begin{equation}\label{Mredgenp}
\widetilde{\mathcal{M}}^{(2)}_{\{p_i\}}(s,t;\alpha,\beta) = \dst \widetilde{\mathcal{M}}_{\{p_i\},st} + \dsu \widetilde{\mathcal{M}}_{\{p_i\},su} +\dtu \widetilde{\mathcal{M}}_{\{p_i\},tu}\;,
\end{equation}
Here the color-stripped partial amplitudes in three channels are given by
\begin{align}\label{eq:mst}
\widetilde{\mathcal{M}}_{\{p_i\},st}&=\sum_{m,n=-\mathcal{E}+2}^{\infty} \frac{c_{m,n}^{st}(\alpha,\beta)}{(s-s_{0}-2m)(t-t_{0}-2n)}\;,\\
\widetilde{\mathcal{M}}_{\{p_i\},su}&=\sum_{m,n=-\mathcal{E}+2}^{\infty} \frac{c_{m,n}^{su}(\alpha,\beta)}{(s-s_{0}-2m)(\tilde{u}-u_{0}-2n)}\;,\\
\widetilde{\mathcal{M}}_{\{p_i\},tu}&=\sum_{m,n=-\mathcal{E}+2}^{\infty} \frac{c_{m,n}^{tu}(\alpha,\beta)}{(t-t_{0}-2m)(\tilde{u}-u_{0}-2n)}\;,
\end{align}
where $\mathcal{E}$ is the extremality defined in \eqref{eq:extremality} and 
\begin{equation}
s_0 = \min(\Sigma_{12},\Sigma_{34})\;,\quad t_0 = \min(\Sigma_{23},\Sigma_{14})\;,\quad u_0 = \min(\Sigma_{24},\Sigma_{13})\;,
\end{equation}
are the lowest twists of two-particle operators in three channels. We will assume here that the numerators $c^{st}_{m,n}$, $c^{su}_{m,n}$, $c^{tu}_{m,n}$ are constants with respect to the Mellin-Mandelstam variables. But for $p_i>2$ the operators have nontrivial weights under $SU(2)_R$ and $SU(2)_L$ and therefore the numerators in general depend on the corresponding cross ratios $\alpha$ and $\beta$.\footnote{In the reduced correlator the $SU(2)_R$ weight is shifted to $p_i-2$, the same value as the $SU(2)_L$ weight.} The three partial amplitudes are related to each other by Bose symmetry
\begin{align}
\widetilde{\mathcal{M}}_{p_1p_2p_3p_4,su}(s,\tilde{u};\alpha,\beta) =&\; \tau^{\frac{p_{21}}{2}} \widetilde{\mathcal{M}}_{p_1p_2p_4p_3,st}\left( s,t\,;\, \frac{\alpha}{\alpha-1},\frac{\beta}{\beta-1} \right)\;,\label{Mgenpst}\\
\widetilde{\mathcal{M}}_{p_1p_2p_3p_4,tu}(t,\tilde{u};\alpha,\beta) =&\; \sigma^{2-\frac{\Sigma_{13}}{2}}\tau^{-2+\frac{\Sigma_{23}}{2}} \widetilde{\mathcal{M}}_{p_1p_4p_2p_3,st}\left( s,t\,;\, \frac{\alpha-1}{\alpha},\frac{\beta-1}{\beta} \right)\;,
\end{align}
which implies the following crossing relations for the numerators
\begin{align}\label{eq:cmncrossing1}
c^{su}_{m,n}(\alpha,\beta)\big|_{p_1p_2p_3p_4} =&\; \tau^{\frac{p_{21}}{2}} c^{st}_{m,n}\!\left.\left( \frac{\alpha}{\alpha-1},\frac{\beta}{\beta-1} \right)\right|_{p_1p_2p_4p_3},\\
c^{tu}_{m,n}(\alpha,\beta)\big|_{p_1p_2p_3p_4} =&\; \sigma^{2-\frac{\Sigma_{13}}{2}}\tau^{-2+\frac{\Sigma_{23}}{2}}c^{st}_{m,n}\!\left.\left(\frac{\alpha-1}{\alpha},\frac{\beta-1}{\beta} \right)\right|_{p_1p_4p_2p_3}.\label{eq:cmncrossing2}
\end{align}

Let us make a few comments about this ansatz. The appearance of simultaneous poles is reasonable to expect in a large region accounting for the majority part of the ansatz. When $m$ is big enough such that the $s$ poles of (\ref{Mgenpst}) overlap with the poles of both s-channel Gamma functions (similar to those in $\widetilde{\mathcal{M}}_{2222}^{(2)}$), the integrand of (\ref{eq:mellindef}) develops triple poles and gives rise to leading $\log^2U$ singularity upon performing the $s$-integral. The leading logarithmic singularity has a rigid structure. It is known in \cite{Alday:2021odx} that the tree-level reduced Mellin amplitude enjoys an eight-dimensional hidden conformal symmetry which is similar to the ten dimensional hidden symmetry of $\mathrm{AdS}_5\times\mathrm{S}^5$ IIB supergravity \cite{Caron-Huot:2018kta}. Reasoning analogous to that of \cite{Caron-Huot:2018kta} shows that the one-loop leading logarithmic singularity for higher KK weights is also determined by the hidden conformal symmetry. This relation can be expressed precisely via the action of a differential operator and ultimately guarantees that the reduced Mellin amplitude has only simultaneous poles in the region responsible for the leading logarithmic singularity. 
	
Compared to the lowest weight case, a major generalization of the ansatz (\ref{Mgenpst}) is that it assumes the above simultaneous pole structure extends to the entire Mellin amplitude. This includes also the poles with are not present in $\widetilde{\mathcal{M}}_{2222}^{(2)}$. As will be illustrated in the next subsection, these poles are tied to the window and below window regions of the CFT spectrum, which includes the exchange of long operators with twists below threshold as well as protected operators. 
	
We should emphasize that at this point the validity of (\ref{Mgenpst}) in the window and below window regions is merely an \emph{assumption}, and we do not have any argument other than it being a natural extension from the above threshold area. However, as we will see, this minimal ansatz turns out to be correct and that only simultaneous poles exist is a general feature of one-loop Mellin amplitudes.

\subsection{Relating ansatz with CFT data}\label{sec:mellinpoles}
	
In this subsection, we explain how the Mellin amplitude ansatz \eqref{Mredgenp} encodes the CFT data. To show this, we compare the integral \eqref{eq:mellindef} with the conformal block decomposition discussed in section \ref{sec:scfblockdecom}. Because the three partial amplitudes in \eqref{Mredgenp} are distinguished by color factors in front and are related to each other by Bose symmetries, we can just focus on one of them, which we choose to be $\widetilde{\mathcal{M}}_{\{p_i\},st}$ in this paper.
	
\begin{figure}[ht]
\centering
		\begin{tikzpicture}
			\definecolor{naturetmpcr1}{HTML}{213363}
			\definecolor{naturetmpred}{HTML}{FA7070}
			\definecolor{naturegreen}{HTML}{00A686}
			\definecolor{cr1}{HTML}{fcf6bd}
			\definecolor{cr2}{HTML}{d0f4de}
			\definecolor{cr3}{HTML}{a9def9}
			\pgfmathsetmacro{\op}{0.7}
			\draw (3.5,4.5) -- ++(0,-0.5) node [anchor=south west] {$s$} -- +(0.5,0) ;
			\draw[->,color=naturetmpcr1 , line width=1pt] (-3,-1.5)--(-3,1.5) node[right] { }; 
			\draw[ color=naturetmpcr1 , line width=1pt] (-3,-1.5)--(-3,3.8) node[above] {initial contour}; 
			\draw[draw=cr3, fill=cr1, opacity=\op] (-2,-0.7) rectangle (0,0) ;
			\draw[draw=cr2, fill=cr2, opacity=\op] (0,-0.7) rectangle (2,0) ;
			\draw[draw=cr3, fill=cr3, opacity=\op] (2,-0.7) rectangle (4.2,0) ;
			\node[below] at (-1,0)  {I: $\log^0(U)$} ;
			\node[below] at (1,0)  {II: $\log^1(U)$} ;
			\node[below] at (3,0)  {$\quad$III: $\log^2(U)$} ;
			\draw[ color=naturegreen,line width=0.7pt,dashed] (-2,-1) node [below] {\color{black}$s_-$} --(-2,3.5) ; 
			\draw[ color=naturegreen,dotted , line width=0.7pt] (0,-1) node [below] {\color{black}$s_0$} --(0,3.5)  ;
			\draw[ color=naturegreen,dotted , line width=0.7pt] (2,-1) node [below] {\color{black}$s_+$} --(2,3.5)   ;
			\draw[-stealth] (-3.5,0)--(5,0) node [right] {$\mathfrak{Re}s$};
			\foreach \x in {-2,-1,0,1,2,3,4}
			\foreach \y in {1}
			{
				\fill[fill=black](\x,\y) circle (0.11cm);
				\fill[fill=cr1](\x,\y) circle (0.1cm);
			}
			\foreach \x in {-2,-1}
			\foreach \y in {0.8}
			{
				\draw[color=naturetmpred,line width=0.7,<-](\x,\y) arc [radius=0.2,start angle=-90,end angle=245];
			}
			\foreach \x in {0,1,2,3,4}
			\foreach \y in {2}
			{
				\fill[fill=black](\x,\y) circle (0.11cm);
				\fill[fill=cr2](\x,\y) circle (0.1cm);
			}
			\foreach \x in {0.2,1.2}
			\foreach \y in {2}
			{
				\draw[color=naturetmpred,line width=0.7,<-](\x,\y) arc [radius=0.2,start angle=0,end angle=180];
			}
			\foreach \x in {0.2,1.2}
			\foreach \y in {1}
			{
				\draw[color=naturetmpred,line width=0.7 ](\x,\y) arc [radius=0.2,start angle=0,end angle=-180];
			}
			\draw[color=naturetmpred,line width=0.7] (-0.2,1) -- (-0.2,2);
			\draw[color=naturetmpred,line width=0.7] (0.2,1) -- (0.2,1.8);
			\draw[color=naturetmpred,line width=0.7] (0.8,1) -- (0.8,2);
			\draw[color=naturetmpred,line width=0.7] (1.2,1) -- (1.2,1.8);
			\foreach \x in {2,3,4}
			\foreach \y in {3}
			{
				\fill[fill=black](\x,\y) circle (0.11cm);
				\fill[fill=cr3](\x,\y) circle (0.1cm);
			}
			\foreach \x in {2.2,3.2,4.2}
			\foreach \y in {3}
			{
				\draw[color=naturetmpred,line width=0.7,<-](\x,\y) arc [radius=0.2,start angle=0,end angle=180];
			}
			\foreach \x in {2.2,3.2,4.2}
			\foreach \y in {1}
			{
				\draw[color=naturetmpred,line width=0.7 ](\x,\y) arc [radius=0.2,start angle=0,end angle=-180];
			}
			\draw[color=naturetmpred,line width=0.7] (1.8,1) -- (1.8,3);
			\draw[color=naturetmpred,line width=0.7] (2.2,1) -- (2.2,2.8);
			\draw[color=naturetmpred,line width=0.7] (2.8,1) -- (2.8,3);
			\draw[color=naturetmpred,line width=0.7] (3.2,1) -- (3.2,2.8);
			\draw[color=naturetmpred,line width=0.7] (3.8,1) -- (3.8,3);
			\draw[color=naturetmpred,line width=0.7] (4.2,1) -- (4.2,2.8);
			\node [naturetmpred,anchor=south,rotate=-90] at (4.3,2) {deformed contours};
		\end{tikzpicture}
\caption{Three infinite series of $s$-poles which extend to the right in the complex $s$ plane. The poles in blue and green come from the Gamma functions and the poles in yellow are from the reduced Mellin amplitude. The poles in the same column actually overlap, and we separate them in order to show their degrees. The overlapping poles have degrees 1, 2, 3, respectively in region I, II, III. By deforming the $s$ integral contour to the right and picking up residues, these poles give rise to different contributions to the small $U$ expansion of the reduced correlator, which can be distinguished by the maximal power of $\log U$ (shown at the bottom of the figure). }
\label{fig:spole}
\end{figure}
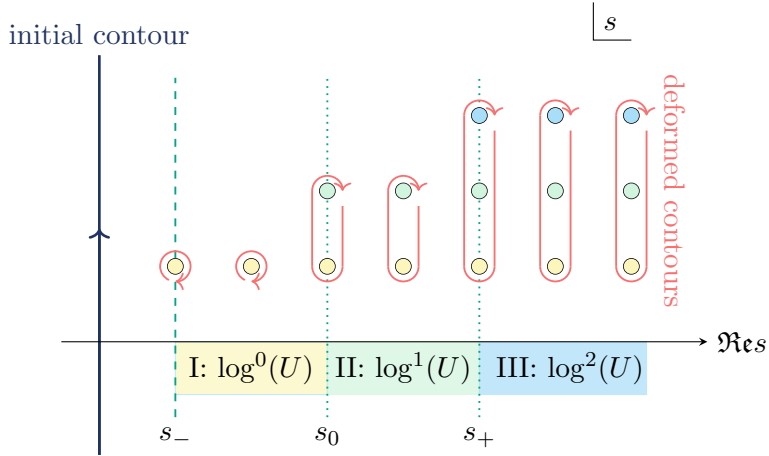

Without loss of generality let us first focus on the $s$-integral. In the integrand of \eqref{eq:mellindef} there are three infinite series of  poles in $s$ which extend to the right, as is illustrated in figure \ref{fig:spole}
\begin{align}
\widetilde{\mathcal{M}}_{\{p_i\},st}\quad &\Rightarrow\quad \frac{1}{s-s_0-2m},\;\quad m=-\mathcal{E}+2,-\mathcal{E}+1,...,\\
\Gamma(\frac{\Sigma_{12}-s}{2})\quad &\Rightarrow\quad \frac{1}{s-\Sigma_{12}-2m},\;\ m=0,1,...,\\
\Gamma(\frac{\Sigma_{34}-s}{2})\quad &\Rightarrow\quad \frac{1}{s-\Sigma_{34}-2m},\;\ m=0,1,....
\end{align}
Thanks to the R-symmetry selection rule and supersymmetry which relates conformal dimensions and R-symmetry charges, the two series of Gamma function poles overlap with the pole series from the Mellin amplitude after certain values of $s$, dividing the poles into three sets (labelled as I, II, III in figure \ref{fig:spole}). 
	
In these three regions, these $s$-poles are triple, double, and simple poles respectively. Let us now close the $s$-contour to the right and pick up their residues. For a pole at a specific location $s=s_*$ of order $k+1$ the resulting residue takes the form
\begin{equation}\label{eq:sintegral}
\int_{\mathcal{C}_*} \frac{\dd s}{2 \pi i}  \frac{U^{\frac{s+2}{2}}}{(s-s_*)^{k+1}} \,h(s)\ = \frac{1}{2^{k+1}}U^{\frac{s_*}{2}+1}\qty(\log^{k} U\,h(s_*)+\cdots),\quad k=0,1,2,
\end{equation}
where $h(s)$ represents factors that are regular in $s$, and $\mathcal{C}_*$ is a contour encircling the pole $s=s_*$. The $\cdots$ on RHS refers to terms with lower powers of $\log U$. Note that in such an integral the location $s_*$ of the pole determines the power of $U$, while the order of the pole determines the maximal power of $\log U$. The summation of all such residues can be treated as an expansion of $\mathcal{H}_{st}^{(2)}$ in the small $U$ limit.
	
To make connections with the CFT data, let us consider the exchange of a long operator with classical twist $\tau_0$ and spin $\ell$ in the s-channel in the same limit (and temporarily put protected operators aside). The corresponding superconformal block (in view of $\mathcal{H}$) and its derivative in the twist behave as
\begin{align}\label{eq:hsmallU}
\partial_{\tau_0}^{k} \mathfrak{h}_{\tau_0,R,\ell}(z,\bar{z};\alpha)& \sim \frac{1}{2^k} U^{\frac{\tau_0}{2}+1}\log^{k}U \,P_R(\alpha)\sum_{p=0}^\infty U^pf_{\tau_0,\ell}^{(p)}(V)+\cdots,\qquad k=0,1,2,
\end{align}
where $f_{\tau_0,\ell}^{(p)}(V)$ are some functions of $V$ that can be obtained by expanding the conformal block $g_{\tau_0+2,\ell}$, and $\cdots$ contains terms with lower powers in $\log U$. 
	
Now we compare \eqref{eq:sintegral} and \eqref{eq:hsmallU}. On the one hand, comparison in the power of $U$ indicates that the residue at $s=s_*$ receives contributions from long operators with twist $\tau_0\leq s_*$. Note that by the fusion rule \eqref{eq:fusion} and the unitarity bound for long multiplets discussed in section \ref{sec:scfblockdecom}, the long operators exchanged in the s-channel should have its twist $\tau\geq s_-$, where $s_- = s_0 -2\mathcal{E}+4=\max(|p_{21}|,|p_{34}|)+4$. This lower bound $s_-$ is in exact agreement with the location of the leftmost pole in our ansatz for Mellin amplitude \eqref{eq:mst}. 
	
On the other hand, the maximal powers of $\log U$ provides a map from the three regions of Mellin amplitude poles to the three regions of operator twists. We concluded in section \ref{sec:scfblockdecom} that contributions to $\mathcal{H}^{(2)}$ from long operators of different twist regions differ in the maximal power of $\log U$. Therefore, the explicit map is as follows: 
\begin{itemize}
\item {\bf Region III ($s\geq s_+ \equiv \max(\Sigma_{12},\Sigma_{34})$):} Poles in this region are triple poles. They necessarily correspond to long operators with twists in the above threshold region $\tau_0\geq s_+$, as their residues include terms with $\log^2U$, which are the leading log singularities at one loop. Note that in the expansion \eqref{eq:H2expansion} these $\log^2U$ singularities only depends on the coefficients $\langle a^{(0)}(\gamma^{(1)})^2 \rangle$. So in principle these data determines the $c_{m,n}^{st}$ coefficients of poles in this region. When matching the two side, each residue in \eqref{eq:sintegral} (with $k=2$) receives contributions in \eqref{eq:hsmallU} only for $s_+\leq\tau_0\leq s_*$.
\item {\bf Region II ($s_0 \leq s < s_+$):} Poles in this region are double poles. They have to correspond to long operators with twists in the window region, since their residues have maximal power one in $\log U$, with leading twists that are below $s_+$. Recall \eqref{eq:a0vanish} that for operators in the window region $a^{(0)}=0$, and so in the expansion \eqref{eq:H2expansion} coefficients for these $\log U$ singularities turn out to depend on $\langle a^{(1)}\gamma^{(1)}\rangle$ only, which serves to constrain the $c_{m,n}^{st}$ coefficients of poles in this region. When matching the two side, each residue in \eqref{eq:sintegral} (with $k=1$) receives contributions in \eqref{eq:hsmallU} only for $s_0\leq\tau_0\leq s_*$.
\item {\bf Region I ($\max(|p_{21}|,|p_{34}|)+4 \equiv s_- \leq s < s_0$):} Poles in this region are simple poles, which do not lead to any $\log U$ singularities in the $s$ integral. They correspond to long operators with twists in the below window region, because their residues have the power of $U$ that matches twists below $s_0$. Since operators in the below window region have both $a^{(0)}=a^{(1)}=0$, in the expansion \eqref{eq:H2expansion} only the first term remains, with coefficients $\langle a^{(2)}\rangle$. These data constrain the $c_{m,n}^{st}$ coefficients of poles in this region. When matching the two side, each residue in \eqref{eq:sintegral} (with $k=0$) receives contributions in \eqref{eq:hsmallU} for $s_-\leq\tau_0\leq s_*$.
\end{itemize}
	
When taking the protected operators into consideration as well, recall that in the perturbative expansion they never lead to $\log U$ singularities. Therefore, when they are present in the s-channel, the poles in region I are associated not only to long operators in the below window region but also to these protected operators. However, unlike the long operators the power of $U$ in the residues cannot be directly interpreted as their twists (see appendix \ref{appx:superblocks}). In this case we also need data of the coefficients $\langle b^{(2)}\rangle$ and $\langle c^{(2)}\rangle$, as can be seen in \eqref{eq:H2protectedblocks}. Their constraints on the coefficients of poles in region I has to be considered whenever there is a non-trivial below window region, corresponding to conditions on the extremality $\mathcal{E}>2$.
	
To summarize the above discussions, along the s-channel the connections between twist regions, operators, CFT data, terms in the small $U$ expansion, and s-channel poles of $\widetilde{\mathcal{M}}_{\{p_i\},st}$, are sketched in table \ref{twist2pole}. A subtlety occurs for the semi-short operators sitting at the lower bound of the window region. Their twists fall into the window region, but the Mellin poles relating to them are in region I. This is because (as mentioned above) the twist region refers to the physical twist of the operators, but the position of the poles corresponds to the parameter $\tau$ of $\mathfrak{h}_{\tau, R, \ell}$ that appears in the decomposition \eqref{eq:Cblock} of semi-short superconformal blocks, which differs from the physical twist of protected multiplets.
\begin{table}[t]
		\begin{tabular}{c|c|c|c|c}
			\toprule
			twist region & operators & data in $\mathcal{H}^{(2)}$ & $U$ expansion of $\mathcal{H}^{(2)}$ & $\frac{1}{(s-s_*)(t-t_*)}\subset\widetilde{\mathcal{M}}_{st}$ \\
			\midrule
			above threshold & long & $\langle a^{(0)}(\gamma^{(1)})^2\rangle$ & $\displaystyle U^{\frac{\tau_0\geq s_+}{2}+1}\log^2U$ & III: $s_*\geq s_+$ \\ 
			\midrule
			window & long & $\langle a^{(1)}\gamma^{(1)}\rangle$ & $\displaystyle U^{\frac{s_0\leq\tau_0< s_+}{2}+1}\log U$ & II: $s_0\leq s_*< s_+$\\
			\midrule
			\multirow{2}{*}{below window} & long & $\langle a^{(2)}\rangle$ & \multirow{2}{*}{$\displaystyle U^{\frac{s_-\leq\tau_0< s_0}{2}+1}$} & \multirow{2}{*}{I: $s_-\leq s_*< s_0$} \\
			& protected & $\langle b^{(2)}\rangle$, $\langle c^{(2)}\rangle$ & & \\
			\bottomrule
		\end{tabular}
\caption{From twist spectrum to poles in Mellin amplitude in the s-channel.}
\label{twist2pole}
\end{table}
	
\begin{figure}[ht]
\centering
		\begin{tikzpicture}
			\definecolor{cr1}{HTML}{fcf6bd}
			\definecolor{cr2}{HTML}{d0f4de}
			\definecolor{cr3}{HTML}{a9def9}
			\pgfmathsetmacro{\op}{0.6}
			\draw[draw=cr1, fill=cr1, opacity=\op] (-2,-2) rectangle (3,3) ;
			\draw[draw=cr2, fill=cr2, opacity=\op] (0,-2) rectangle (3,3) ;
			\draw[draw=cr3, fill=cr3, opacity=\op] (2,-2) rectangle (3,3) ;
			\draw[color=black, thin, dashed] (-2,-2) --(-2,3.35) ;
			\draw[color=black, thin, dashed] (-2,-2) --(3.35,-2) ;
			\draw[-stealth] (-2.5,0)--(3.5,0) node[right] {$s$} ;
			\draw[-stealth] (0,-2.5)--(0,3.5) node[above] {$t$}; 
			\node [anchor=north] at (-0.05,0) {$(s_0,t_0)$};
			\foreach \x in {-2,-1,0,1,2,3}
			\foreach \y in {-2,-1,0,1,2,3}
			{
				\fill[fill=black](\x,\y) circle (0.11cm);
				\fill[fill=white](\x,\y) circle (0.1cm);
			}
			\foreach \x in {-2,-1}
			\foreach \y in {-2,-1,0,1,2,3}
			{
				\fill[fill=cr1](\x,\y) circle (0.1cm);
			}
			\foreach \x in {0,1}
			\foreach \y in {-2,-1,0,1,2,3}
			{
				\fill[fill=cr2](\x,\y) circle (0.1cm);
			}
			\foreach \x in {2,3}
			\foreach \y in {-2,-1,0,1,2,3}
			{
				\fill[fill=cr3](\x,\y) circle (0.1cm);
			}
			\node[below] at (-2,-2.5) {$s_-$} ;
			\node[below] at (0,-2.5) {$s_0$} ;
			\node[below] at (2,-2.5) {$s_+$} ;
		\end{tikzpicture}
		~
		\begin{tikzpicture}
			\definecolor{cr1}{HTML}{fcf6bd}
			\definecolor{cr2}{HTML}{d0f4de}
			\definecolor{cr3}{HTML}{a9def9}
			\pgfmathsetmacro{\op}{0.6}
			\fill [cr1, opacity=\op] (-2,-2) rectangle (3,3) ;
			\fill [cr2, opacity=0.3] (0,-2) rectangle (3,0) ;
			\fill [cr2  ,opacity=0.3](-2,0) rectangle (0,3);
			\fill [cr2  ,opacity=\op](0,0) rectangle (3,3);
			\fill [cr3, opacity=0.2] (2,-2) rectangle (3,0) ;
			\fill [cr3  ,opacity=0.2](-2,2) rectangle (0,3);
			\fill [cr3, opacity=0.3] (2,0) rectangle (3,2) ;
			\fill [cr3, opacity=0.3] (0,2) rectangle (2,3) ;
			\fill [cr3  ,opacity=\op](2,2) rectangle (3,3);
			\draw[color=black,thin,dashed] (-2,-2) --(-2,3.35) ;
			\draw[color=black,thin,dashed] (-2,-2) --(3.35,-2) ;
			\draw[-stealth] (-2.5,0)--(3.5,0) node[right] {$s$};
			\draw[-stealth] (0,-2.5)--(0,3.5) node[above] {$t$}; 
			\node [anchor=north] at (-0.05,0) {$(s_0,t_0)$};
			\foreach \x in {-2,-1,0,1,2,3}
			\foreach \y in {-2,-1,0,1,2,3}
			{
				\fill[fill=black](\x,\y) circle (0.11cm);
				\fill[fill=white](\x,\y) circle (0.1cm);
			}
			\foreach \x in {-2,-1,...,3} \foreach \y in {-2,-1,...,3} \fill [cr1,opacity=\op] (\x,\y) circle (0.1cm);
			\foreach \x in {0,1,2,3} \foreach \y in {-2,-1} \fill [cr2,opacity=0.3] (\x,\y) circle (0.1cm);
			\foreach \x in {-2,-1} \foreach \y in {0,1,2,3} \fill [cr2,opacity=0.3] (\x,\y) circle (0.1cm);
			\foreach \x in {0,1,2,3} \foreach \y in {0,1,2,3} \fill [cr2,opacity=\op] (\x,\y) circle (0.1cm);
			\foreach \x in {2,3} \foreach \y in {-2,-1} \fill [cr3,opacity=0.2] (\x,\y) circle (0.1cm);
			\foreach \x in {-2,-1} \foreach \y in {2,3} \fill [cr3,opacity=0.2] (\x,\y) circle (0.1cm);
			\foreach \x in {2,3} \foreach \y in {0,1} \fill [cr3,opacity=0.3] (\x,\y) circle (0.1cm);
			\foreach \x in {0,1} \foreach \y in {2,3} \fill [cr3,opacity=0.3] (\x,\y) circle (0.1cm);
			\foreach \x in {2,3} \foreach \y in {2,3} \fill [cr3,opacity=\op] (\x,\y) circle (0.1cm);
			\node[below] at (-2,-2.41) {$s_-$} ;
			\node[below] at (0,-2.5) {$s_0$} ;
			\node[below] at (2,-2.5) {$s_+$} ;
			\node[left] at (-2.3,2) {$\ t_+$} ;
			\node[left] at (-2.3,0) {$t_0\ $} ;
			\node[left] at (-2.3,-2) {$\ t_-$} ;
		\end{tikzpicture}
\caption{The pole structures of $\widetilde{M}_{\{p_i\},st}$ on $s$-$t$ plane, with $p_1=8,\ p_2=8,\ p_3=4,\ p_4=8$ as an example. The dots represent the possible locations of the simultaneous poles in the ansatz. In the left figure, the below-window region (yellow) $s_- \leq s < s_0$ contains only simple poles in $s$, while the window region (green) $s_0 \leq s < s_+$ has double poles, and the above-threshold region (blue) $s \geq s_+$ has triple poles. In the right figure, similar regions are defined for $t$ poles and combined with the regions for $s$ poles.}
\label{fig:stpole}
\end{figure}
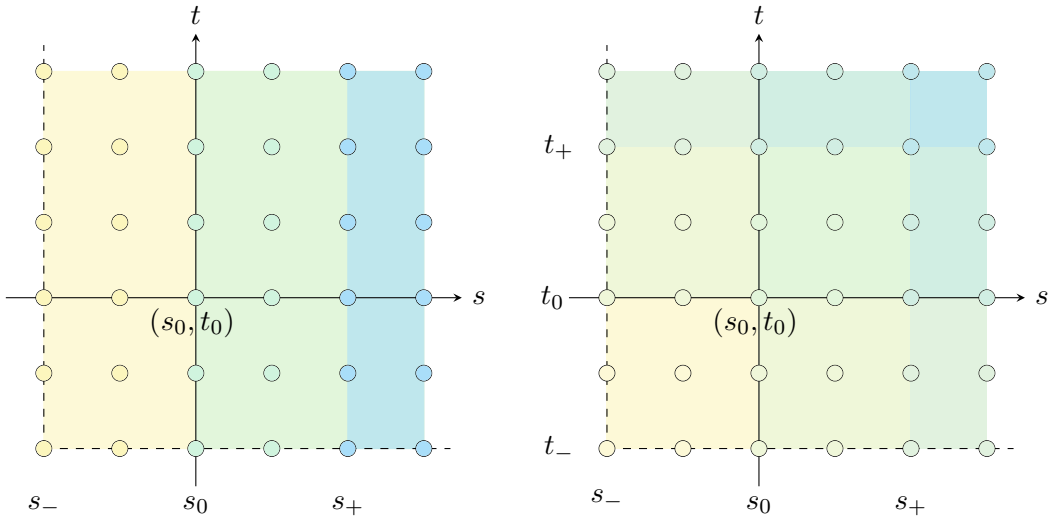

Let us now turn to the $t$-integral. Analogous to the s-channel, we define
\begin{equation}
t_-=\max(|p_{23}|,|p_{14}|)+4\;,\quad t_0=\min(\Sigma_{23},\Sigma_{14})\;, \quad t_+=\max(\Sigma_{23},\Sigma_{14})\;,
\end{equation}
and they separate t-channel poles into three sets as well (see figure \ref{fig:stpole}). In addition to $U\ll 1$, when $V\ll 1$ we can further close the $t$ contour to the right and pick up the residues. The reduced correlator $\mathcal{H}_{\{p_i\},st}^{(2)}$, represented as a two-fold inverse Mellin integral \eqref{eq:mellindef} thus turns into a series expansion with respect to small $U$ and $V$. Clearly, the residue of each simultaneous pole in $\widetilde{\mathcal{M}}_{\{p_i\},st}$ contains a term with maximal power of $\log U$ and $\log V$ and leading twist in the power of $U$ and $V$, following the correspondence
\begin{equation}\label{eq:pole2uv}
\widetilde{\mathcal{M}}_{st}\supset\frac{1}{(s-s_0-2m)(t-t_0-2n)}\ \Leftrightarrow\  \mathcal{H}_{st}^{(2)}\supset U^{\frac{s_0}{2}+m+1}V^{\frac{t_0}{2}-\frac{\Sigma_{23}}{2}+n}\log^p(U)\log^q(V)\;,
\end{equation}
where $p=0,1,2$ for the pole in s-channel region I, II, III, respectively, and similarly for $q$ in different t-channel parts. As will be discussed in section \ref{sec:recursion}, the terms on the RHS of \eqref{eq:pole2uv} can be independently worked out by the recursive use of CFT data at lower orders of the bulk loop expansion. These terms then serve to determine the $c_{m,n}^{st}$ coefficients in our ansatz. This observation forms the basis of the main strategy that we use in bootstrapping the Mellin amplitudes at the one-loop level, which will be further described in the next section. 

\section{Bootstrap algorithm}\label{sec:bootstrap}
	
In this section, we describe how to systematically bootstrap the one-loop Mellin amplitudes for arbitrary KK modes, starting from our ansatz \eqref{eq:mst}
\begin{equation}\label{eq:mst2}
\widetilde{\mathcal{M}}_{\{p_i\},st} = \sum_{m,n=-\mathcal{E}+2}^{\infty} \frac{c_{m,n}^{st}(\alpha,\beta)}{(s-s_{0}-2m)(t-t_{0}-2n)}\;.
\end{equation}
Explicit examples illustrating this algorithm will be provided in the next section. Readers who are eager to see the general structures of the resulting one-loop Mellin amplitudes can directly go to section \ref{sec:oneloopgeneral}.  To simplify the discussion, we will temporarily ignore the color group $G_F$ and flavor group $SU(2)_L$ in our theory at first, and will incorporate them at the end of section \ref{sec:recursion}.

\subsection{Outline of the algorithm}\label{sec:algorithm}
	
As we saw in the previous section, to determine all the residues in \eqref{eq:mst}, we must identify terms with the maximal power of $\log U$ in three different regions of twists, i.e., $\log^2U$ in the above-threshold, $\log^1 U$ in the window, and $\log^0U$ in the below-window region. Correspondingly we need to obtain three types of OPE data in the s-channel
\begin{subequations}\label{eq:dataforbootstrap}
\begin{alignat}{2}
&\langle a^{(0)}(\gamma^{(1)})^2 \rangle\;,\qquad&& \tau_0 \geq s_+\;,\\
&\langle a^{(1)}\gamma^{(1)} \rangle,\qquad&&  s_0 \leq \tau_0 < s_+\;,\\
&\langle a^{(2)} \rangle\;,\langle b^{(2)} \rangle\;,\langle c^{(2)} \rangle\;,\qquad&&  s_- \leq \tau_0 < s_0\;.
\end{alignat}
\end{subequations}
These data can be extracted from the lower-level correlators, namely disconnected and tree-level correlators. We will discuss this in section \ref{sec:recursion}. For the moment let us assume that the data \eqref{eq:dataforbootstrap} are already available, and we will outline the algorithm that utilize them to completely solve the ansatz \eqref{eq:mst2}.
	
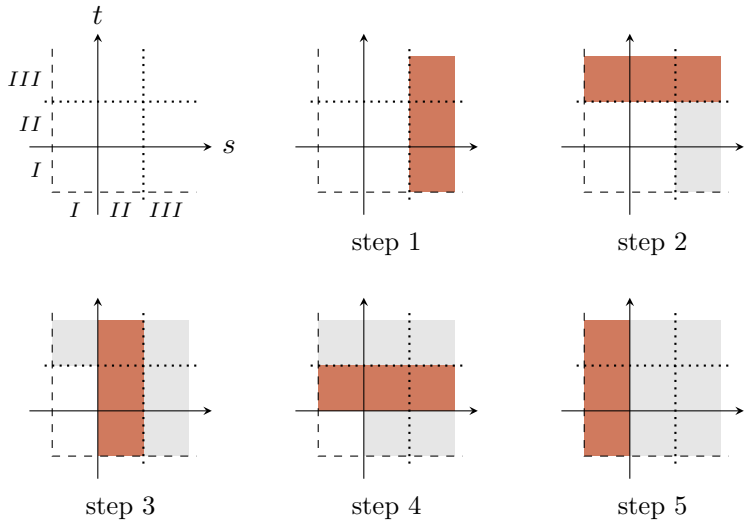
\begin{figure}[ht]
\centering
		\begin{tikzpicture}
			\begin{scope}
				\draw [-stealth] (-0.9,0) -- (1.5,0) node [right] {$s$};
				\draw [-stealth] (0,-0.9) -- (0,1.5) node [above] {$t$};
				\draw [dashed] (-0.6,1.3) -- (-0.6,-0.6) -- (1.3,-0.6);
				\draw [thick,dotted] (-0.7,0.6) -- (1.3,0.6) (0.6,-0.7) -- (0.6,1.3);
				\node [anchor=north] at (-0.3,-0.6) {\scriptsize $I$};
				\node [anchor=north] at (0.3,-0.6) {\scriptsize $II$};
				\node [anchor=north] at (0.9,-0.6) {\scriptsize $III$};
				\node [anchor=east] at (-0.6,-0.3) {\scriptsize $I$};
				\node [anchor=east] at (-0.6,0.3) {\scriptsize $II$};
				\node [anchor=east] at (-0.6,0.9) {\scriptsize $III$};
			\end{scope}
			\begin{scope}[xshift=3.5cm]
				\fill [BrickRed!60] (0.6,-0.6) rectangle (1.2,1.2);
				\draw [-stealth] (-0.9,0) -- (1.5,0);
				\draw [-stealth] (0,-0.9) -- (0,1.5);
				\draw [dashed] (-0.6,1.3) -- (-0.6,-0.6) -- (1.3,-0.6);
				\draw [thick,dotted] (-0.7,0.6) -- (1.3,0.6) (0.6,-0.7) -- (0.6,1.3);
				\node [anchor=north] at (0.3,-1) {\small step 1};
			\end{scope}
			\begin{scope}[xshift=7cm]
			\fill [gray!20] (0.6,-0.6) rectangle (1.2,1.2);
			\fill [BrickRed!60] (-0.6,0.6) rectangle (1.2,1.2);
			\draw [-stealth] (-0.9,0) -- (1.5,0);
			\draw [-stealth] (0,-0.9) -- (0,1.5);
			\draw [dashed] (-0.6,1.3) -- (-0.6,-0.6) -- (1.3,-0.6);
			\draw [thick,dotted] (-0.7,0.6) -- (1.3,0.6) (0.6,-0.7) -- (0.6,1.3);
			\node [anchor=north] at (0.3,-1) {\small step 2};
			\end{scope}
			\begin{scope}[yshift=-3.5cm]
			\fill [gray!20] (0.6,-0.6) rectangle (1.2,1.2);
			\fill [gray!20] (-0.6,0.6) rectangle (1.2,1.2);
			\fill [BrickRed!60] (0,-0.6) rectangle (0.6,1.2);
			\draw [-stealth] (-0.9,0) -- (1.5,0);
			\draw [-stealth] (0,-0.9) -- (0,1.5);
			\draw [dashed] (-0.6,1.3) -- (-0.6,-0.6) -- (1.3,-0.6);
			\draw [thick,dotted] (-0.7,0.6) -- (1.3,0.6) (0.6,-0.7) -- (0.6,1.3);
			\node [anchor=north] at (0.3,-1) {\small step 3};
			\end{scope}
			\begin{scope}[xshift=3.5cm,yshift=-3.5cm]
			\fill [gray!20] (0.6,-0.6) rectangle (1.2,1.2);
			\fill [gray!20] (-0.6,0.6) rectangle (1.2,1.2);
			\fill [gray!20] (0,-0.6) rectangle (0.6,1.2);
			\fill [BrickRed!60] (-0.6,0) rectangle (1.2,0.6);
			\draw [-stealth] (-0.9,0) -- (1.5,0);
			\draw [-stealth] (0,-0.9) -- (0,1.5);
			\draw [dashed] (-0.6,1.3) -- (-0.6,-0.6) -- (1.3,-0.6);
			\draw [thick,dotted] (-0.7,0.6) -- (1.3,0.6) (0.6,-0.7) -- (0.6,1.3);
			\node [anchor=north] at (0.3,-1) {\small step 4};
			\end{scope}
			\begin{scope}[xshift=7cm,yshift=-3.5cm]
			\fill [gray!20] (0.6,-0.6) rectangle (1.2,1.2);
			\fill [gray!20] (-0.6,0.6) rectangle (1.2,1.2);
			\fill [gray!20] (0,-0.6) rectangle (0.6,1.2);
			\fill [gray!20] (-0.6,0) rectangle (1.2,0.6);
			\fill [BrickRed!60] (-0.6,-0.6) rectangle (0,1.2);
			\draw [-stealth] (-0.9,0) -- (1.5,0);
			\draw [-stealth] (0,-0.9) -- (0,1.5);
			\draw [dashed] (-0.6,1.3) -- (-0.6,-0.6) -- (1.3,-0.6);
			\draw [thick,dotted] (-0.7,0.6) -- (1.3,0.6) (0.6,-0.7) -- (0.6,1.3);
			\node [anchor=north] at (0.3,-1) {\small step 5};
			\end{scope}
		\end{tikzpicture}
\caption{Regions of poles solved in each step of the algorithm. The red regions contain poles that are newly solved in each step (including those overlapping with the previous steps), while the gray regions contain poles that are solved in previous steps, and the white regions contain poles that are not yet determined.}
\label{fig:algorithm}
\end{figure}

The algorithm consists of six steps and each step solves a part of the ansatz. This is schematically illustrated in figure \ref{fig:algorithm}, and the details are as follows.
\begin{enumerate}
\item We first compute the simultaneous pole coefficients in region III in the s-channel. Taking residues at $s=s_*\geq s_+$ gives rise to terms proportional to $U^{\frac{s_*}{2}+1}\log^2U$. From the perturbative expansion \eqref{eq:H2expansion} and the small $U$ approximation \eqref{eq:hsmallU} we see that these terms can be computed from the CFT side by (for each $R$ representation)
\begin{equation}\label{eq:logU2cftsumpre}
U^{\frac{s_*}{2}+1}\log^2U\,\sum_{\tau_0=s_+}^{s_*}\sum_{\ell=0}^\infty\frac{1}{8}\langle a^{(0)}(\gamma^{(1)})^2\rangle_{\tau_0,R,\ell}\,f_{\tau_0,\ell}^{(s_*-\tau_0)}(V)\;.
\end{equation}
However, deriving each $f_{\tau_0,\ell}^{(s_*-\tau_0)}$ from $\mathfrak{h}_{\tau_0,R,\ell}$ is usually difficult  and performing different summations for different values of $s_*$ is very inefficient. In practice we can alternatively perform the summation
\begin{equation}\label{eq:logU2cftsum}
\log^2U\,\sum_{\tau_0=s_+}^{\infty}\sum_{\ell=0}^\infty\frac{1}{8}\langle a^{(0)}(\gamma^{(1)})^2\rangle_{\tau_0,R,\ell}\,\mathfrak{h}_{\tau_0,R,\ell}\;,
\end{equation}
and its order $U^{\frac{s_*}{2}+1}$ piece in the small $U$ expansion is equivalent to \eqref{eq:logU2cftsumpre}. This allows us to organize the computations for all $s_*\geq s_+$ into a single resummation of conformal blocks and a single small $U$ expansion (the feasibility of the above infinite summation will be discussed soon). Similar strategy will be applied in subsequent steps as well. To move on with the computation, note that on the Mellin side we have infinitely many poles in $t$, which also fall into three different regions in the t-channel. Through the $t$ integral a pole at $t=t_*$ will develop terms with maximal power in $\log V$ of the form
\begin{equation}\label{eq:logU2Vpattern}
V^{\frac{t_*-\Sigma_{23}}{2}}
\begin{cases}
\log^2V,&t_*\geq t_+\;,\\
\log^1V,&t_0\leq t_*<t_+\;,\\
\log^0V,&t_-\leq t_*<t_0\;.
\end{cases}
\end{equation}
As mentioned in section \ref{sec:mellinpoles} the presence the t-channel poles is tied to the infinite support on spin in the s-channel OPE. Hence in order to correctly extract the $V$ dependence in \eqref{eq:logU2Vpattern}, we first need to perform the resummation in spin $\ell$ in \eqref{eq:logU2cftsum} and then expand the result in the small $V$ limit. Matching each pattern of \eqref{eq:logU2Vpattern} then determines the coefficient $c_{\frac{s_*-s_0}{2},\frac{t_*-t_0}{2}}^{st}$, for the range $s_*\geq s_+$ and $t_*\geq t_-$.
\item Note that  $\widetilde{\mathcal{M}}_{\{p_i\},st}$ has Bose symmetry under $p_1\leftrightarrow p_3$, which interchanges $s$ and $t$. So we can also perform the same analysis as the previous step but on $\widetilde{\mathcal{M}}_{p_3p_2p_1p_4,st}$. Through crossing this determines the coefficients in the original $\widetilde{\mathcal{M}}_{p_1p_2p_3p_4,st}$ for the range $s_*\geq s_-$ and $t_*\geq t_+$. The explicit relation is
\begin{equation}\label{eq:Mstcrossing}
\widetilde{\mathcal{M}}_{p_1p_2p_3p_4,st}(s,t;\alpha,\beta)=\sigma^{-\frac{\Sigma_{12}}{2}+1}\tau^{\frac{\Sigma_{23}}{2}-2}\widetilde{\mathcal{M}}_{p_3p_2p_1p_4}(t,s;1-\alpha,1-\beta)\;,
\end{equation}
which can be derived from \eqref{eq:Hcrossing} ($\sigma,\tau$ being the cross ratios defined in \eqref{eq:defst}). Note that there is an overlapping range, i.e., $s_*\geq s_+$ and $t_*\geq t_+$, which both steps independently determines. This provides a first consistency check on the computation. By now these two steps already solve most of the ansatz, leaving only finitely many coefficients unfixed.
\item We then consider the simultaneous poles that belong to region II in the s-channel, $s_0\leq s< s_+$. Taking residues at $s=s_*$ in this range yields terms proportional to $U^{\frac{s_*}{2}+1}\log U$. On the CFT side such terms are produced by, following the same treatment as in \eqref{eq:logU2cftsum}, the order $U^{\frac{s_*}{2}+1}$ piece in the small $U$ expansion of the summation (for each $R$ representaion)
\begin{equation}\label{eq:logU1cftsum}
\log U\,\sum_{\tau_0=s_0}^{s_+-2}\sum_{\ell=0}^\infty\frac{1}{2}\langle a^{(1)}\gamma^{(1)}\rangle_{\tau_0,R,\ell}\,\mathfrak{h}_{\tau_0,R,\ell}\;.
\end{equation}
The analysis on the accompanying t-channel poles and the $t$ integral is the same as that in step 1. In other words, we again need to resum the spin in \eqref{eq:logU1cftsum} and expand the result for small $V$, and then search and match terms with the pattern in \eqref{eq:logU2Vpattern}. This determines the Mellin coefficients $c_{\frac{s_*-s_0}{2},\frac{t_*-t_0}{2}}^{st}$ for the range $s_0\leq s_*<s_+$ and $t_*\geq t_-$. Among these the range $t_*\geq t_+$ also provides another consistency check against the result from step 2.
\item Similar to step 2, we can perform the same analysis as step 3 but for $\widetilde{\mathcal{M}}_{p_3p_2p_1p_4,st}$. Then by crossing this determines the coefficients in the original $\widetilde{\mathcal{M}}_{p_1p_2p_3p_4,st}$ for the range $s_*\geq s_-$ and $t_0\leq t_*<t_+$. Further consistency checks also apply here, but we will refrain from listing them in detail.
\item To fix the remaining coefficients for the range $s_-\leq s<s_0$ and $t_-\leq t<t_0$, we take the residues in both $s$ and $t$ at these simultaneous poles. There is no s-channel logarthimic singularity and the CFT calculation involves the OPE coefficients $\langle a^{(2)} \rangle$, $\langle b^{(2)} \rangle$, $\langle c^{(2)} \rangle$. To be explicit,  by \eqref{eq:H2expansion} and \eqref{eq:H2protectedblocks} here we need to compute the summation
\begin{equation}\label{eq:step5}
\underbrace{\frac{z\bar{z}\alpha^2}{(z-\alpha)(\bar{z}-\alpha)}\qty(\sum_R \langle b^{(2)}\rangle_R\mathcal{B}_R+\sum_{R,\ell}\langle c^{(2)}\rangle_{R,\ell}\mathcal{C}_{R,\ell})}_{\rm protected}+\underbrace{\sum_{\tau_0=s_-}^{s_0-2}\sum_{R,\ell}\langle a^{(2)}\rangle_{\tau_0,R,\ell}\mathfrak{h}_{\tau_0,R,\ell}}_{\rm long}\;,
\end{equation}
and extract terms with the power $U^{\frac{s_*}{2}+1}V^{\frac{t_*-\Sigma_{23}}{2}}\log^kV$  in the small $U$ and $V$ limit, which determines the coefficient at $s=s_*$ ($s_-\leq s_*<s_0$) and $t=t_*\geq t_-$. Among these, the comparison at $t_*<t_0$ ($k=0$) solves the remaining coefficients, while that at $t_*\geq t_0$ ($k>0$) provides further checks against previous steps. Of course, one can alternatively carry out a similar analysis in the t-channel.
\item Finally, note that the resummation of s-channel blocks \eqref{eq:logU2cftsum}, \eqref{eq:logU1cftsum} and \eqref{eq:step5} contains terms subleading in the power of $\log V$ in the t-channel apart from the maximal ones listed in \eqref{eq:logU2Vpattern}. For example, in the t-channel above threshold region $t_*\geq t_+$ there are also terms with $\log^{<2}V$. In order to further confirm the bootstrap results from the previous steps we should also check that these subleading terms in the small $V$ expansion are correctly produced by the Mellin amplitude as well (In this check we need to consider the entire contribution of the Mellin amplitude, instead of picking up only a particular simultaneous pole as we did for the leading terms). Since the presence of any extra individual poles of the form $\frac{1}{s-s_*}$ (without dependence on $t$) will affect these subleading terms, this check also serves to confirm that there are actually no individual poles of this type (and by crossing no individual $\frac{1}{t-t_*}$ poles either), and so the Mellin amplitude indeed only contains simultaneous poles.
\end{enumerate}
	
A few comments are in order. First, an important part of the algorithm is to compute the logarithmic singularities. As indicated by the angle brackets, to use the tree-level data at one loop one needs to take into account the phenomenon of operator mixing. The details of how this data is unmixed from the tree-level correlators and used in the CFT calculation will be discussed in section \ref{sec:recursion}. 
	
Second, so far we have only presented the algorithm in its most basic version. There are additional structures of the Mellin amplitudes which allow us to further streamline the algorithm. In  step 1 and 2 of the algorithm the CFT computation of the leading logarithmic $\log^2U$ can be facilitated by using an eight-dimensional hidden conformal symmetry of the tree-level amplitude \cite{Alday:2021odx}. The use of this hidden symmetry allows us to avoid the computation of data for infinitely many twists in the above threshold region to determine the infinitely many coefficients in $\widetilde{\mathcal{M}}_{\{p_i\},st}$. We will show in section \ref{sec:hiddenconfsymm} how hidden symmetry simplifies the implementation of the above algorithm by following a similar strategy used in \cite{Alday:2019nin} for supergravity one-loop amplitudes. We also mention in advance that quite surprisingly, the hidden conformal symmetry provides a further simplification by determining a larger part of the ansatz. We will postpone explaining the precise meaning of this statement until we consider the explicit examples in section \ref{sec:examples}. This extra simplification will be taken into account in section \ref{sec:oneloopgeneral} when we discuss the structure of the general one-loop Mellin amplitudes. 
	
Third, let us note that there are overlaps between the set of poles determined in each step. While this overlap provides non-trivial consistency checks of the computation, in practice if the purpose is to solve the ansatz as quickly as possible, then it is not necessary to strictly follow all the steps. The coefficient of a pole can be solved either in the s-channel or in the t-channel. Depending on the specific situation, one way is usually more convenient than the other. We will see this in the explicit examples in the next section.
	
Finally, let us mention that the checks in step 6 only rule out the existence of other singularities. Additional regular terms in the Mellin amplitude are not excluded.  However, we can constrain them  by considering its flat-space limit. The discussion follows exactly that of \cite{Alday:2021ajh} for the $p_i=2$ case and the only regular part is a constant ambiguity which corresponds to a UV counter term.

\subsection{Operator mixing and unitarity recursion}\label{sec:recursion}
	
The discussion of operator mixing and unitarity recursion of SYM in AdS$_5\times$S$^3$ largely follows that for supergravity in $\mathrm{AdS}_5\times\mathrm{S}^5$ in \cite{Aprile:2019rep}. As we mentioned earlier, the OPE data needed is  contributed by double-particle operators $[pq]_{\tau,R,\ell}$. These double-particle operators are dual to double-particle states formed by the single-particle states $\mathcal{O}_p$ and $\mathcal{O}_q$ and fall into three types
\begin{equation}
[pq]_{\tau,R,\ell} = \begin{cases}
\text{half-BPS } \mathcal{B}_{R},&\quad R = \frac{p+q}{2}\;,\\
\text{semi-short } \mathcal{C}_{R,\ell},&\quad R = \frac{p+q}{2}-1\;,\\
\text{long } \mathcal{A}_{\tau,R,\ell},&\quad \frac{|p-q|}{2}\leq R\leq \frac{p+q}{2}-2\ \text{  and  }\ \tau \geq p+q\;.
\end{cases}
\end{equation}
The three-point coefficients of the double-particle operators $[pq]_{\tau,R,\ell}$ with operators $\mathcal{O}_p$ and $\mathcal{O}_q$ is non-vanishing at the disconnected level
\begin{equation}
C_{pq,[pq]} = C^{(0)}_{pq,[pq]} + a_F\, C^{(1)}_{pq,[pq]} + a_F^2\, C^{(2)}_{pq,[pq]} + \cdots\;,
\end{equation}
since we can form double-particle states by combining two single-particle states even when there is no interaction. Additionally, we have
\begin{equation}
C^{(0)}_{pq,[rs]} = 0\;,\quad p,q\neq r,s\;.
\end{equation}
	
There are several double-particle operators with the same classical quantum numbers, such as $[pq]_{\tau_0,R,\ell}$ and $[(p-1)(q+1)]_{\tau_0,R,\ell}$. When these double-particle operators are long operators, they will generally not be eigenstates of the dilation operator and will heavily mix with each other.\footnote{By ``heavily mix'', we mean that the elements of the mixing matrix among these double-particle operators are $\order{1}$.} If we represent the true eigenstates of the dilation operator by $(\mathcal{K}_i)_{\tau_0,R,\ell}$, where $i$ labels these different eigenstate operators with the same classical quantum numbers, then the three-point coefficients of these $(\mathcal{K}_i)_{\tau_0,R,\ell}$ will be
\begin{equation}\label{eq:Cseries}
C_{pq,\,\mathcal{K}_i} = C^{(0)}_{pq,\,\mathcal{K}_i} + a_F\, C^{(1)}_{pq,\,\mathcal{K}_i} + \cdots\;, \qquad C^{(0)}_{pq,\,\mathcal{K}_i} = 0\, \text{ for }\, \tau_0 < p+q\;.
\end{equation}
Here $C^{(0)}_{pq,\,\mathcal{K}_i}$ vanishes for $\tau_0 < p+q$ because the operator $(\mathcal{K}_i)_{\tau_0,R,\ell}$ with $\tau_0 < p+q$ cannot mix with the double-particle operator $[pq]$, whose twist is at least $p+q$. Therefore, the three-point coefficient must vanish at the disconnected level. We can impose this information in the OPE data by writing the OPE data as the product of two three-point coefficients. Explicitly, assuming $\Sigma_{12}\leq\Sigma_{34}$, we can express the OPE data we want in terms of three-point coefficients as follows
\begin{align}
\langle a^{(0)} (\gamma^{(1)})^2 \rangle =& \sum_i C^{(0)}_{p_1p_2,\,\mathcal{K}_i}\! \left(\gamma^{(1)}_{\,\mathcal{K}_i} \right)^2 C^{(0)}_{p_3p_4,\,\mathcal{K}_i}\;,&& \tau_0\geq \Sigma_{34}\;,&&\label{eq:a0g1sq}\\
\langle a^{(1)} \gamma^{(1)} \rangle =& \sum_i C^{(0)}_{p_1p_2,\,\mathcal{K}_i}\, \gamma^{(1)}_{\,\mathcal{K}_i}\; C^{(1)}_{p_3p_4,\,\mathcal{K}_i}\;,&& \Sigma_{12} \leq \tau_0 < \Sigma_{34}\;,&&\label{eq:a1g1}\\
\langle a^{(2)} \rangle =& \sum_i C^{(1)}_{p_1p_2,\,\mathcal{K}_i}\, C^{(1)}_{p_3p_4,\,\mathcal{K}_i}\;,&& s_-\leq \tau_0 < \Sigma_{12}\;.&&
\end{align}
Here we can see that when $\tau_0<\Sigma_{34}$ in \eqref{eq:a0g1sq} and $\tau_0<\Sigma_{12}$ in \eqref{eq:a1g1}, the corresponding data become zero due to \eqref{eq:Cseries}, as we expect from the analysis of Mellin amplitudes. Furthermore, we can observe that these data only depend on quantities with superscript $(0)$ and $(1)$, which correspond to the disconnected and tree level respectively.
	
We can now describe how to obtain these OPE data using unitarity recursion. As we have seen, there are many double-particle operators that mix with each other. For a fixed set of quantum numbers $\tau_0$, $R$, and $\ell$, we consider the pairs $(p,q)$ of $\mathcal{O}_p$ and $\mathcal{O}_q$ that can form double-particle operators with these quantum numbers and denote the set of these pairs by $\mathcal{S}_{\tau_0,R,\ell}$. For example, for $\tau_0=8$, $R=1$, and arbitrary $\ell$, the set of pairs are given by
\begin{equation}
\mathcal{S}_{8,1,\ell} = \{(2,4),(3,3),(3,5),(4,4)\}\;.
\end{equation}
The OPE data $\langle a^{(0)} (\gamma^{(1)})^2 \rangle$ at one loop can be computed from the lower-level data by
\begin{align}
\langle a^{(0)} (\gamma^{(1)})^2 \rangle_{p_1p_2p_3p_4} =& \sum_i C^{(0)}_{p_1p_2,\,\mathcal{K}_i}\! \left(\gamma^{(1)}_{\,\mathcal{K}_i} \right)^2 C^{(0)}_{p_3p_4,\,\mathcal{K}_i}\nonumber\\
=& \sum_{i,(p,q)\in \mathcal{S}} C^{(0)}_{p_1p_2,\,\mathcal{K}_i}\, \gamma^{(1)}_{\,\mathcal{K}_i}\, C^{(0)}_{pq,\,\mathcal{K}_i} \left(C^{(0)}_{pq,\,\mathcal{K}_i}\, C^{(0)}_{pq,\,\mathcal{K}_i} \right)^{-1} C^{(0)}_{pq,\,\mathcal{K}_i}\, \gamma^{(1)}_{\,\mathcal{K}_i}\,  C^{(0)}_{p_3p_4,\,\mathcal{K}_i}\nonumber\\
=& \sum_{(p,q)\in \mathcal{S}} \langle a^{(0)} \gamma^{(1)} \rangle_{p_1p_2pq}\, \langle a^{(0)} \rangle_{pqpq}^{-1}\, \langle a^{(0)} \gamma^{(1)} \rangle_{pqp_3p_4}\;.\label{eq:a0g1sqrec}
\end{align}
In the second line, we used the fact that the number of labels $i$ equals the number of pairs $(p,q)$, so $C^{(0)}_{pq,\mathcal{K}i}$ is a square matrix and is invertible. In the third line, we employed the fact that $\langle a^{(0)} \rangle_{pqpq}$ is a diagonal matrix, and its inverse is simply the inverse of its diagonal elements. Following similar reasoning, we have 
\begin{align}
\langle a^{(1)} \gamma^{(1)} \rangle_{p_1p_2p_3p_4} =& \sum_{(p,q)\in \mathcal{S}} \langle a^{(0)} \gamma^{(1)} \rangle_{p_1p_2pq}\, \langle a^{(0)} \rangle_{pqpq}^{-1}\, \langle a^{(1)} \rangle_{pqp_3p_4},&& \Sigma_{12} \leq \tau_0 < \Sigma_{34},&&\label{eq:a1g1rec}\\
\langle a^{(2)} \rangle_{p_1p_2p_3p_4} =& \sum_{(p,q)\in \mathcal{S}} \langle a^{(1)} \rangle_{p_1p_2pq}\, \langle a^{(0)} \rangle_{pqpq}^{-1}\, \langle a^{(1)} \rangle_{pqp_3p_4},&& s_-\leq \tau_0 < \Sigma_{12}.&&\label{eq:a2rec}
\end{align}
Note that we have assumed $\Sigma_{12}\leq\Sigma_{34}$. If $\Sigma_{12}>\Sigma_{34}$, we should interchange $\langle a^{(0)} \gamma^{(1)} \rangle$ and $\langle a^{(1)} \rangle$ in \eqref{eq:a1g1rec} to obtain the correct expressions.
	
Apart from these considerations, there are a few extra subtleties regarding two aspects.
\begin{itemize}
\item {\bf Protected operators}\\
According to the fusion rule \eqref{eq:fusion}, the semi-short operators $\mathcal{C}_{R,\ell}$ that can appear in the OPE of $\langle p_1p_2p_3p_4 \rangle$ are those with $s_-/2 - 2 \leq R \leq s_0/2-1$. The unitarity recursion method works for semi-short operators $\mathcal{C}_{R,\ell}$ with $2R+2 < s_0$ (recall that the twist of $\mathcal{C}_{R,\ell}$ is $2R+2$). Therefore, we can obtain the OPE data $\langle c^{(2)} \rangle$ of these semi-short operators using unitarity recursion
\begin{equation}
\langle c^{(2)} \rangle_{p_1p_2p_3p_4} = \sum_{(p,q)\in \mathcal{S}'} \langle c^{(1)} \rangle_{p_1p_2pq}\, \langle c^{(0)} \rangle_{pqpq}^{-1}\, \langle c^{(1)} \rangle_{pqp_3p_4},\ \ s_-\! - 4 \leq 2R+2 < s_0, \label{eq:c2rec}
\end{equation}
where $\mathcal{S}'_{R,\ell} = \{(p,q)\,|\,p,q\geq2,\ q=2R+2-p\}$ is the set of pairs that can generate semi-short double-particle operators with quantum number $R$ and $\ell$. However, there are also semi-short operators with $2R+2=s_0$, and the OPE data of these operators cannot be computed by unitarity recursion directly. To obtain this data, we impose the constraint that protected operators should be able to reproduce $\mathcal{G}^{(2)}_\text{protected}$, i.e., the protected part of the correlator at one-loop level. This results in a sum rule among the OPE data of all semi-short operators
\begin{equation}\label{eq:identityc}
\langle c^{(2)} \rangle_{R,\ell} + \langle c^{(2)} \rangle_{R-1,\ell+1} + \dots +  \langle c^{(2)} \rangle_{R_0,\ell+R-R_0} = 0,
\end{equation}
where $R=s_0/2-1=\min(\Sigma_{12},\Sigma_{34})/2-1$ and $R_0=s_-/2 - 2= \max(|p_{21}|,|p_{34}|)/2$. Thus, we can obtain the data $\langle c^{(2)} \rangle_{R,\ell}$ with $s_0=2R+2$ by using this linear relation. For the half-BPS operators, their three-point coefficients are protected by supersymmetry and can be directly computed in the free theory. In fact, these coefficients can also be expressed in terms of the semi-short OPE data by setting $\ell=-1$ in the data
\begin{equation}\label{eq:b2c}
\langle b^{(2)} \rangle_R = -\langle c^{(2)} \rangle_{R-1,-1}.
\end{equation}
For more details on the discussion of the protected sector and the derivation of \eqref{eq:identityc}, see appendix \ref{appx:Gprotected}.
\item {\bf Color structures and $SU(2)_L$ dependence}\\
In the discussion above, we have ignored the color group $G_F$ and the $SU(2)_L$ flavor group dependence in the OPE data. Here, we provide a brief explanation for how to handle these extra structures. For the $SU(2)_L$ group, we should decompose all the OPE data discussed above into $SU(2)_L$ harmonics (labled by $L$ in the following)
\begin{equation}
    \langle a^{(2)} \rangle_{\tau_0,R,\ell} = \sum_{L} \langle a^{(2)} \rangle_{\tau_0,R,L,\ell} P_L(\beta).
\end{equation}
The unitarity recursion \eqref{eq:a0g1sqrec}, \eqref{eq:a1g1rec}, \eqref{eq:a2rec} and \eqref{eq:c2rec} should be understood as working with each specific $SU(2)_L$ channel. For the color structures, we can decompose them into different color channels and then work with each channel. However, there is an equivalent but more intuitive diagrammatic method to deal with these structures, based on the gluing identities described in appendix \ref{appx:colordecp}. Schematically, at lower levels there are four different color structures $\bt$, $\bu$, $\ct$, and $\cu$.\footnote{The color structure $\bs$ is trivial because it only relates to the identity operator, so we omit it here. The color structure $\cs$ can be written in $\ct$ and $\cu$ via the Jacobi identity $\cs+\ct+\cu=0$.} We can treat these color structures as algebraic variables and introduce the following rules for inversion and multiplication 
\begin{equation}
    (\bt)^{-1} = \bt,\quad (\bu)^{-1} = \bu,\quad (\bt + (-1)^\ell \bu)^{-1} = \frac{1}{4}(\bt + (-1)^\ell \bu),
\end{equation}
\begin{equation}
    (\ct)^2 = \dst,\quad (\cu)^2 = \dst,\quad \ct\cu = -\dsu,
\end{equation}
\begin{equation}
    \dst\bt=\dst,\quad \dst\bu=\dsu,\quad \dsu\bt=\dsu,\quad \dsu\bu=\dst.
\end{equation}
Using these rules, we can obtain OPE data at one-loop level which contains the expected color structures $\dst$ and $\dsu$ in the s-channel. The multiplication of color structures in these rules can be interpreted as ``gluing'' two color structures along the s-channel. The validity of these rules is explained in appendix \ref{appx:colordecp}.
\end{itemize}

\subsection{Hidden conformal symmetry}\label{sec:hiddenconfsymm}
	
The eight-dimensional hidden conformal symmetry discovered in \cite{Alday:2021odx} provides a much more convenient strategy for determining the $\log^2(U)$ singularities of the one-loop correlator, which are more difficult to obtain by directly using unitarity recursion. This symmetry allows us to organize all tree-level correlators into a single generating function, from which we can derive the tree-level correlators $\mathcal{H}^{(1)}_{p_1p_2p_3p_4}$ by applying a differential operator $\mathcal{D}_{p_1p_2p_3p_4}$. Additionally, the hidden conformal symmetry automatically diagonalizes the mixing matrices, allowing for the determination of the tree-level anomalous dimensions $\gamma^{(1)}$ for all double-particle operators as a simple rational function of the 8D conformal spin $\ell_{\text{8d}}$ \cite{Drummond:2022dxd}. This makes it possible to find a universal formula for the leading logarithmic singularities of the reduced correlator with arbitrary KK modes at arbitrary loops \cite{Huang:2023oxf}, in particular the $\log^2U$ singularities at one loop.
	
More precisely, at tree level we should restore the kinematic factors for the reduced correlators in the following way \cite{Alday:2021odx}
\begin{equation}
\begin{split}
    H^{(1)}_{\{p_i\}} = & \left(\frac{t_{12}}{x_{12}^2}\right)^{\Sigma_{12}/2}\left(\frac{t_{34}}{x_{34}^2}\right)^{\Sigma_{34}/2}\left(\frac{x_{14}^2 t_{24}}{x_{24}^2 t_{14}}\right)^{p_{21}/2}\left(\frac{x_{14}^2 t_{13}}{x_{13}^2 t_{14}}\right)^{p_{34}/2}  \\
    & \times (x_{12})^{-2} (x_{34})^{-2}(v_{12})^{-2} (v_{34})^{-2}(\bar{v}_{12})^{-2} (\bar{v}_{34})^{-2}\, \mathcal{H}^{(1)}_{\left\{p_i\right\}}(U,V; \alpha, \beta) \;,
\end{split}
\end{equation}
such that they are four-point correlators with shifted conformal dimensions $p_i$ and $SU(2)_{L,R}
$ spins $\frac{p_i-2}{2}$. The lowest level correlator $H^{(1)}_{2222}$, which is a function of $x_{ij}^2$ only, can be promoted into a generating function by replacing the four dimensional distances $x_{ij}^2$ by the eight-dimensional distances $x_{ij}^2-t_{ij}$
\begin{equation}
    \mathbf{H}^{(1)}(x_i,t_i)=H^{(1)}_{2222}(x_{ij}^2-t_{ij})\;.
\end{equation}
All the other higher level correlators can be obtained by expanding the generating function in $t_{ij}$ and collect the correct powers that can appear in $H^{(1)}_{\{p_i\}}$. This expresses the redcued correlators $\mathcal{H}^{(1)}_{\{p_i\}}$ in terms of differential operators $\mathcal{D}_{p_1p_2p_3p_4}$ acting on $\mathcal{H}^{(1)}_{2222}$. For example, one finds
\begin{align}
    & \mathcal{D}_{2222}=1\;, \\
    & \mathcal{D}_{2332}=-\frac{\sqrt{U}}{\sqrt{\sigma}} \tau \partial_V\;, \\
    & \mathcal{D}_{2233}=3-U \partial_U\;.
\end{align}
	
With the differential operators $\mathcal{D}_{p_1p_2p_3p_4}$, we can express the $\dst$ part of the $\log^2(U)$ singularities for arbitrary KK modes at one-loop level as \cite{Huang:2023oxf}
\begin{equation}\label{eq:leadinglog}
    \left. \mathcal{H}^{(2)}_{\{p_i\}}\right\vert_{\dst\log^2(U)} = \Delta^{(4)} \mathcal{D}_{p_1p_2p_3p_4}\mathcal{D}_{(2)}\,  h^{(2)}(z)\;,
\end{equation}
where the seed function $h^{(2)}(z)$ is given by
\begin{equation}
    h^{(2)}(z) = \frac{61 z^2-135 z+72}{36 z^2}+\frac{(z-1)^2 }{z^3}\log(1-z)+\frac{(z-1)^3}{2z^3}\left( 2\text{Li}_2(z) + \log ^2(1-z) \right)\;.
\end{equation}
We also define the second-order differential operator $\mathcal{D}_{(2)}$ as
\begin{equation}
    \begin{split}
		\mathcal{D}_{(2)}f(z) = &\left[ \left(\frac{z \zb}{\zb - z}\right)^5 f(z) + \left(\frac{z \zb}{\zb - z}\right)^4 \frac{z^2}{2}\partial_z f(z)+ \left(\frac{z \zb}{\zb - z}\right)^3 \frac{z^3}{12}\partial^2_z \left(z f(z)\right)  \right] \\&+ (z\leftrightarrow\zb)\;,
    \end{split}
\end{equation}
and the forth-order differential operator $\Delta^{(4)}$ as
\begin{equation}\label{eq:Delta4}
    \Delta^{(4)}= \frac{z\bar z}{\bar z-z} (\mathcal{D}^+_z-\mathcal{D}^-_{\alpha})( \mathcal{D}^+_{\zb}-\mathcal{D}^-_{\alpha}) \frac{\bar z-z}{z\bar z}\;,
\end{equation}
where
\begin{equation}\label{eq:Dxpm}
    \mathcal{D}_x^\pm=x^2\partial_x (1-x)\partial_x \mp \frac{1}{2} (p_{21}+p_{34})x^2\partial_x-\frac{1}{4} p_{21} p_{34} x\;,
\end{equation}
is the Casimir of $SL(2)$ group. 
	
Now let us explore the implication of hidden conformal symmetry by rewriting \eqref{eq:leadinglog} in Mellin space. The idea is to define a {\it pre-amplitude} on which hidden conformal symmetry has the same simple action as at tree level. This corresponds to the object $\mathcal{D}_{(2)}\,  h^{(2)}(z)$ which we act with the tree-level operators $\mathcal{D}_{p_1p_2p_3p_4}$. More precisely, let us consider a function $\lambda_{2222}(z,\zb)$, whose $\log^2(U)$ singularity is given by
\begin{align}\label{eq:lambda2222}
    \lambda_{2222}(z,\zb)\big|_{\log^2(U)} = \mathcal{D}_{(2)}\,  h^{(2)}(z)\;,
\end{align}
and we consider its corresponding Mellin amplitude $\mathcal{L}_{2222}(s,t)$
\begin{equation}
    \lambda_{2222}(z,\zb)=\int_{-i \infty}^{i \infty} \frac{d s d t}{(2 \pi i)^2} U^{\frac{s+2}{2}} V^{\frac{t-4}{2}} \mathcal{L}_{2222}(s,t)\, \Gamma_{2222}(s,t)\;.
\end{equation}
Following the analysis presented in section \ref{sec:algorithm} we can use the spacetime formula \eqref{eq:lambda2222} to determine poles of $\mathcal{L}_{2222}$ in the range $s\geq 4$ (which produce $\log^2U$) as well as their coefficients. It turns out that in this range there are only simultaneous poles in $\mathcal{L}_{2222}$, and this Mellin amplitude can be written as
\begin{align}\label{eq:L2222}
    \mathcal{L}_{2222}(s,t) = \sum_{m,n=0}^{\infty}\frac{b_{m, n} }{(s-4-2 m)(t-4-2 n)} + \mathcal{P}_{2222}(s,t)\;,
\end{align}
with coefficients
\begin{equation}\label{eq:bmnnotruncation}
    b_{m,n}=\frac{n}{6 (m+1) (m+n+1) (m+n+2)}\;.
\end{equation}
The term $\mathcal{P}_{2222}(s,t)$ represents other contributions that do not affect the $\log^2(U)$ singularity. It particular, it can include $s$ poles located at $s<4$ with degree no higher than $2$.
	
In Mellin space, the diffrential operators in (\ref{eq:leadinglog}) become difference operators. This follows straightforwardly from the definiton of the Mellin representation. For example, acting on (\ref{eq:mellindef}) we have the following rules for translating the building blocks of the differential operator
\begin{equation}\label{eq:diffop1}
    \widehat{U\partial_U} \circ \widetilde{\mathcal{M}}(s,t) =  \left(\frac{s}{2}+1\right) \widetilde{\mathcal{M}}(s,t),\quad \widehat{V\partial_V} \circ \widetilde{\mathcal{M}}(s,t) =  \left(\frac{t}{2}-\frac{\Sigma_{23}}{2}\right) \widetilde{\mathcal{M}}(s,t)\;,
\end{equation}
and
\begin{equation}\label{eq:diffop2}
    \widehat{U^m V^n} \circ \widetilde{\mathcal{M}}(s,t)=\frac{\Gamma_{\{p_i\}}(s-2m,t-2n)}{\Gamma_{\{p_i\}}(s,t)}\widetilde{\mathcal{M}}(s-2m,t-2n)\;,
\end{equation}
A hat has been added to emphasize that they are difference operators. Note that in (\ref{eq:leadinglog}) we start with a Mellin representation with weights 2 and end with a representation with weights $\{p_i\}$. Therefore, the difference of the Gamma factor and powers of $U$ and $V$ in the definition should also be taken into account. We find the differential operator $\mathcal{D}_{p_1p_2p_3p_4}$ becomes
\begin{equation}\label{eq:Dpqrs}
    \widehat{\mathcal{D}}_{p_1p_2p_3p_4}\circ \widetilde{\mathcal{M}}(s,t)=\sum _{i=-s_0/2-2}^{-s_-/2+2} \, \sum _{j=(\Sigma_{23}-\Sigma_{14}+\kappa_{u})/4 }^{(\Sigma_{23}-\Sigma_{14}-\kappa_{u})/4-i} \frac{ \sigma ^i\tau ^j}{a\hspace{1pt} !\hspace{1pt} b\hspace{1pt} !\hspace{1pt}c\hspace{1pt}!\hspace{1pt}d\hspace{1pt}!\hspace{1pt}e\hspace{1pt}!\hspace{1pt}f\hspace{1pt}!} \widetilde{\mathcal{M}}(s+2\, i,t+2\, k)\;,
\end{equation}
where $k=(2 j-\Sigma_{23}+4)/2$, and
\begin{subequations}\label{eq:abcdef}
\begin{align}
    a&=\frac{1}{2} (2 i+\Sigma_{12}-4) \;,\quad   &b&=\frac{1}{2} (2 i+\Sigma_{34}-4)\;,\\
    c&=\frac{1}{2} (2 k+\Sigma_{14}-4) \;,\quad   &d&=\frac{1}{2} (2 k+\Sigma_{23}-4)\;,\\
    e&=\frac{1}{2} (-2 i-2 k-\Sigma_{13}+4) \;, \quad  &f&=\frac{1}{2} (-2 i-2 k-\Sigma_{24}+4)\;.
\end{align}
\end{subequations}
Following this procedure, it is also not difficult to derive $\widehat{\Delta}^{(4)}$. However, the explicit expression is too long and we refrain from presenting it here. All in all, the hidden symmetry formula \eqref{eq:leadinglog} for the $\log^2(U)$ singularity can be represented in Mellin space as
\begin{equation}\label{eq:deltaL2222}
    \widetilde{\mathcal{M}}_{\{p_i\},st}\hspace{3px} \approxeq \hspace{3px} \widehat{\Delta}^{(4)} \circ \widehat{\mathcal{D}}_{p_1 p_2 p_3 p_4} \circ \mathcal{L}_{2222}\;,
\end{equation}
where the ``$\approxeq$'' indicates that the equality holds only for $s$ poles with $s\geq s_+$, i.e., the poles in region III. 

A few comments are in order. First, it is {\it a priori} unclear that the RHS of \eqref{eq:deltaL2222} does not contain single poles in $s$ due to the complicated action of these difference operators. This, however, turns out to be true. By contrast, the action does lead to single poles in $t$. These single poles can be cancelled by adding additional singularities in the range $m<0$ corresponding to the ansatz \eqref{eq:L2222}. Nevertheless, the pre-amplitude \eqref{eq:L2222} is just an intermediate tool and we are not interested in determining its full expression. Second, the action of the difference operator shifts the location of poles. Therefore, when we focus on specific simultaneous poles of $\widetilde{\mathcal{M}}_{\{p_i\},st}$, it involves a finite collection of $b_{m,n}$ coefficients with shifted indices. In the generic case when the simultaneous poles are deep in the bulk region, all such $b_{m,n}$ coefficients are defined. However, when approaching the boundary some of the $b_{m,n}$ coefficients have negative indices. In this situation, we have
\begin{equation}\label{eq:bmntruncation}
    b_{m,n}=0,\qquad n<0\;,
\end{equation}
which is consistent with the range of summation in \eqref{eq:L2222}. Related to the first comment, the coefficients $b_{m,n}$ with $m<0$ can also be set to zero if we just want to obtain the simultaneous pole coefficients in $\widetilde{\mathcal{M}}_{\{p_i\},st}$. The poles in $\mathcal{L}_{2222}$ with $m<0$ are just to cancel the single poles in $t$ after the action and do not affect the simultaneous poles. 

The formula \eqref{eq:deltaL2222} is the main result which improves our original algorithm presented in section \ref{sec:algorithm}. Using it we can avoid following the unitarity recursion to generate CFT data for steps 1 and 2 and therefore make the algorithm more streamlined for computing more complicated correlators.

\section{Explicit examples at one loop}\label{sec:examples}
	
In this section we demonstrate the bootstrap algorithm by working out a few explicit examples. These examples are selected to disentangle and showcase different features in the algorithm and are presented in an order of increasing complexity.  The first example is the class of correlators of $\langle 22pp\rangle$. These correlators have a window region of which the size grows with $p$ but has no below window region. We then consider $\langle 3355\rangle$ which has both window and below window regions. From these examples we observe interesting patterns which persist in more general correlators. As an illustration, we present the explicit check of these patterns in the more complicated example of $\langle 4455\rangle$.

\subsection{Warm up: $\langle {\mathcal{O}}_2 {\mathcal{O}}_2{\mathcal{O}}_p {\mathcal{O}}_p\rangle$}\label{sec:o2o2opop}
	
Let us first consider the class of correlators $\langle22pp\rangle$. The extremality of these correlators is always $\mathcal{E}=2$. Consequently there is no below window region. Because the weights are pairwise equal, the Mellin amplitude is invariant under $t\leftrightarrow \tilde{u}$ leaving two independent color stripped partial amplitudes 
\begin{equation}\label{symm22pp}
    \widetilde{\mathcal{M}}_{22pp,{st}}=\widetilde{\mathcal{M}}_{22pp,{su}}\big|_{t\leftrightarrow\tilde{u}}\;,\quad \widetilde{\mathcal{M}}_{22pp,{tu}}=\widetilde{\mathcal{M}}_{22pp,{tu}}\big|_{t\leftrightarrow\tilde{u}}\;.
\end{equation}
Alternatively, equivalent to considering these partial amplitudes, we can consider only the $st$ component of the four-point function with different orderings. This is more convenient and we will do it in the following.
	
Let us first consider the ordering $\langle22pp\rangle$. As we discussed in section \ref{sec:hiddenconfsymm}, the consequence of the hidden conformal symmetry is conveniently captured by the action of the difference operator $\widehat{\mathcal{D}}_{22pp}$ on the pre-amplitude
\begin{equation}
    \mathcal{L}_{22pp}=\widehat{\mathcal{D}}_{22pp} \circ \mathcal{L}_{2222}= \frac{1}{\Gamma (p-1)}\sum_{m,n=0}^{\infty}\frac{b_{m, n} }{(s-4-2 m)(t-p-2-2 n)}+\mathcal{P}_{22pp}\;.
\end{equation}
To obtain the simultaneous pole coefficients of region III in s-channel (step 1 in the algorithm), we further need to act with the difference operator $\widehat{\Delta}^{(4)}$. We find the coefficients are given by (Here we introduce an extra label to the coefficients to distinguish different weight orderings.)
\begin{equation}\label{res22pp}
    c_{m,n}^{st}[22pp]=\frac{p \left(m (n+1) (m+n+2)(p+1) -(m-n) (2 n+m+3)\right)}{6 (m+n)_3\Gamma (p-1)}\;.
\end{equation}
Note that this expression is only expected to be valid for the range $m\geq p-2$ and $n\geq 0$. However, we will find that the validity of this result extends all the way to $m\geq 0$, modulo a small subtlety at the point $m=n=0$ where the expression becomes ill defined.\footnote{More precisely, one finds different answers depending on whether $m$ or $n$ is first set to zero.} This can be confirmed by considering the $st$ component in the ordering $\langle p22p\rangle$, which is the step 2 in the algorithm. By computing $\widehat{\Delta}^{(4)}\circ\widehat{\mathcal{D}}_{p22p}\circ\mathcal{L}_{2222}$ we obtain
\begin{equation}\label{resp22p}
    c_{m,n}^{st}[p22p]=(\alpha\beta)^{\frac{2-p}{2}}\frac{ p \left(n (m+1)(m+n+2)(p+1) -(n-m) (2 m+n+3)\right)}{6 (m+n)_3  \Gamma (p-1)}\;.
\end{equation}
As expected, through the crossing relation \eqref{eq:Mstcrossing} the expression \eqref{resp22p} recovers the result in \eqref{res22pp}. But the essential difference is that this result is now valid for the entire range of poles $m\geq0$ and $n\geq0$, since there is no window region in the s-channel of $\langle p22p\rangle$ (or equivalently the t-channel of $\langle 22pp\rangle$). For this reason all simultaneous poles in the ansatz for $\widetilde{\mathcal{M}}_{22pp,st}$ are determined by hidden conformal symmetry. This also fixes the aforementioned ambiguity at $(m,n)=(0,0)$, where one should first take $m=0$ and then $n=0$ in (\ref{resp22p}) as how these coefficients are obtained. Explicitly, we have
\begin{equation}\label{eq:c0022pp}
    c_{0,0}^{st}[p22p]=(\alpha\beta)^{\frac{2-p}{2}}\frac{p (p+1) }{12 \Gamma (p-1)}=(\alpha\beta)^{\frac{2-p}{2}}c_{0,0}^{st}[22pp]\;.
\end{equation}
	
We have now completely determined $\widetilde{\mathcal{M}}_{22pp,st}$. To follow the next step of the algorithm we need to study how the terms proportional to $\log U$ in the window region correctly reproduce the poles in region II of s-channel, with $0\leq m<p-2$. This will provide consistency checks on the result \eqref{res22pp}. We will not do it here for general $p$. However, we will perform this check explicitly for $p=3,4$ below, to illustrate the computation in the window region. 
	
Via the crossing relation \eqref{symm22pp},  $\widetilde{\mathcal{M}}_{22pp,st}$ also determines $\widetilde{\mathcal{M}}_{22pp,su}$. In order to obtain the full Mellin amplitude $\widetilde{\mathcal{M}}_{22pp}^{(2)}$ we also need to work out $\widetilde{\mathcal{M}}_{22pp,tu}$. By analysis similar to the step 2 above, we can consider the $st$ component of the correlator $\langle 2p2p\rangle$, whose s-channel contains no window region as well. We find
\begin{equation}\label{res2p2p}
    \begin{split}
        c_{m,n}^{st}[2p2p]&=(\alpha\beta)^{\frac{2-p}{2}}\frac{ p \left((m+1)(n+1)(m+n) (p+1)-(m-n)^2\right)}{6 (m+n)_3  \Gamma (p-1)}\\
        &=(\alpha\beta)^{\frac{2-p}{2}} c_{m,n}^{tu}[22pp]\;,
    \end{split}
\end{equation}
which is valid again for all poles $m,n\geq0$.
	
Having obtained the simultaneous pole coefficients for general $p$, we now verify the results in a few examples. The focus is to show that s-channel window region data are consistent with the prediction \eqref{res22pp}.~\\

\paragraph{\underline{$\langle {\mathcal{O}}_2{\mathcal{O}}_2{\mathcal{O}}_3{\mathcal{O}}_3\rangle $}}~\\
	
\noindent In this case the prediction of the leading logarithm from $\widehat{\Delta}^{(4)}\circ\widehat{\mathcal{D}}_{2233}\circ\mathcal{L}_{2222}$ is
\begin{equation}\label{res2233}
    c_{m,n}^{st}[2233]=\frac{4 m^2 n+3 m^2+4 m n^2+11 m n+5 m+2 n^2+3 n}{2 (m+n)_3}\;.
\end{equation}
By the previous argument this formula is valid for all poles in $\widetilde{\mathcal{M}}_{2233,st}$, including poles in the s-channel region II. The s-channel window region of $\langle2233\rangle$ contains a single twist  $\tau_{0}=4$. The corresponding data $\langle a^{(1)}\gamma^{(1)}\rangle_{2233}^{\tau_{0}=4}$ can be computed by unitarity recursion following section \ref{sec:recursion} (here $R=L=0$)
\begin{align}
    \langle a^{(1)}\gamma^{(1)}\rangle_{2233}^{\tau_{0}=4}&=\langle a^{(0)}\gamma^{(1)}\rangle_{2222}\,\langle a^{(0)}\rangle_{2222}^{-1}\,\langle a^{(1)}\rangle_{2233}=-\frac{6 \Gamma (\ell+3)^2 }{(\ell+1) (\ell+4) \Gamma (2 \ell+5)}\;,
\end{align}
where the necessary ingredients $\langle a^{(0)}\gamma^{(1)}\rangle_{2222}$, $\langle a^{(0)}\rangle_{2222}$ and $\langle a^{(1)}\rangle_{2233}$ are provided in appendix \ref{appx:data}. Following step 3 of the algorithm we need to compute the block sum \eqref{eq:logU1cftsum}, specifically
\begin{equation}\label{eq:resummation2233}
    \mathcal{H}_{2233}^{(2)}\supset\log U\,\sum_{\ell=0}^{\infty} \frac{1}{2}\langle a^{(1)}\gamma^{(1)}\rangle_{2233}^{\tau_{0}=4}\, \mathfrak{h}_{4,0,\ell}\;,
\end{equation}
and then extract pieces at order $U^3$ in the small $U$ expansion. The resummation in spin in \eqref{eq:resummation2233} can be conveniently carried out, and we obtain
\begin{equation}\label{eq:H2233U3logU}
    \mathcal{H}_{2233}^{(2)}\big\vert_{U^3\log U}=-\frac{3 (V+1) W_2-4 V+\log (V) (5 V+11)+4}{2 (V-1)^4}\;,
\end{equation}
where the function $W_2$ is
\begin{equation}
    W_2  =2\text{Li}_2(1-V)+\log ^2(V).
\end{equation}
In order to compare with the Mellin amplitude we also need to expand this result with respect to small $V$. Because the t-channel only contains above threshold region, it suffices to keep terms proportional to $\log^2V$ only, which is
\begin{equation}\label{eq:H2233U3logUlog2V}
    \mathcal{H}_{2233}^{(2)}\big\vert_{U^3\log U\log^2V}=-\frac{3}{2}-\frac{5}{12}V-21V^2-45V^3-\frac{165}{2}V^4+\mathcal{O}(V^5)\;.
\end{equation}
The term at order $V^j$ should be reproduced by the simultameous pole at $s=4$ and $t=5+2j$, or equivalently at $m=0$ and $n=j$. By computing the residues
\begin{equation}\label{eq:2233mellinres}
    \underset{t=5+2n}{\mathrm{Res}}\underset{s=4}{\mathrm{Res}}\,U^{\frac{s+2}{2}}V^{\frac{t-\Sigma_{23}}{2}}\frac{c_{0,n}^{st}[2233]}{(s-4)(t-5-2n)}\underbrace{\Gamma(\frac{s-4}{2})\Gamma(\frac{s-6}{2})\Gamma(\frac{t-5}{2})^2\Gamma(\frac{s+t-3}{2})^2}_{\Gamma_{2233}}\;,
\end{equation}
for $n\in\mathbb{N}$ and extracting the $\log^2V$ piece, we find the result exactly matches \eqref{eq:H2233U3logUlog2V}. In particular, in the case $n=0$, we can see that $c_{0,n}^{st}[2233]$ should be substituted by the $p=3$ case of \eqref{eq:c0022pp} in order to match the first term in \eqref{eq:H2233U3logUlog2V}. Equivalently, we can obtain the value of $c_{0,0}^{st}[2233]$ by first setting $m=0$ and then $n=0$.

Of course, the above computation is further subject to the consistency checks in step 6 to make sure that the Mellin amplitude contains only simultaneous poles. So far, we have only used $\log^2 U\log^2V$ terms in the above threshold region and $\log U\log^2V$ terms in the window region. We should also check that all the terms subleading in $\log V$ are  correctly reproduced. This check can be straightforwardly performed and we conclude that no further singularities are present.

\paragraph{\underline{$\langle {\mathcal{O}}_2{\mathcal{O}}_2{\mathcal{O}}_4{\mathcal{O}}_4\rangle $}}~\\
	
\noindent Let us move on to this slightly more complicated example, where operator mixing starts to occur in the window region. In this example the leading logarithmic singularity gives  $\widehat{\Delta}^{(4)}\circ\widehat{\mathcal{D}}_{2244}\circ\mathcal{L}_{2222}$ generates
\begin{equation}
    c_{m,n}^{st}[2244]=\frac{\left(5 m^2 n+4 m^2+5 m n^2+14 m n+7 m+2 n^2+3 n\right)}{3 (m+n)_3}\;.
\end{equation}
Again as a special case of $\langle 22pp\rangle$ the general argument guarantees that this result is valid for all poles in $\widetilde{\mathcal{M}}_{2244,st}$. This in particular includes the s-channel window region, which now contains two twists, namely $\tau_0=4$ and $\tau_0=6$.
	
Let us first consider $\tau_0=4$, which involves no operator mixing and is thus simpler. By unitarity recursion (with the ingredients again provided in appendix \ref{appx:data}) we have
\begin{align}
    \langle a^{(1)}\gamma^{(1)}\rangle_{2244}^{\tau_{0}=4}&=\langle a^{(0)}\gamma^{(1)}\rangle_{2222}\,\langle a^{(0)}\rangle_{2222}^{-1}\,\langle a^{(1)}\rangle_{2244}=-\frac{4 \Gamma (\ell+3)^2 }{(\ell+1) (\ell+4) \Gamma (2 \ell+5)}\;,
\end{align}
Notice that we still have $R=L=0$. 	The subsequent resummation are completely similar to that in $\langle 2233\rangle$. Following step 3 of the algorithm we compute
\begin{equation}\label{eq:resummation2244}
    \mathcal{H}_{2244}^{(2)}\supset\log U\,\sum_{\ell=0}^{\infty} \frac{1}{2}\langle a^{(1)}\gamma^{(1)}\rangle_{2244}^{\tau_{0}=4}\, \mathfrak{h}_{4,0,\ell}\;,
\end{equation}
and extract the coefficient at order $U^3$. The resummation in spin yields a result almost the same as that in \eqref{eq:H2233U3logU} except for an overall factor
\begin{equation}
    \mathcal{H}_{2244}^{(2)}\big\vert_{U^3\log(U)}=-\frac{3 (V+1)  W_2 -4 V+(5 V+11) \log (V)+4}{3 (V-1)^4}\;.
\end{equation}
The subsequent analysis is again similar. We expand in $V$, extract the $\log^2V$ terms, and then compare them with the corresponding contributions from the Mellin amplitude (i.e., $m=0$ and arbitrary $n$). The details of this computation are left to the reader. Here we observed that as long as we first set $m=0$ and then $n=0$ for the Mellin coefficient at $(m,n)=(0,0)$, the outcome of the resummation again perfectly coincides with the prediction of the Mellin amplitude.
	
Next let us focus on $\tau_{0} =6$, where we are required to solve the operator mixing. Here in the unitarity recursion we need to identify all pairs that allow the propagation of $\tau_{0} =6$ operators, that is\footnote{Note that in the recursion we need to consider each $SU(2)_R \times SU(2)_L$ representation.}
\begin{equation}
    \begin{split}
		\langle a^{(1)}\gamma^{(1)}\rangle_{2244}^{\tau_{0}=6}&=
		\begin{bmatrix}
		\langle a^{(0)}\gamma^{(1)}\rangle_{2222}\, & \langle a^{(0)}\gamma^{(1)}\rangle_{2233}  
		\end{bmatrix}
		\begin{bmatrix}
		\langle a^{(0)}\rangle_{2222}^{-1} & 0\\
		0 &\langle a^{(0)}\rangle_{3333}^{-1}
		\end{bmatrix}
		\begin{bmatrix}
		\langle a^{(1)}\rangle_{2244}\, \\ \langle a^{(1)}\rangle_{3344}  
		\end{bmatrix}\\
		&=\frac{4 (13 \ell  (\ell +7)+178) \Gamma (\ell +4)^2 }{(\ell +1) (\ell +2) (\ell +5) (\ell +6) \Gamma (2 \ell +7)}\;.
    \end{split}
\end{equation}
In summing over the conformal blocks one should be careful with the fact that now the $\tau_{0} =4$ data also contributes, as this part can also generate $U^4$ terms in the small $U$ expansion. Specifically,  following step 3 of the algorithm the summation reads
\begin{equation}
    \mathcal{H}_{2244}^{(2)}\supset\log U\,\sum_{\ell=0}^{\infty}\frac{1}{2} \left(\langle a^{(1)}\gamma^{(1)}\rangle_{2244}^{\tau_{0}=4}\,\mathfrak{h}_{4,0,\ell}+\langle a^{(1)}\gamma^{(1)}\rangle_{2244}^{\tau_{0}=6}\,\mathfrak{h}_{6,0,\ell}\right).
\end{equation}
Extracting the $U^4$ after the spin summation gives
\begin{equation}
    \mathcal{H}_{2244}^{(2)}\big\vert_{U^4\log(U)}=\frac{(V (79 V+428)+91) \log (V)}{3 (V-1)^6}+\frac{67 V+175}{3 (V-1)^5}+\frac{(V (35 V+94)+11) W_2}{(V-1)^6}.
\end{equation}
In the small $V$ expansion this further yields the $\log^2V$ pieces
\begin{equation}\label{eq:H2244U4logUlog2V}
    \mathcal{H}_{2244}^{(2)}\big\vert_{U^4\log U\log^2V}=11+160V+830V^2+2800V^3+7385V^4+\mathcal{O}(V^5).
\end{equation}
One can explicitly check that this agrees with the Mellin prediction (by using the simultaneous poles with $m=1$ and arbitrary $n$)
\begin{equation}\label{eq:2244mellinres}
    \underset{t=6+2n}{\mathrm{Res}}\underset{s=6}{\mathrm{Res}}\,U^{\frac{s+2}{2}}V^{\frac{t-6}{2}}\frac{c_{1,n}^{st}[2244]}{(s-6)(t-6-2n)}\underbrace{\Gamma(\frac{s-4}{2})\Gamma(\frac{s-8}{2})\Gamma(\frac{t-6}{2})^2\Gamma(\frac{s+t-4}{2})^2}_{\Gamma_{2244}}.
\end{equation}

\subsection{$\langle {\mathcal{O}}_3{\mathcal{O}}_3{\mathcal{O}}_5{\mathcal{O}}_5\rangle$}\label{sec:o3o3o5o5}
	
We choose the example of $\langle 3355\rangle$ for two reasons. First, it allow us to see how the prediction of hidden symmetry extends to the entire set of poles with $m>0$ but not to $m=0$. While in $\langle 22pp\rangle$ we could think about this extension as the consequence of the fact that the t-channel above-threshold region accounts for all poles in the Mellin amplitude, this is not the case in the current example. The reason is that here for the first time we will encounter the below window region, such that we will find poles that obey the above extension but do not belong to region III in any channel.\footnote{More specifically, in this example there is a unique pole of this type, locating at $(m,n)=(1,-1)$. While the simpler examples $\langle3333\rangle$ and $\langle3344\rangle$ also have below window region, their Mellin amplitudes do not have such type of poles.} Second, using this example we will demonstrate how the semi-short operators in this below window region affect the coefficients of region I poles in the Mellin amplitude. 
	
This example has extremality $\mathcal{E}=3$, and so the minimal value of $m$ and $n$ is $-1$. In step 1 of the algorithm, by $\widetilde{\Delta}^{(4)}\circ\widetilde{\mathcal{D}}_{3355}\circ\mathcal{L}_{2222}$, leading logarithms in the above-threshold region yields for $m\geq 2$ and $n\geq0$
\begin{align}\label{3355res}
    c_{m,n}^{st}&=\frac{F^{0,0}_{m,n}}{6 (m+n+1)_3}-\frac{1}{\alpha}\frac{ F^{1,0}_{m,n}}{6 (m+n+1)_3}-\frac{1}{\beta}\frac{ F^{0,1}_{m,n}}{6 (m+n)_4} +\frac{1}{\alpha\beta}\frac{   F^{1,1}_{m,n}}{6 (m+n)_4}, 
\end{align}
where $F^{i,j}_{m,n}$ are polynomials of $m$ and $n$,
\begin{align}
    F^{0,0}_{m,n} =&\, 46 + 124 m + 39 m^2 + 74 n + 136 m n + 28 m^2 n + 21 n^2 + 28 m n^2,\\
    F^{1,0}_{m,n} =&\,32 + 98 m + 33 m^2 + 48 n + 102 m n + 21 m^2 n + 12 n^2 + 21 m n^2,\\
    F^{0,1}_{m,n} =&\,34 m + 103 m^2 + 33 m^3 + 28 n + 146 m n + 135 m^2 n + 21 m^3 n + 43 n^2 \nonumber\\
    &+ 114 m n^2 + 42 m^2 n^2 + 12 n^3 + 21 m n^3,\\
    F^{1,1}_{m,n} =&\,118 m + 181 m^2 + 51 m^3 + 76 n + 352 m n + 270 m^2 n + 42 m^3 n + 96 n^2 \nonumber\\
    &+ 243 m n^2 + 84 m^2 n^2 + 24 n^3 + 42 m n^3.
\end{align}
For $m\geq2$ and $n<0$, the hidden symmetry instead predicts
\begin{align}\label{3355nres}
    c_{m,-1}^{st}=\frac{5 (5 m+9)}{12 (m+1)_2}-\frac{1}{\alpha}\frac{5 (5 m+8)}{12 (m+1)_2}-\frac{1}{\beta }\frac{5  (5 m+9)}{12 (m+1)_2}+\frac{1}{\alpha\beta}\frac{5 (5m+8)}{12(m+1)_2}.
\end{align}
Equation \eqref{3355nres} is not an analytic continuation of \eqref{3355res} but they are closely related because they are both determined by the $\log^2U$ coefficient which is dictated by hidden conformal symmetry. The reason for a different formula for $c_{m,-1}^{st}$ is also quite simple. To compute it from $\mathcal{L}_{2222}$ we encounter $b_{m',n'}$ with negative indices, which according to the truncation condition \eqref{eq:bmntruncation} vanishes instead of following the formula \eqref{eq:bmnnotruncation}. Therefore, $c_{m,-1}^{st}$ does not follow from the naive analytic continuation of $c_{m,n}^{st}$.
	
Following step 2 of the algorithm we can also determine the coefficients for $-1\leq m\leq 1$ and $n\geq 0$ in addition to those in \eqref{3355res} and \eqref{3355nres}. In particular, we can expect that the poles with $m=0,1$ and $n\geq 0$ share the same result as \eqref{3355res}. Here we choose to skip this step, and will instead show how these data can come from the below window analysis in step 5 later on.
	
For step 3, the calculation in the window region is similar to that of the previous example. Therefore, we will be brief and mostly focus on the result. The s-channel window region contains two twists $\tau_0=6$ and $\tau_0=8$. We first examine $\tau_0=6$, which constrains poles at $m=0$. The relevant data are(see appendix \ref{appx:data})
\begin{align}
    \langle a^{(1)}\gamma^{(1)}\rangle_{6,0,\ell}= &-P_0(\beta)\frac{5  \left(\ell ^4+14 \ell ^3+93 \ell ^2+308 \ell +388\right) \Gamma (\ell +4)^2}{2(\ell +1) (\ell +2) (\ell +5) (\ell +6) \Gamma (2 \ell +7)} \nonumber\\
    &+P_1(\beta)\frac{5  (\ell +1) (\ell +6) \Gamma (\ell +4)^2}{(\ell +2) (\ell +5) \Gamma (2 \ell +7)}\;,\\
    \langle a^{(1)}\gamma^{(1)}\rangle_{6,1,\ell}=&+P_0(\beta)\frac{2  (\ell +1) (\ell +6)  \Gamma (\ell +4)^2}{(\ell +3) (\ell +4) \Gamma (2 \ell +7)} \nonumber\\
    &-P_1(\beta)\frac{4  \left(\ell ^4+14 \ell ^3+105 \ell ^2+392 \ell +540\right) \Gamma (\ell +4)^2}{(\ell +1) (\ell +3) (\ell +4) (\ell +6) \Gamma (2 \ell +7)}\;.
\end{align}
We resum the blocks at this twist following \eqref{eq:logU1cftsum} and extract the $U^4\log U$ piece. The first several orders of its small $V$ expansion read
\begin{equation}\label{eq:H3355U4logU}
    \begin{split}
		\mathcal{H}_{3355}^{(2)}\big\vert_{U^4\log(U)}
		=&+V^{-1}\qty(-\frac{(-4+7\alpha)(-1+\beta)}{2\alpha\beta})\\
		&+V^0\qty(-\frac{38 - 14 \alpha - 16 \beta + 23 \alpha \beta}{\alpha\beta}\log^2V+\cdot)\\
		&+V^1\qty(-\frac{2(196 - 83 \alpha - 92 \beta + 141 \alpha \beta)}{\alpha\beta}\log^2V+\cdots)\\
		&+V^2\qty(-\frac{10(182 - 81 \alpha - 88 \beta + 139\alpha \beta)}{\alpha\beta}\log^2V+\cdots)+\mathcal{O}(V^3),
    \end{split}
\end{equation}
where ``$\cdots$'' refers to terms with subleading powers of $\log V$ at each order. Note that differing from $\langle22pp\rangle$, the t-channel of this example contains both the above threshold region as well as the below window region, in the above expansion we clearly observe the distinction between patterns at different orders. At order $V^{\geq0}$ the expression starts with $\log^2V$, in correspondence to twists above threshold, while at order $V^{-1}$ there is only a rational function, which is consistent with our expectation for the below window region. By comparing terms at each order $V^{n\geq 0}$ and the Mellin amplitude prediction from the simultaneous pole at $m=0$ and $n\geq 0$, we find the coefficients of these poles are 
\begin{align}
    c_{0,n}^{st}&=\frac{F^{0,0}_{0,n}}{6 (n+1)_3}-\frac{1}{\alpha}\frac{ F^{1,0}_{0,n}}{6 (n+1)_3}-\frac{1}{\beta}\frac{ F^{0,1}_{0,n}}{6 (n)_4} +\frac{1}{\alpha\beta}\frac{   F^{1,1}_{0,n}}{6 (n)_4},
\end{align}
which indeed obeys the formula \eqref{3355res} (by setting $m=0$). On the other hand, using the $V^{-1}$ term in \eqref{eq:H3355U4logU} we find the coefficient with $n<0$ to be
\begin{align}\label{eq:3355c0m1}
    c_{0,-1}^{st}=\frac{7}{8}-\frac{1}{2\alpha} -\frac{7}{8\beta}+\frac{1}{2\alpha \beta}.
\end{align}
Note that if we naively take the $m\to0$ limit of $c_{m,-1}^{st}$ in \eqref{3355nres}, we would have concluded  $c_{0,-1}^{st}=\frac{15}{8}-\frac{5}{3\alpha}-\frac{15}{8\beta}+\frac{5}{3\alpha\beta}$, which differs from the correct result above. Therefore, we see that the prediction of hidden symmetry fails to extend to this location.
	
Next we move on to study $\tau_0=8$, for which we have the data (see appendix \ref{appx:data})
\begin{align}
    &{\langle a^{(1)}\gamma^{(1)}\rangle_{8,0,\ell}}=\nonumber\\
    &+P_0(\beta)\frac{15  \left(3\ell^6+81\ell^5+1011\ell ^4+7263 \ell ^3+30698 \ell ^2+70704 \ell +69856\right) \Gamma (\ell +5)^2}{2(\ell +1) (\ell +2)(\ell+3) (\ell +6) (\ell +7)(\ell+8) \Gamma (2 \ell +9)} \nonumber\\
    &-P_1(\beta)\frac{45  (\ell +1) (\ell +8)(\ell^2+9\ell+22) \Gamma (\ell +5)^2}{(\ell +2)(\ell+3) (\ell +6)(\ell+7) \Gamma (2 \ell +9)}\;,\\
    &{\langle a^{(1)}\gamma^{(1)}\rangle_{8,1,\ell}}=\nonumber\\
    &-P_0(\beta)\frac{  (\ell +1) (\ell +8)(71\ell^2+639\ell+1594)  \Gamma (\ell +5)^2}{2(\ell +2) (\ell +4)(\ell+5)(\ell+7) \Gamma (2 \ell +9)} \nonumber\\
    &+P_1(\beta)\frac{  \left(71\ell^6+1917\ell^5+23415\ell ^4+162675 \ell ^3+661554 \ell ^2+1461888 \ell +1364800\right) \Gamma (\ell +4)^2}{(\ell +1) (\ell +2)(\ell+4)(\ell+5) (\ell +7) (\ell +8) \Gamma (2 \ell +9)}\;.
\end{align}
To correctly extract the piece $U^5\log U$ we need to sum over conformal blocks with both $\tau_0=6$ and $\tau_0=8$, as was shown in \eqref{eq:logU1cftsum}. The first few orders of this piece in the small $V$ expansion are
\begin{equation}\label{eq:H3355U5logU}
    \begin{split}
		\mathcal{H}_{3355}^{(2)}\big\vert_{U^5\log(U)}
		=&+V^{-1}\qty(\frac{5 (-13 + 14 \alpha) (-1 + \beta)}{2\alpha\beta})\\
		&+V^0\qty(\frac{2 (350 - 170 \alpha - 163 \beta + 209 \alpha \beta)}{\alpha\beta}\log^2V+\cdots)\\
		&+V^1\qty(\frac{10(1579 - 732 \alpha - 734 \beta + 992 \alpha \beta)}{\alpha\beta}\log^2V+\cdots)\\
		&+V^2\qty(\frac{90(1350 - 630 \alpha - 637 \beta + 881 \alpha \beta)}{\alpha\beta}\log^2V+\cdots)+\mathcal{O}(V^3).
    \end{split}
\end{equation}
By comparing terms at each order $V^{n\geq 0}$ and the Mellin amplitude prediction from the simultaneous pole at $m=1$ and $n\geq 0$, we find the coefficients of these poles are 
\begin{align}
    c_{1,n}^{st}&=\frac{F^{0,0}_{1,n}}{6 (n+2)_3}-\frac{1}{\alpha}\frac{ F^{1,0}_{1,n}}{6 (n+2)_3}-\frac{1}{\beta}\frac{ F^{0,1}_{1,n}}{6 (n+1)_4} +\frac{1}{\alpha\beta}\frac{   F^{1,1}_{1,n}}{6 (n+1)_4},
\end{align}
which again perfectly agrees with the formula \eqref{3355res} (by setting $m=1$). Moreover, the term at order $V^{-1}$ in \eqref{eq:H3355U5logU} dictates that
\begin{equation}
    c_{1,-1}^{st}=\frac{35}{36}-\frac{65}{72\alpha}-\frac{35}{36\beta}+\frac{65}{72\alpha\beta}.
\end{equation}
Unlike \eqref{eq:3355c0m1} for the pole at $(m,n)=(0,-1)$, this coefficient turns out to be exactly the $m\to1$ limit of the hidden symmetry prediction \eqref{3355nres}. As we already mentioned at the beginning, this result is quite non-trivial since the pole at $(m,n)=(1,-1)$  is not tied to above threshold twist in any channel!
	
\begin{figure}[t]
    \centering
		\begin{tikzpicture}
			\definecolor{cr1}{HTML}{fcf6bd}
			\definecolor{cr2}{HTML}{d0f4de}
			\definecolor{cr3}{HTML}{a9def9}
			\definecolor{naturegreen}{HTML}{00A686}
			\definecolor{naturered}{HTML}{E64B35}
			\definecolor{naturecolor3}{HTML}{A9DEF9}
			
			\draw[-stealth] (-2,0)--(6,0) node[right] {$s$};
			\draw[-stealth] (0,-2) node [anchor=north] {$s_0$} --(0,3) node[above] {$t$}; 
			\draw [gray,dashed] (-1,2.5) -- (-1,-1)--(5.5,-1) ;
			\draw [dotted] (4,-2) node [anchor=north] {$s_+$} -- (4,2.5);
			\draw [{latex}-{latex}] (0,-1.5) -- (4,-1.5);
			\node [anchor=north] at (2,-1.5) {\scriptsize $2(p-3)$};
			\node [anchor=east] at (-2,0) {$t_+\equiv t_0$};
			
			\node [anchor=north] at (-0.05,0) {$(s_0,t_0)$};
			
			\foreach \x in {-1,0,1}
			\foreach \y in {-1,0,1,2}
			{
				\fill[fill=black](\x,\y) circle (0.11cm);
				\fill[fill=naturecolor3](\x,\y) circle (0.1cm);
			}
			\foreach \y in {-0.5,0.5,1.5} \node [anchor=center] at (2,\y) {$\cdots$};
			\foreach \x in {3,4,5}
			\foreach \y in {-1,0,1,2}
			{
				\fill[fill=black](\x,\y) circle (0.11cm);
				\fill[fill=naturecolor3](\x,\y) circle (0.1cm);
			}
			\foreach \x in {1,3}
			{
				\fill[fill=black](\x,-1) circle (0.11cm);
				\fill[fill=yellow](\x,-1) circle (0.1cm);
			}
			
			\foreach \x in {-1,0}
			\foreach \y in {-1,-1}
			{
				\fill[fill=naturered!70] (\x,\y) circle (0.1cm);
			}
			\fill[draw=gray,fill=white] (-1,-1) circle (0.1cm);
		\end{tikzpicture}
    \caption{Poles of $\widetilde{\mathcal{M}}_{33pp,st}$ ($p\geq 3$), of which $\widetilde{\mathcal{M}}_{3355,st}$ is a special case. Coefficients of poles marked blue directly follow from hidden conformal symmetry. Coefficients of poles marked yellow do not have to obey the same result in principle, but by detailed recursive computation it turns out they do, so that in some sense the hidden symmetry receives an extension in this region. In contrast, coefficient of the pole marked red violates the hidden symmetry result and have to be separately worked out. The pole marked white turns out to vanish.}
    \label{fig:33pp}
\end{figure}
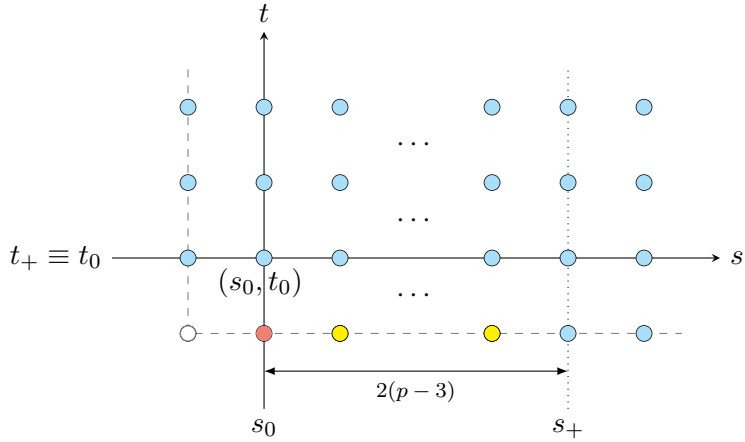

Consequently, at this point we are able to conclude that for $\langle3355\rangle$ the structure of coefficients descending from hidden symmetry essentially extends to any poles with $m>0$ or with $n\geq0$ (the inclusion of the bound $n=0$ is because the boundary of the t-channel threshold region is related to poles at $n=0$ in this example). We also find this extension continues to hold for more general situation of $\langle 33pp\rangle$, as shown in figure \ref{fig:33pp}, where it has to be non-trivially confirmed by recursive computation at $p-4$ simultaneous poles (marked yellow).
	
Going back to the example of $\langle3355\rangle$, let us mention that there is no need to deal with step 4 of the algorithm since there is no t-channel window region in this example. So we directly move on to step 5, where we compute the contributions from the below-window data and illustrate how they determine the remaining poles at $m=-1$ and any $n$. Recall the fusion rule for $\langle3355\rangle$
\begin{equation}
    \mathcal{B}_{\frac{3}{2}} \times \mathcal{B}_{\frac{3}{2}} \simeq \bigoplus_{R=0}^{3} \mathcal{B}_R + \bigoplus_{R=0}^{2}\mathcal{C}_{R,\ell} + \bigoplus _{R=0}^{1}\,\mathcal{A}_{\tau,R,\ell}.
\end{equation}
From the perspective of the super multiplets (see \eqref{eq:step5}), the undetermined coefficients account for this part of the reduced correlator is
\begin{equation}\label{fusion3344}
    \mathcal{H}_{3355}^{(2)}\supset\frac{z\bar{z}\alpha^2}{(z-\alpha)(\bar{z}-\alpha)}\sum_{R=0}^2 \, \sum_{l\geq-1}^{\infty} \, \langle c^{(2)}\rangle_{R,l} \, \mathcal{C}_{R, \ell} +\sum_{R=0}^{1}\ \sum_{l\geq0}^{\infty} \,  \sum_{\tau_0\geq R+4}^{4} \, \langle a^{(2)}\rangle_{\tau_0,R,l}\,  \mathfrak{h}_{\tau_0,R, \ell}.
\end{equation}
By modifying the lower limit of the spin summation in the semi-short operator, we incorporate the half-BPS operator as well, thanks to the relation $\mathcal{C}_{R,-1}\cong \mathcal{B}_{R+1}$ (see appendix \ref{appx:superblocks}). For the long operator, there exists a lower bound on the twist due to unitarity. Hence here we only need to consider the coefficient in front of the long operator term $\mathcal{A}_{4,0, \ell}$. These contributions are identified as
\begin{align}
    \langle a^{(2)} \rangle_{4,0,\ell}&=\langle a^{(1)} \rangle_{3322}\,\langle a^{(0)} \rangle_{2222}^{-1}\,\langle a^{(1)} \rangle_{2255}\;.
\end{align}
Here $R=L=0$, hence the contribution to $\langle a^{(2)}\rangle$ from the long multiplet is given by
\begin{equation}
    \langle a^{(2)} \rangle_{4,0,\ell} =\frac{5 \Gamma (\ell +3)^2 }{2(\ell +1) (\ell +4) \Gamma (2 \ell +5)}.
\end{equation}
The inclusion of semi-short contributions into $\langle c^{(2)}\rangle $ requires a detailed analysis. Following identity \eqref{eq:identityc}, we have 
\begin{equation}
    -\langle c^{(2)}\rangle_{2,\ell}=\langle c^{(2)}\rangle_{1,\ell+1}.
\end{equation}
This precisely allows two semi-short multiplets to combine together, giving rise to a contribution resembling that of a long multiplet. Moreover, it enables us to obtain $\langle c^{(2)}\rangle _{2,\ell}$ by calculating $\langle c^{(2)}\rangle_{1,\ell}$—where $\langle c^{(2)}\rangle_{1,\ell}$ can be glued together through unitary recursion using the twist-$2$ data. When extracting the data, we focus specifically on the $P_0 (\alpha)$ representation. The required data is 
\begin{align}
    \langle c^{(1)} \rangle_{3322}^{\tau_{0}=2}&=\frac{P_0 (\beta) \Gamma (\ell +3)^2 \left(\ct+(-1)^{\ell } \cu\right)}{\Gamma (2 \ell +5)}, \\
    \langle c^{(0)} \rangle_{2222}^{\tau_{0}=2}&=-\frac{P_0 (\beta)  (\ell +2) (\ell +3) \Gamma (\ell +3)^2 \left(\bt+(-1)^{\ell +1}\bu\right)}{\Gamma (2 \ell +5)},\\
    \langle c^{(1)} \rangle_{2255}^{\tau_{0}=2}&=\frac{ P_0 (\beta)  \Gamma (\ell +3)^2 \left(\ct+(-1)^{\ell } \cu\right)}{\Gamma (2 \ell +5)}.
\end{align}
Thus, the unitary recursion is given by
\begin{align}
    \langle c^{(2)} \rangle_{1,l}&=\langle c^{(1)} \rangle_{3322}^{\tau_{0}=2} \left(\langle c^{(0)} \rangle_{2222}^{\tau_{0}=2}\right)^{-1}\langle c^{(1)} \rangle_{2255}^{\tau_{0}=2}.
\end{align}
After a simple calculation, we obtain
\begin{equation}
    -\langle c^{(2)} \rangle_{2,l-1}=\langle c^{(2)} \rangle_{1,l}=-\frac{ \Gamma (\ell +3)^2  }{(\ell +2) (\ell +3) \Gamma (2 \ell +5)}.
\end{equation}
Now that we have found all the data contributing to $\tau_0=4$ and their dependencies on $SU(2)_L$ is $P_0(\beta)=1$, the next step is to perform the following resummation
\begin{equation}
    \sum_{\ell=0}^{\infty} \left(\langle a^{(2)} \rangle_{4,0,\ell} \mathfrak{h}_{4,0,\ell}  -\langle c^{(2)} \rangle_{2,l-1} \mathfrak{h}_{4,1,\ell} \right) +\mathcal{O}\left(U^4\right).
\end{equation}
Finally, the expression for $\mathcal{H}_{3355}^{(2)}\big\vert_{U^3}$ is:
\begin{equation}
    \mathcal{H}_{3355}^{(2)}\big\vert_{U^3}=\frac{ (2 V \alpha + 3 \alpha + V-1)W_2}{2 \alpha  (V-1)^4}-\frac{19 \alpha -18}{6 \alpha  (V-1)^3}+\frac{(14 V \alpha + 35 \alpha - 3 V-15) \log (V)}{6 \alpha  (V-1)^4}.
\end{equation}
By comparing it with the Mellin amplitude, we can determine all the residues along the column with $m=-1$. The result can be divided into two parts
\begin{equation}
    c^{st}_{-1,-1}=0,
\end{equation}
and 
\begin{equation}\label{3344mm1}
    c^{st}_{-1,n}=\frac{5 n+9}{12 (n+1)_2}-\frac{1}{\alpha}\frac{1 }{4 (n+1)_2},\quad n\geq 0.
\end{equation}
The vanishing of $c^{st}_{-1,-1}$ turns out to be a universal phenomenon at extremality $\mathcal{E}=3$. In particular, for $\langle33pp\rangle$ we  verified that $c^{st}_{-1,-1}=0$ holds for any $p$.

\subsection{$\langle {\mathcal{O}}_4{\mathcal{O}}_4{\mathcal{O}}_5{\mathcal{O}}_5\rangle$}
	
Finally we examine the example of $\langle 4455\rangle$, which has  extremality $\mathcal{E}=4$. We will be brief about the implementation of the algorithm and will mainly discuss the results. In paricular, we will use this example to illustrate a few features of the poles in region I. 
	
Same as in previous examples, the analysis begins by using the s-channel leading logarithmic singularities to determine the coefficients of all simultaneous poles in region III, with $m\geq 1$ and any $n$. Likewise, we also consider the leading logarithmic singularities of $\langle 5445\rangle$, and by crossing they fix coefficients of all t-channel region III simultaneous poles, with $n\geq 0$ and any $m$. Since these results are quite lengthy, we will not present them here but collect them in the ancillary file of the arXiv submission.
	
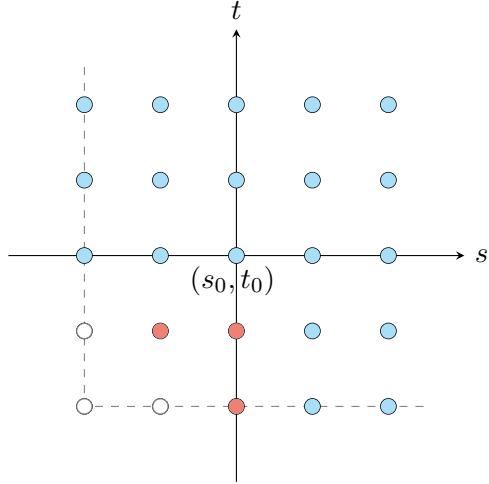
\begin{figure}[hb]
    \centering
		\begin{tikzpicture}
			\definecolor{cr1}{HTML}{fcf6bd}
			\definecolor{cr2}{HTML}{d0f4de}
			\definecolor{cr3}{HTML}{a9def9}
			\definecolor{naturegreen}{HTML}{00A686}
			\definecolor{naturered}{HTML}{E64B35}
			\definecolor{naturecolor3}{HTML}{A9DEF9}
			\draw[-stealth] (-3,0)--(3,0) node[right] {$s$};
			\draw[-stealth] (0,-3)--(0,3) node[above] {$t$}; 
			\node [anchor=north] at (-0.05,0) {$(s_0,t_0)$};
			\draw[gray,dashed] (-2,2.5)--(-2,-2)--(2.5,-2) ;
			\foreach \x in {-2,-1,0,1,2}
			\foreach \y in {-2,-1,0,1,2}
			{
				\fill[fill=black](\x,\y) circle (0.11cm);
				\fill[fill=naturecolor3](\x,\y) circle (0.1cm);
			}
			\foreach \x in {-2,-1,0}
			\foreach \y in {-2,-1}
			{
				\fill[fill=naturered!70] (\x,\y) circle (0.1cm);
			}
			\fill[draw=gray,fill=white] (-2,-1) circle (0.1cm);
			\fill[draw=gray,fill=white] (-2,-2) circle (0.1cm);
			\fill[draw=gray,fill=white] (-1,-2) circle (0.1cm);
		\end{tikzpicture}
    \caption{Poles of $\widetilde{\mathcal{M}}_{44pp,st}$ ($p\geq 4$), of which $\widetilde{\mathcal{M}}_{4455,st}$ is a special case. Coefficients of poles at $s\geq s_0$ and $t_0\geq t_0$ obey a common formula. Those at $s>t_0$ and $t<t_0$ and those at $s<s_0$ and $t_0\geq t_0$ follow different formulas separately (as a result of truncation in $b_{mn}$). All these formulas have the common origin from hidden conformal symmetry, so we mark these poles with the same color blue. Coefficients of the poles marked red have to be worked out from recursive data from the window and the below window twist regions, while the poles marked white turn out to be absent from $\widetilde{\mathcal{M}}_{44pp,st}$.}
    \label{figue4455}
\end{figure}

At this point we are left with only the unknown coefficients of 6 simultaneous poles, as shown in figure \ref{figue4455}.
Two of these poles locate at $m=0$, and are determined by data from the s-channel window region. The explicit results for these coefficients are
\begin{align}
    c_{0,-1}^{st}&=\frac{(\beta -1) \left(291 \alpha ^2 \beta -115 \alpha ^2-378 \alpha  \beta +658 \alpha +99 \beta -447\right)}{144 \alpha ^2 \beta ^2}\;, \\
    c_{0,-2}^{st}&=\frac{(\alpha -1) (11 \alpha -3) (\beta -1)^2}{16 \alpha ^2 \beta ^2}\;. 
\end{align}
The remaining four poles have $m,n<0$, and are determined by data from the s-channel below window region following computations similar to that of the previous example. We find that their coefficients are
\begin{align}
    c_{-1,-1}^{st}&=\frac{\left(5 \alpha ^2-6 \alpha +4\right) (\beta -1)}{4 \alpha ^2 \beta }\;, \\
    c_{-2,-1}^{st}&=c_{-1,-2}^{st}=c_{-2,-2}^{st}=0\;.
\end{align}
Very interestingly, we see again that the simultaneous poles with $m+n<2-\mathcal{E}$ are absent, as in the case of $\langle 3355\rangle$. While this phenomenon still seems somewhat special, we have in fact calculated the results for all $\langle 44pp\rangle$ correlators, and their analytic structures share exactly the same pattern as $\langle 4455\rangle$, as shown in figure \ref{figue4455}. This structure can be extended to the general situation, as will be discussed in the next section.

\section{General Structure of one-loop Mellin amplitudes}\label{sec:oneloopgeneral}
	
From the analysis of the explicit examples in the previous section, a clear pattern emerges for the general structure of the one-loop Mellin amplitudes. We summarize its main features below.\footnote{To make sure it is not just a fluke, we have checked this pattern in many other examples. This includes the correlators $\langle 22pp\rangle$ $\langle 33pp\rangle$ and $\langle 44pp\rangle$ for $p=2,\cdots, 10$, and also some random cases $\langle 2468\rangle$, $\langle 3357\rangle$, $\langle 3557\rangle$, $\langle 4268\rangle$, $\langle 4424\rangle$, $\langle 4464\rangle$, $\langle 5555\rangle$, $\langle 6248\rangle$, $\langle 6626\rangle$, $\langle 6646\rangle$, $\langle 8848\rangle$.}
\begin{itemize}
\item {\it Analytic structure:} The one-loop reduced Mellin amplitude for arbitrary KK weights contains only simultaneous poles as in (\ref{eq:2222Bst}) with numerators which are constants with respect to the Mellin variables. 
\item {\it Hidden conformal symmetry:} The eight-dimensional hidden conformal symmetry observed at tree level determines most of the one-loop amplitude, and is only mildly broken inside a small portion of the ansatz with finitely many simultaneous poles. Naively, the hidden conformal symmetry is only expected to determine the $s$-poles with $s\geq s_+$ which contributes to the $\log^2U$ singularity. However, we find that its implication extends to a larger region and allows us to determine all poles with $s>s_0$. This phenomenon non-trivially occurs whenever the window region size $s_+-s_0\geq 4$. Using the Bose symmetry between $s$ and $t$, implemented by exchanging $p_1$ and $p_3$, hidden conformal symmetry also determines the simultaneous poles whenever $t>t_0$. 
\item {\it Bulk, edge and corner:} To facilitate the discussion, let us introduce the following terminologies which divide the regions of poles into three parts
\begin{equation}
\begin{split}
    {\rm Bulk}:={}&   \{(s,t)|s\geq s_0 \cap t\geq t_0 \}\backslash{\{(0,0)\}}\;,\\
    {\rm Edge}:={}&\{(s,t)|s>s_0\cap t<t_0\}\cup \{(s,t)|s<s_0\cap t>t_0\}\;,\\
    {\rm Corner}:={}& \{(s,t)|s\leq s_0 \cap t\leq t_0\}\;.
\end{split}
\end{equation}
Hidden conformal symmetry determines the poles in both the bulk and edge regions. The corner region does not follow from this hidden structure, but it contains only finitely many poles.
\item {\it Zeros:} Although the coefficients of the poles in the corner region need to be computed separately using CFT data from unitarity recursion, we observe a general feature which further simplifies the analytic structure. The numerators of the simultaneous poles in the lower-left triangular region bounded by $s+t<s_0+t_0-2\mathcal{E}+4$ are zero.
\end{itemize}

\subsection{Bulk region and edge region}\label{sec:bulkandedge}

Using the implication of the hidden symmetry \eqref{eq:deltaL2222} we can derive a general expression for the coefficients of the simultaneous poles in the bulk region. To carry out this computation, let us first recall that in \eqref{eq:Dpqrs} the difference operator $\widehat{\mathcal{D}}_{p_1p_2p_3p_4}$ acts on $\mathcal{L}_{2222}(s,t)$ as
\begin{equation}\label{eq:DL2222}
    \widehat{\mathcal{D}}_{p_1p_2p_3p_4}\circ \mathcal{L}_{2222}(s,t)=\sum _{i=-s_0/2-2}^{-s_-/2+2} \, \sum _{j=(\Sigma_{23}-\Sigma_{14}+\kappa_{u})/4 }^{(\Sigma_{23}-\Sigma_{14}-\kappa_{u})/4-i} \frac{ \sigma ^i\tau ^j}{a\hspace{1pt} !\hspace{1pt} b\hspace{1pt} !\hspace{1pt}c\hspace{1pt}!\hspace{1pt}d\hspace{1pt}!\hspace{1pt}e\hspace{1pt}!\hspace{1pt}f\hspace{1pt}!}\mathcal{L}_{2222}(s+2\, i,t+2\, k)\;,
\end{equation}
where $k=(2 j-\Sigma_{23}+4)/2$, $\sigma$, $\tau$ are related to $\alpha$, $\beta$ as in (\ref{eq:defst}), and the variables $a$ to $f$ are defined in \eqref{eq:abcdef}. In \eqref{eq:PRdef} we introduced the harmonic polynomials $P_R(\alpha)$ for the R-symmetry $SU(2)_R$. We can similarly define the polynomials $P_L(\alpha)$ for the flavor symmetry $SU(2)_L$ as well. With these we can decompose the R-symmetry and flavor symmetry cross ratios monomial $\sigma^i\tau^j$ as
\begin{equation}\label{eq:sigmatauPP}
    \sigma ^i \tau ^j=\sum _{R=s_-/2-2}^{-i}\  \sum _{L=s_-/2-2}^{-i}f_R\, f_L\, Y_{R,L}(\alpha,\beta)\;, \quad Y_{R,L}(\alpha,\beta)=P_R(\alpha)P_L(\beta)\;,
\end{equation}
where the coefficients are given by 
\begin{align}
    f_K=& (-1)^{-i-K}\frac{\Gamma (i+1)}{\Gamma (-j-K+1) \Gamma (i+j+K+1)}\nonumber\\
    &\, _3F_2\left(i+K,\frac{p_{21}}{2}+K+1,\frac{p_{34}}{2}+K+1;i+j+K+1,2 K+2;1\right)\;.
\end{align}
The virtue of this decomposition is that when we further act with the operator \eqref{eq:Delta4}
\begin{equation}
    \Delta^{(4)}= \frac{z\bar z}{\bar z-z} (\mathcal{D}^{+}_z-\mathcal{D}^{-}_{\alpha})( \mathcal{D}_{\zb}^{+}-\mathcal{D}^{-}_{\alpha}) \frac{\bar z-z}{z\bar z}\;,
\end{equation}
the derivative $\mathcal{D}^{-}_\alpha$ (defined in \eqref{eq:Dxpm}) acts diagonally on $P_R(\alpha)$
\begin{equation}
    \mathcal{D}^{-}_\alpha P_R(\alpha)=R (R+1) P_R(\alpha)\;.
\end{equation}
	
For fixed values of $i$, $j$, $R$ and $L$ in \eqref{eq:DL2222} and \eqref{eq:sigmatauPP}, the expression of $\widehat{\Delta}^{(4)}$ acting on $M(s+2i,t+2k)$ can be easily determined. As a result, the corresponding contribution to the coefficient $c_{m,n}^{st}$ of the simultaneous poles  in the bulk region can be computed to be
\begin{align}\label{eq:CijRL}
    C_{i,j,R,L}=&-\mathcal{P}_1^{p_1,p_3,p_2,p_4}\left(s',s'+t'+2\right) B\left(s'+2i,t'+2k+2\right) \nonumber \\
    &-\mathcal{P}_1^{p_3,p_2,p_1,p_4}\left(s',t'\right) B\left(s'+2i,t'+2k-2\right)\nonumber \\
    &+\mathcal{P}_3\left(s'\right) B\left(s'+2i-4,t'+2k+2\right) \nonumber \\
    &+\left[\mathcal{P}_1^{p_1,p_3,p_2,p_4}\left(s',s'+t'+2\right)+\mathcal{P}_1^{p_3,p_2,p_1,p_4}\left(s',t'\right)+\mathcal{P}_2(s')\right] B\left(s'+2i,t'+2k\right)\nonumber \\
    &+\left[\mathcal{P}_1^{p_1,p_2,p_3,p_4}\left(s',s'\right)+\mathcal{P}_3 \left(s'\right)-\mathcal{P}_4 \left(s',t'\right)\right] B\left(s'+2i-2,t'+2k\right) \nonumber \\
    &+\left[\mathcal{P}_1^{p_1,p_2,p_3,p_4}\left(s',s'\right)-\mathcal{P}_3 \left(s'\right)+\mathcal{P}_4 \left(s',t'\right)\right] B\left(s'+2i-2,t'+2k+2\right)\;.
\end{align}
Here the polynomials $\mathcal{P}$ are defined as
\begin{align}
    \mathcal{P}_1^{p_1,p_2,p_3,p_4}\left(s,t\right) &=\frac{1}{16} \left(\Sigma_{12}-t\right) \left(\Sigma_{34}-t\right) (2 R-s+2) (2 R+s)\;,\\
    \mathcal{P}_2\left(s \right) &=\frac{1}{16} (2 R-s) (2 R-s+2) (2 R+s) (2 R+s+2)\;,\\
    \mathcal{P}_3\left(s \right) &=\frac{1}{16} \left(\Sigma_{12}-s\right) \left(\Sigma_{12}-s+2\right) \left(\Sigma_{34}-s\right) \left(\Sigma_{34}-s+2\right)\;,\\
    \mathcal{P}_4\left(s ,t\right) &=\frac{1}{8} \left(\Sigma_{12}-s\right) \left(\Sigma_{34}-s\right) \left(p_1 p_3+p_2 p_4-(s-2) (t+2)\right)\;,
\end{align}
and the variables $s'$ and $t'$ are related to $s_0$ and $t_0$ respectively,
\begin{equation}
    s'=\frac{1}{2}(s_0+2m-4),\quad t'=\frac{1}{2}(t_0+2n-4)\;.
\end{equation}
The function $B(x,y)$ serves as an indicator of the shift applied to $\mathcal{L}_{2222}$
\begin{equation}
    B(x,y)=b_{x/2-2,\, y/2-2}=\frac{2y-8}{3 (x-2) (x+y-5) (x+y-4)}\;.
\end{equation}
In other words, it is related to the coefficients appearing in $\mathcal{L}_{2222}$ but with indices shifted. Here it is important to be reminded again that $b_{m,n}$ is defined to be zero for negative indices \eqref{eq:bmntruncation}
\begin{equation}\label{eq:bmntruntion}
    b_{m,n}=0 \;, \quad   n<0\;.
\end{equation}
All in all, we find the following final expression for the coefficients of the simultaneous poles in the bulk region which can be written as a quadruple sum\footnote{We also collect this formula in the ancillary file for the reader's convenience.}
\begin{equation}\label{generalcstpqrs}
    c^{st}_{m,n}=\sum _{i=-s_0/2-2}^{-s_-/2+2} \, \sum _{j=(\Sigma_{23}-\Sigma_{14}+\kappa_{u})/4 }^{(\Sigma_{23}-\Sigma_{14} -\kappa_{u})/4-i}  \sum _{R=s_-/2-2}^{-i}  \sum _{L=s_-/2-2}^{-i}\frac{ f_R\, f_L\, C_{i,j,R,L}}{a\hspace{1pt} !\hspace{1pt} b\hspace{1pt} !\hspace{1pt}c\hspace{1pt}!\hspace{1pt}d\hspace{1pt}!\hspace{1pt}e\hspace{1pt}!\hspace{1pt}f\hspace{1pt}!}  Y_{R,L}(\alpha,\beta)\;.
\end{equation}
	
We can follow exactly the same procedure to obtain the coefficients of poles in the s-channel edge region (with $s>s_0$ and $t<t_0$). The only subtlety here is that, due to the truncation of $b_{m,n}$ \eqref{eq:bmntruntion}, the resulting analytic expressions in this region differ from that of \eqref{eq:CijRL} in the bulk region. Nevertheless, the results in both regions root from the same origin in terms of the action $\widehat{\Delta}^{(4)}\circ\widehat{\mathcal{D}}_{p_1p_2p_3p_4}\circ\mathcal{L}_{2222}$. Additionally, by crossing from $\langle p_3p_2p_1p_4\rangle$ we can similarly derive the corresponding result for coefficients of poles in the t-channel edge region (with $t>t_0$ and $s<s_0$). 
	
It is in the above sense that we conclude the eight-dimensional hidden symmetry dictates the structure of the Mellin amplitude in both the bulk and the edge regions (marked blue in figure \ref{pqrs}). Therefore we have successfully obtained the general expression for $\langle p_1p_2p_3p_4\rangle$, with the exception of the corner region defined by $s\leq s_0$ and $t\leq t_0$.

\subsection{Corner region}
	
Deriving a general formula for the simultaneous pole coefficients in the corner region presents a greater challenge as it requires finding a  complete expression for $\langle a^{(0)}\gamma^{(1)}\rangle_{p_1p_2p_3p_4}$, $\langle a^{(1)}\rangle_{p_1p_2p_3p_4}$, $\langle c^{(1)}\rangle_{p_1p_2p_3p_4}$ (and $\langle b^{(1)}\rangle_{p_1p_2p_3p_4}$). However, the size of the corner region is determined by extremality. At each extremality, the computational difficulty does not vary much for different weight distributions. Therefore, we can compute these coefficients case by case with fixed extremality and then try to find the general pattern. In this way, we have found the general formulas for the coeffcients for $\mathcal{E}=2$ and $\mathcal{E}=3$, which will be presented in the next subsection. 
 
\begin{figure}[ht]
    \centering
		\begin{tikzpicture}
			\definecolor{cr1}{HTML}{fcf6bd}
			\definecolor{cr2}{HTML}{d0f4de}
			\definecolor{cr3}{HTML}{a9def9}
			\definecolor{naturegreen}{HTML}{00A686}
			\definecolor{naturered}{HTML}{E64B35}
			\definecolor{naturecolor3}{HTML}{A9DEF9}
			\draw[draw=cr3, fill=cr3, opacity=0.5] (-3,0) rectangle (2,2) ;
			\draw[draw=cr3, fill=cr3, opacity=0.5] (0,-3) rectangle (2,0) ;
			\draw[draw=naturered, fill=naturered, opacity=0.3] (-3,0) -- (0,0) -- (0,-3) -- cycle;
			\draw[->] (-3.5,0) node [anchor=east] {$t_0$} -- (2.5,0) node[right] {$s$};
			\draw[->] (0,-3.5) node [anchor=north] {$s_0$} -- (0,2.5) node[above] {$t$}; 
			\draw[color=black, thin, opacity=0.8, dashed] (-3,-3.5) node [anchor=north] {$s_-$} --(-3,2.5) ;
			\draw[color=black, thin, opacity=0.8, dashed] (-3.5,-3) node [anchor=east] {$t_-$} --(2.5,-3) ;
			\foreach \x in {-3,-2,-1,0,1,2}
			\foreach \y in {-3,-2,-1,0,1,2}
			{
				\fill[fill=black](\x,\y) circle (0.11cm);
			}
			\foreach \x in {-3,-2,-1,0}
			\foreach \y in {-3,-2,-1,0}
			{
				\fill[fill=white](\x,\y) circle (0.1cm);
			}
			\foreach \x in {-3,-2,-1,0,1,2}
			\foreach \y in {1,2}
			{
				\fill[fill=cr3](\x,\y) circle (0.1cm);
			}
			\foreach \x in {1,2}
			\foreach \y in {-3,-2,-1,0}
			{
				\fill[fill=cr3](\x,\y) circle (0.1cm);
			}
			\foreach \x in {-3,-2,-1,0}
			\foreach \y in {0}
			{
				\fill[fill=naturered, opacity=0.5] (\x,\y) circle (0.1cm);
			}
			\foreach \x in {-2,-1,0}
			\foreach \y in {-1}
			{
				\fill[fill=naturered, opacity=0.5] (\x,\y) circle (0.1cm);
			}
			\foreach \x in {-1,0}
			\foreach \y in {-2}
			{
				\fill[fill=naturered, opacity=0.5] (\x,\y) circle (0.1cm);
			}
			\foreach \x in {0}
			\foreach \y in {-3}
			{
				\fill[fill=naturered, opacity=0.5] (\x,\y) circle (0.1cm);
			}
			\foreach \x in {-3,-2,-1}
			\foreach \y in {-3,-2}
			\foreach \z in {-3}
			{
				\fill[fill=white,dashed](\x,-4-\x) circle (0.12cm);
				\fill[fill=white,dashed](\y,-5-\y) circle (0.12cm);
				\fill[fill=white,dashed](\z,-6-\z) circle (0.12cm);
				\fill[fill=gray,dashed](\x,-4-\x) circle (0.11cm);
				\fill[fill=gray,dashed](\y,-5-\y) circle (0.11cm);
				\fill[fill=gray,dashed](\z,-6-\z) circle (0.11cm);
				\fill[fill=white,dashed](\x,-4-\x) circle (0.1cm);
				\fill[fill=white,dashed](\y,-5-\y) circle (0.1cm);
				\fill[fill=white,dashed](\z,-6-\z) circle (0.1cm);
			}
			\draw [{Latex}-{Latex}] (-3,0.3) -- +(3,0);
			\node [anchor=south] at (-1.5,0.3) {\scriptsize $2\mathcal{E}-4$};
			\node [anchor=north] at (-0.05,0) {$(s_0,t_0)$};
			\draw[draw=cr3, fill=cr3, opacity=0.8] (3.5,1.8) rectangle (3.8,2.1) ;
			\draw[draw=naturered, fill=naturered, opacity=0.3] (3.5,1.2) rectangle (3.8,1.5) ;
			\node[right] at (3.8,1.95) {Bulk $+$ Edge Region};
			\node[right] at (3.8,1.3)  {Corner Region};
		\end{tikzpicture}
    \caption{The general pattern of poles in $\widetilde{\mathcal{M}}_{p_1p_2p_3p_4,st}$.}
    \label{pqrs}
\end{figure}
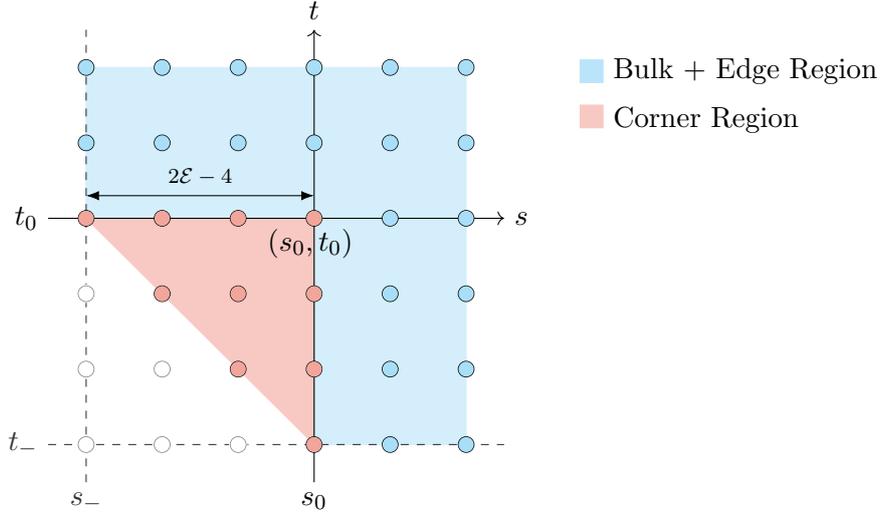

By continuing this computation to even higher extremalities we also observe an interesting pattern among the coefficients in this region. This pattern is illustrated in figure \ref{pqrs}, where the poles in the corner region are further separated in to two groups, marked by red and white respectively. The pole coefficients at the white points always turn out to vanish. We have verified this vanishing phenomenon in various examples with extremality up to $\mathcal{E}=5$. Via crossing, the vanishing of poles also happen in other channels. We have
\begin{align}\label{zerocondition}
    c^{st}_{m,n}&=0,\qquad s+t  <s_0+t_0-2\mathcal{E}+4\;,\\
    c^{su}_{m,n}&=0,\qquad s+\tilde{u}<s_0+u_0-2\mathcal{E}+4\;,\\
    c^{tu}_{m,n}&=0,\qquad t+\tilde{u}<t_0+u_0-2\mathcal{E}+4\;.
\end{align}

\subsection{Explicit results with generic weights at fixed extremalities}
	
Let us now present the results for the simultaneous pole coefficients for the lowest two values of the extremality. It is convenient to define the following parameters
\begin{equation}
    \kappa_s=|\Sigma_{12}-\Sigma_{34}|\;, \qquad \kappa_t=|\Sigma_{23}-\Sigma_{14}|\;, \qquad \kappa_u=|\Sigma_{24}-\Sigma_{13}|\;.
\end{equation}
They give an alternative parameterization than $\{p_i\}$ when used together with the extremality $\mathcal{E}$. The advantage of using this parameterization is that these $\kappa$ parameters directly measures the size of the windows in each OPE channel.

\paragraph{\underline{$\langle{\mathcal{O}}_{p_1}{\mathcal{O}}_{p_2}{\mathcal{O}}_{p_3}{\mathcal{O}}_{p_4}\rangle ,\quad \mathcal{E}=2$}}~\\
	
\noindent In the simplest case of $\mathcal{E}=2$, the summation in \eqref{generalcstpqrs} can be easily performed, so that coefficients of poles in the bulk region become
\begin{equation}
    c_{m,n}^{st}= (\alpha\beta)^{-\frac{\kappa_t+\kappa_u}{4}} \frac{(\Sigma -4) F_{m,n} }{24  (m+n)_3 \left( \frac{\kappa_s}{2}\right) ! \left( \frac{\kappa_t}{2}\right) !\left( \frac{\kappa_u}{2}\right) !}\;,
\end{equation}
where
\begin{align}
    F_{m,n}=&(\kappa_u  +2) (m+1) (n+1) (m+n)\nonumber \\
    &+( \kappa_t  +2) \,n\, (m+1)  (m+n+2)+(\kappa_s +2) \,m\, (n+1) (m+n+2)\;.
\end{align}
Special care need to be taken for the simultaneous pole at $(m,n)=(0,0)$ in the corner region. The correct coefficient is determined to be
\begin{equation}
    c^{st}_{0,0}= (\alpha\beta)^{-\frac{\kappa_t+\kappa_u}{4}}\frac{(\kappa_u+6) (\Sigma -4)}{48  \left( \frac{\kappa_s}{2}\right) ! \left( \frac{\kappa_t}{2}\right) !\left( \frac{\kappa_u}{2}\right) !}\;.
\end{equation}~\\

\paragraph{\underline{$\langle {\mathcal{O}}_{p_1}{\mathcal{O}}_{p_2}{\mathcal{O}}_{p_3}{\mathcal{O}}_{p_4}\rangle ,\quad \mathcal{E}=3$}}~\\
	
\noindent For the case of $\mathcal{E}=3$, we attach the bulk and edge results in the ancillary file, and we will only focus on the corner region here. The simultaneous pole at $(m,n)=(-1,-1)$ is absent as we mentioned above
\begin{equation}
    c^{st}_{-1,-1}=0\;.
\end{equation}
For simplicity, let us define
\begin{equation}
    C^{st}_{m,n}= (\alpha\beta)^{-\frac{\kappa_t+\kappa_u}{4}} c^{st}_{m,n}\;.
\end{equation}
For $(m,n)=(-1,0)$, the coefficient is
\begin{equation}
    C^{st}_{-1,0}=-\frac{-12 \kappa _s-2 \kappa _s \kappa _u-6 \kappa _t+\kappa _t \kappa _u-\kappa _u^2-16 \kappa _u-60}{48  \left(\frac{\kappa _s}{2}+1\right)!  \left(\frac{\kappa _t}{2}\right)!  \left(\frac{\kappa _u}{2}\right)!}-\frac{3 \kappa _t+\kappa _u+6}{24 \alpha   \left(\frac{\kappa _s}{2}\right)! \left(\frac{\kappa _t}{2}\right)! \left(\frac{\kappa _u}{2}\right)!}.
\end{equation}
For $(m,n)=(0,-1)$, the result is
\begin{align}
    C^{st}_{0,-1}=&-\frac{(\beta -1) \left(-6 \kappa _s+\kappa _s \kappa _u-12 \kappa _t-2 \kappa _t \kappa _u-\kappa _u^2-16 \kappa _u-60\right)}{48 \beta   \left(\frac{\kappa _s}{2}\right)!  \left(\frac{\kappa _t}{2}+1\right)!  \left(\frac{\kappa _u}{2}\right)!}\nonumber\\
    &+\frac{(\beta -1) \left(3 \kappa _s \kappa _t+\kappa _s \kappa _u-6 \kappa _t-\kappa _t \kappa _u-\kappa _u^2-14 \kappa _u-48\right)}{48 \alpha  \beta   \left(\frac{\kappa _s}{2}\right)!  \left(\frac{\kappa _t}{2}+1\right)!  \left(\frac{\kappa _u}{2}\right)!}\;,
\end{align}
and finally
\begin{align}
    C^{st}_{0,0}=&-\frac{\kappa _s \left(2 \kappa _t+\kappa _u+12\right)+\left(\kappa _t+\kappa _u+8\right) \left(\kappa _t+\kappa _u+10\right)}{72 \alpha   \left(\frac{\kappa _s}{2}\right)!  \left(\frac{\kappa _t}{2}+1\right)!  \left(\frac{\kappa _u}{2}\right)!}\nonumber \\
    &-\frac{\kappa _s \left(\kappa _u+8\right)+3 \kappa _t^2+2 \kappa _t \left(\kappa _u+11\right)+\left(\kappa _u+8\right) \left(\kappa _u+10\right)}{72 \beta   \left(\frac{\kappa _s}{2}\right)!  \left(\frac{\kappa _t}{2}+1\right)!  \left(\frac{\kappa _u}{2}\right)!}\nonumber \\
    &+\frac{\kappa _s \left(3 \kappa _t+\kappa _u+14\right) \left(\kappa _s+\kappa _t+\kappa _u+10\right)+\kappa _t^2 \left(\kappa _u+14\right)+\left(\kappa _u+10\right) \left(14 \kappa _t+\kappa _t \kappa _u+4 \kappa _u+32\right)}{144  \left(\frac{\kappa _s}{2}+1\right)!  \left(\frac{\kappa _t}{2}+1\right)!  \left(\frac{\kappa _u}{2}\right)!}\nonumber \\
    &+\frac{\left(\kappa _t+\kappa _u+8\right) \left(3 \kappa _t \left(\kappa _u+6\right)+\kappa _u \left(\kappa _u+16\right)+52\right)-\kappa _s \left(\kappa _t-\kappa _u-6\right) \left(3 \kappa _t+\kappa _u+8\right)}{144 \alpha  \beta   \left(\frac{\kappa _s}{2}\right)!  \left(\frac{\kappa _t}{2}+1\right)!  \left(\frac{\kappa _u}{2}+1\right)!}\;.
\end{align}

\section{Outlook}
	
In this paper we carried out a systematic study in Mellin space of one-loop correlators of four super gluons with arbitrary KK modes in $\mathrm{AdS}_5\times\mathrm{S}^3$. In addition to obtaining the Mellin amplitudes for several infinite families of correlators, we also made a few preliminary observations on their general structures. In particular, we observed that the validity of the universal formula descending from the eight-dimensional hidden conformal symmetry extends to a much larger set of Mellin poles than those that are known to be tied to this symmetry. This leaves only a finite set of coefficients to be fixed at each extremality, of which the number is independent of the distribution of the weights. Our results pointed out that the super gluon scattering amplitudes in $\mathrm{AdS}_5\times\mathrm{S}^3$ enjoy remarkable simplicity. Our work leads to many interesting directions for future explorations. We list a few of these directions below.

First of all, the observed simplicity in the super gluon amplitudes still needs to be better understood. The extension of the validity of the hidden conformal symmetry at one-loop level is quite unexpected. It would be interesting to find an intuitive explanation for this phenomenon. At tree level, hidden conformal symmetry can be elegantly manifested by writing down a generating function of reduced correlators which is obtained by promoting the four dimensional distances in the lowest weight correlator into eight-dimensional distances. Ideally, we would like to also have a similar position space description at one-loop level that captures this extension. Finding it will be very useful for understanding the physical implications of this hidden symmetry as well as its breaking in the corner region. 
 
In this paper we derived universal formulas for the coefficients of Mellin poles in the bulk and edge regions as was discussed in section \ref{sec:oneloopgeneral}. These results are valid for correlators with arbitrary external KK weights. However, we do not yet have a full control of the pole coefficients in the corner region, except for the special cases with $\mathcal{E}=2$ and $\mathcal{E}=3$. For higher extremalities, we are only able to work out the coefficients case by case at the moment. To obtain general expressions for these corner coefficients, it will be important to perform a more detailed study for the data in the below window region which are responsible for these coefficients. Related to this, it would also be interesting to understand physically why half of the pole coefficients vanish in this region, which we observed empirically.

Last but not least, one can also perform similar analysis for one-loop amplitudes in other theories. Two noticeable examples are IIB supergravity in AdS$_5\times$S$^5$ and AdS$_3\times$S$^3$ which also enjoy the same kind of hidden conformal symmetry as SYM on AdS$_5\times$S$^3$. The presence of the hidden conformal symmetry should similarly simplify the analysis of the Mellin amplitudes and it would be interesting to see if the implication of the hidden conformal symmetry also receives an analogous extension. Compared to the super gluon case, the  supergravity amplitudes are simpler due to the absence of color and flavor symmetries. On the other hand, precisely because of their absence processes which are otherwise distinguished by colors are all mixed. This makes the analysis of the unitarity more complicated and might lead to different analytic properties for the Mellin amplitudes.

\acknowledgments
	
ZH, BW and EYY are supported by National Science Foundation of China under Grant No.~12175197 and Grand No.~12147103. EYY is also supported by National Science Foundation of China under Grant No.~11935013, and by the Fundamental Research Funds for the Chinese Central Universities under Grant No.~226-2022-00216. XZ is supported by funds from University of Chinese Academy of Sciences (UCAS), funds from the Kavli Institute for Theoretical Sciences (KITS), the Fundamental Research Funds for the Central Universities, and the NSFC Grant No.~12275273.

\appendix
\section{Color decomposition}\label{appx:colordecp}
	
In this appendix, we present a detailed discussion about how to handle the color structures. Typically, at each perturbative order the correlator $\mathcal{G}^{I_1I_2I_3I_4}$ can be decomposed into different color factors in the form of
\begin{equation}\label{eq:CIIIIexpansion}
    \mathcal{G}^{I_1I_2I_3I_4}=\sum_CC^{I_1I_2I_3I_4}\mathcal{G}_C\;,
\end{equation}
For instance, in the formula \eqref{eq:G0pqqp} for the diconnected correlator, the terms $\bs$, $\bt$, and $\bu$ are the color factors $C^{I_1I_2I_3I_4}$
\begin{equation}
    \mathcal{G}^{(0),\,I_1 I_2 I_3 I_4}_{pqpq} = \delta_{pq}\,\mathtt{b}_s +  \delta_{pq}\,\beta^2 \left(\frac{U}{\sigma}\right)^{\frac{p+q}{2}} \mathtt{b}_t +\frac{\beta^2}{(1-\beta)^2}\left(\frac{U}{\sigma}\right)^{p}\left(\frac{\tau}{V}\right)^{\frac{p+q}{2}}\mathtt{b}_u \;.
\end{equation}
An equivalent but sometimes more convenient option is to use projectors which are in one-to-one correspondence with the irreducible representations appearing in the tensor product of two adjoint representations. The above color factors can be written as 
\begin{equation}\label{eq:cons}
    C^{I_1 I_2 I_3 I_4} = \sum_{\mathbf{a}\in \mathbf{adj}\otimes\mathbf{adj}} P^{I_1 I_2 \vert I_3 I_4}_{\mathbf{a}} C_{\mathbf{a}}\;,
\end{equation}
where $P_{\mathbf{a}}^{I_1I_2\vert I_3I_4}$ represents the s-channel projector associated with the irreducible representation $\mathbf{a}$. The s-channel projectors obey several symmetry properties
\begin{equation}\label{eq:Pparity}
    P^{I_1 I_2 \vert I_3 I_4}_{\mathbf{a}} = (-1)^{\abs{R_{\mathbf{a}}}} P^{I_2 I_1 \vert I_3 I_4}_{\mathbf{a}}\;,\qquad P^{I_1 I_2 \vert I_3 I_4}_{\mathbf{a}} = P^{I_3 I_4 \vert I_1 I_2}_{\mathbf{a}}\;,
\end{equation}
where $\abs{R_{\mathbf{a}}}$ denotes the parity of the representation $\mathbf{a}$. Additionally, the projectors are idempotent
\begin{equation}\label{eq:Pidempotency}
    P^{I_1 I_2 \vert I_5 I_6}_{\mathbf{a}} P^{I_6 I_5 \vert I_3 I_4}_{\mathbf{b}} = \delta_{{\mathbf{a}}{\mathbf{b}}}P^{I_1 I_2 \vert I_3 I_4}_{\mathbf{a}}\;.
\end{equation}
Particularly, from \eqref{eq:Pidempotency}, we have $P^{I_1 I_2 \vert I_3 I_4}_{\mathbf{a}} P^{I_1 I_2 \vert I_3 I_4}_{\mathbf{b}} = \delta_{{\mathbf{a}}{\mathbf{b}}}{\rm dim}(R_{\mathbf{a}})$. By utilizing the properties mentioned above, we can calculate the coefficient $C_{\mathbf{a}}$ of the color factor in \eqref{eq:cons} as follows
\begin{equation}
    C_{\mathbf{a}} = \text{dim}(R_{\mathbf{a}})^{-1} P^{I_1 I_2 \vert I_3 I_4}_{\mathbf{a}} C^{I_1 I_2 I_3 I_4}\;.
\end{equation}
	
Different color factors can also be related by crossing relations. We define $C_s \equiv C^{I_1 I_2 I_3 I_4}$. When we exchange $p_1$ and $p_3$, $C_s$ becomes $C_t \equiv C^{I_3 I_2 I_1 I_4}$. Similarly, by changing $p_1$ and $p_4$, $C_s$ transforms into $C_u \equiv C^{I_4 I_2 I_3 I_1}$. Using projectors we find
\begin{subequations}\label{eq:Cscrossing}
\begin{align}
    (C_t)_{\mathbf{a}} &= ({\rm F}_t)_{\mathbf{a}}^{\ {\mathbf{a}}'} (C_s)_{{\mathbf{a}}'}\;,\\
    (C_u)_{\mathbf{a}} &= ({\rm F}_u)_{\mathbf{a}}^{\ {\mathbf{a}}'} (C_s)_{{\mathbf{a}}'}\;,
\end{align}
\end{subequations}
where
\begin{subequations}
\begin{align}
    {({\rm F}_t)_{\mathbf{a}}}^{{\mathbf{a}}'} & \equiv \sum_{I_1,I_2,I_3,I_4} \text{dim}(R_{\mathbf{a}})^{-1} P^{I_3 I_2 \vert I_1 I_4}_{\mathbf{a}} P^{I_1 I_2 \vert I_3 I_4}_{{\mathbf{a}}'}\;,\\
    ({\rm F}_u)_{\mathbf{a}}^{\ {\mathbf{a}}'} & \equiv \sum_{I_1,I_2,I_3,I_4} \text{dim}(R_{\mathbf{a}})^{-1} P^{I_4 I_2 \vert I_3 I_1}_{\mathbf{a}} P^{I_1 I_2 \vert I_3 I_4}_{{\mathbf{a}}'}\;.
\end{align}
\end{subequations}
To demonstrate the use of the projectors, let us now consider the specific example of the $E_8$ group where these matrices can be explicitly found in \cite{Chang:2017xmr}
\begin{equation}\label{eq:E8FtFu}
    {\rm F}_t=\left(
		\begin{array}{ccccc}
			\frac{1}{248} & \frac{125}{8} & \frac{3375}{31} & 1 & \frac{245}{2} \\
			\frac{1}{248} & -\frac{3}{8} & \frac{27}{31} & \frac{1}{5} & -\frac{7}{10} \\
			\frac{1}{248} & \frac{1}{8} & \frac{23}{62} & -\frac{1}{30} & -\frac{7}{15} \\
			\frac{1}{248} & \frac{25}{8} & -\frac{225}{62} & \frac{1}{2} & 0 \\
			\frac{1}{248} & -\frac{5}{56} & -\frac{90}{217} & 0 & \frac{1}{2} \\
		\end{array}
		\right)\;,\quad {\rm F}_u=\left(
		\begin{array}{ccccc}
			\frac{1}{248} & \frac{125}{8} & \frac{3375}{31} & -1 & -\frac{245}{2} \\
			\frac{1}{248} & -\frac{3}{8} & \frac{27}{31} & -\frac{1}{5} & \frac{7}{10} \\
			\frac{1}{248} & \frac{1}{8} & \frac{23}{62} & \frac{1}{30} & \frac{7}{15} \\
			-\frac{1}{248} & -\frac{25}{8} & \frac{225}{62} & \frac{1}{2} & 0 \\
			-\frac{1}{248} & \frac{5}{56} & \frac{90}{217} & 0 & \frac{1}{2} \\
		\end{array}
		\right).\;
\end{equation}
Let us first consider the tree-level color factors $\mathtt{c}_{s}$, $\mathtt{c}_{t}$, and $\mathtt{c}_{u}$, which are expressed in terms of the structure constants as follows
\begin{equation}\label{app:csdef}
    \mathtt{c}_s = f^{I_1 I_2 J} f^{J\, I_3 I_4}\;,\quad \mathtt{c}_t= f^{I_1 I_4 J} f^{J\, I_2 I_3}\;,\quad \mathtt{c}_u = f^{I_1 I_3 J} f^{J\, I_4 I_2}\;.
\end{equation}
In the example of $E_8$, the tensor product $\mathbf{adj}\otimes\mathbf{adj}$ includes in total five irreducible representations: $\mathbf{1}$, $\mathbf{3875}$, $\mathbf{27000}$, $\mathbf{248}$ ($\mathbf{adj}$), and $\mathbf{30380}$. Note that $\cs$ already represents the exchange of the adjoint representation ${\bf 248}$ in the s-channel. Therefore, it is proportional to the projector $P_{\mathbf{248}}^{I_1I_2\vert I_3I_4}$, and we have
\begin{equation}\label{eq:csdef}
    (\mathtt{c}_s)_{\mathbf{a}}\equiv P_{\mathbf{248}}^{I_1I_2\vert I_3I_4} (\mathtt{c}_s)^{I_1I_2I_3I_4} = \psi^2 h^\vee (\underset{\substack{\scriptsize\uparrow\\\mathbf{1}}}{0},\underset{\substack{\scriptsize\uparrow\\\mathbf{3875}}}{0},\underset{\substack{\scriptsize\uparrow\\\mathbf{27000}}}{0},\underset{\substack{\scriptsize\uparrow\\\mathbf{248}}}{1},\underset{\substack{\scriptsize\uparrow\\\mathbf{30380}}}{0})^T\;.
\end{equation}
Here $\psi^2$ represents the squared length of the longest root and $h^\vee$ is the dual Coxeter number. On the other hand, the decomposition of $\mathtt{c}_t$ and $\mathtt{c}_u$ involves a mixture of different s-channel projectors\footnote{Note that $(-\mathtt{c}_t)_{\mathbf{a}} = ({\rm F}_t)_{\mathbf{a}}^{\ {\mathbf{a}}'} (\mathtt{c}_s)_{{\mathbf{a}}'}$ and $(-\mathtt{c}_u)_{\mathbf{a}} = ({\rm F}_u)_{\mathbf{a}}^{\ {\mathbf{a}}'} (\mathtt{c}_s)_{{\mathbf{a}}'}$. The additional minus sign come from the discrepancy between $\mathtt{c}_{s/t/u}$ and $C_{s/t/u}$.}
\begin{align}\label{eq:ctudef}
    (\mathtt{c}_t)_{\mathbf{a}}= -\psi^2 h^\vee \left(1,\frac{1}{5},-\frac{1}{30},\frac{1}{2},0 \right)^T,\quad 
    (\mathtt{c}_u)_{\mathbf{a}} = \psi^2 h^\vee \left(1,\frac{1}{5},-\frac{1}{30},-\frac{1}{2},0 \right)^T.
\end{align}	
	
Let us turn to the color structures at one loop, which are denoted as $\dst$, $\dsu$ and $\dtu$ in (\ref{eq:dst}). We can obtain these one-loop color factors by joining together two tree-level color factors
\begin{equation}
    (C_1C_2)^{I_1I_2I_3I_4}=C_1^{I_1I_2I_5I_6}C_2^{I_6I_5I_3I_4}\;.
\end{equation}
For example, the structure $\dst$ can be viewed as the product of two t-channel adjoint projectors. Similarly, $\dsu$ is the product of two adjoint projectors in the t-channel and u-channel. To write these structures in the s-channel, we can use the crossing relation to express each crossed channel projector as a linear combination of the s-channel projectors. The idempotent property \eqref{eq:Pidempotency} then allows us to evaluate the product and express the result as a sum of s-channel projectors. Following these steps, we find 
\begin{subequations}\label{appxeq:iddstandc}
\begin{align}
    (\mathtt{d}_{st})_{\mathbf{a}} &= ((-\mathtt{c}_t)^2)_{\mathbf{a}} = \left( \psi^2 h^\vee \right)^2 \left(1,\frac{1}{25},\frac{1}{900},\frac{1}{4},0 \right),\\
    (\mathtt{d}_{su})_{\mathbf{a}} &= (-\mathtt{c}_t \mathtt{c}_u)_{\mathbf{a}} = \left( \psi^2 h^\vee \right)^2 \left(1,\frac{1}{25},\frac{1}{900},-\frac{1}{4},0 \right).
\end{align}
\end{subequations}
The structure $\dtu$ can be obtained by crossing
\begin{align}
    (\mathtt{d}_{tu})_{\mathbf{a}} = {({\rm F}_u)_{\mathbf{a}}}^{\mathbf{a}'} (\mathtt{d}_{st})_{\mathbf{a}'} &= \left( \psi^2 h^\vee \right)^2 \left(\frac{1}{2},-\frac{3}{50},\frac{4}{225},0,0 \right).
\end{align}
\begin{figure}[h]
    \centering
		\begin{tikzpicture}
			\draw [line width=1.5pt] (-0.5,0.5) -- (-0.5,-0.5);
			\draw [line width=1.5pt](-0.5,-0.5) -- (-1.5,-0.5);
			\draw [line width=1.5pt](-1.5,-0.5) -- (-1.5,0.5);
			\draw [line width=1.5pt](-1.5,0.5) -- (-0.5,0.5);
			
			\draw [line width=1.5pt] (-1.5,-0.5) -- (-2.0,-1.0);
			\draw [line width=1.5pt] (-1.5,0.5) -- (-2.0,1.0);
			\draw [line width=1.5pt] (-0.5,0.5) -- (0,1.0);
			\draw [line width=1.5pt] (-0.5,-0.5) -- (0,-1.0);
			
			\draw [draw=blue, densely dashed ](-1,0.8) -- (-1,-0.8);
			
			\node [anchor=east] at (-2.0,1.0) {$I_1$};
			\node [anchor=west] at (0,1.0) {$I_4$};
			\node [anchor=west] at (0,-1.0) {$I_3$};
			\node [anchor=east] at (-2.0,-1.0) {$I_2$};
			\node at (-3.5,0) {$\dst=(-{\mathtt{c}}_{t})^{2} = $};
		\end{tikzpicture} 
		\begin{tikzpicture}
			\draw [line width=1.5pt] (-0.5,0.5) -- (-0.5,-0.5);
			\draw [line width=1.5pt](-0.5,-0.5) -- (-1.5,-0.5);
			\draw [line width=1.5pt](-1.5,-0.5) -- (-1.5,0.5);
			\draw [line width=1.5pt](-1.5,0.5) -- (-0.5,0.5);
			
			\draw [line width=1.5pt] (-1.5,-0.5) -- (-2.0,-1.0);
			\draw [line width=1.5pt] (-1.5,0.5) -- (-2.0,1.0);
			\draw [line width=1.5pt] (-0.5,0.5) -- (0,-1.0);
			\draw [draw=white,line width=2pt] (-0.37,0.11) -- (-0.28,-0.16);
			\draw [line width=1.5pt] (-0.5,-0.5) -- (0,1.0);
			\draw [draw=blue, densely dashed ](-1,0.8) -- (-1,-0.8);
			
			\node at (-3.6,0) {$\dsu=(-{\mathtt{c}}_{t}){\mathtt{c}}_{u} = $};
			
			\node [anchor=east] at (-2.0,1.0) {$I_1$};
			\node [anchor=west] at (0,1.0) {$I_4$};
			\node [anchor=west] at (0,-1.0) {$I_3$};
			\node [anchor=east] at (-2.0,-1.0) {$I_2$};
		\end{tikzpicture}
    \caption{The color structures of one loop diagrams, generated by gluing 2 exchange diagrams together. }
    \label{fig:leadinglog}
\end{figure}
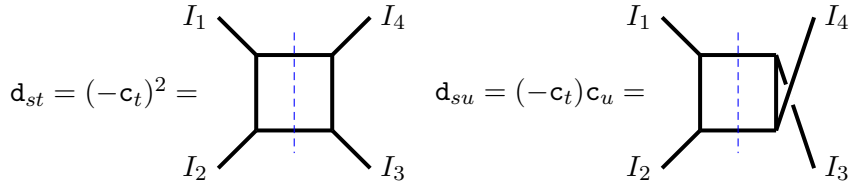

\section{Superconformal blocks of half-BPS and semi-short multiplets}\label{appx:superblocks}
	
In section \ref{sec:scfblockdecom} we briefly reviewed the decomposition of the four-point correlator \eqref{eq:solscfWI2} into a chiral part $f(z;\beta)$ and a dynamical part $\mathcal{F}(z,\bar{z};\alpha,\beta)$ which takes the form
\begin{equation}\label{eq:fFdecp}
\begin{split}
    \mathcal{G}(z,\zb;\alpha,\beta) = &\ \frac{z-\alpha}{z-\zb}\chi(\zb;\alpha)f(z;\beta) + \frac{\zb-\alpha}{\zb-z}\chi(z;\alpha)f(\zb;\beta) \\
    &+ \frac{(z-\alpha)(\zb-\alpha)}{z\zb\alpha^2} \mathcal{F}(z,\zb;\alpha,\beta)\;,
\end{split}
\end{equation}
with the function $\chi(z;\alpha)$ defined in \eqref{eq:chidef}. We mentioned that this decomposition has the virtue that it simplifies the expressions of superconformal blocks. In particular, the long superconformal blocks receive contributions only from $\mathcal{F}$. In this appendix we also collect the results of superconformal blocks for other multiplets
	\begin{subequations}\label{eq:superblocks}
		\begin{align}
			\mathcal{B}_R(z,\zb;\alpha) =& \begin{cases}
				f(z) = k_R(z)\;,\\[5px]
				\displaystyle \mathcal{F}(z,\zb;\alpha) = \sum_{i=i_0}^{R-2} g_{2i+4,R-i-2}(z,\zb)P_{i}(\alpha)\;,
			\end{cases}\\
			\label{eq:Cblock}\mathcal{C}_{R,\ell}(z,\zb;\alpha) =& \begin{cases}
				f(z) = -k_{R+\ell+2}(z)\;,\\[5px]
				\displaystyle \mathcal{F}(z,\zb;\alpha) = -\sum_{i=i_0}^{R-1} g_{2i+4,R-i+\ell}(z,\zb)P_{i}(\alpha)\;,
			\end{cases}\\
			\mathcal{A}_{\tau,R,\ell}(z,\zb;\alpha) =& \begin{cases}
				f(z) = 0\;,\\[5px]
				\displaystyle \mathcal{F}(z,\zb;\alpha) = g_{\tau+2,\ell}(z,\zb)P_R(\alpha)\;,
			\end{cases}
		\end{align}
	\end{subequations}
where $i_0 = \max(|p_{21}|,|p_{34}|)/2$ is the lowest possible $R$-charge in the OPE. The chiral part and the dynamical part of the superconformal blocks are just the bosonic conformal blocks in 2d and 4d. Therefore, we can directly relate the coefficients of the conformal block decomposition of the chiral part and the dynamical part of a correlator to the OPE data of the superconformal decomposition of that correlator. In particular, because the long superconformal blocks have vanishing chiral part, the chiral correlator $f(z;\beta)$
	\begin{equation}\label{eq:chiralG}
		f(z;\beta) = \mathcal{G}(z, \zb ; \zb ,\beta),
	\end{equation}
encodes the information of only the protected multiplets.
	
The expressions of the superconformal blocks are compatible with the recombination rule in $\mathcal{N}=2$ superconformal algebra \cite{Dolan:2002zh,Cordova:2016emh}
	\begin{equation}
		\mathcal{C}_{R,\ell} = \mathcal{C}_{R+1,\ell-1} + \mathcal{A}_{2R+2,R,\ell}\;.
	\end{equation}
The half-BPS block can be viewed as a semi-short block with $\ell=-1$
	\begin{equation}\label{eq:BCrelation}
		\mathcal{B}_R = -\mathcal{C}_{R-1,-1}\;.
	\end{equation}
	
For completeness, let us also give the expressions for the superconformal blocks as a linear combination of bosonic conformal blocks in correspondence with the super primary and super descendants exchanged in a superconformal multiplet \cite{Nirschl:2004pa,Cordova:2016emh,Baume:2019aid}. For the half-BPS multiplet, we have
	\begin{align}\label{eq:halfB1}
		\mathcal{B}_R = g_{2R,0}P_R - \zeta_{R}\, g_{2R,1}P_{R-1} + \zeta_{R-1}\zeta_{R}\, g_{2R+2,0}P_{R-2}\;.
	\end{align}
For the semi-short multiplet we have
	\begin{align}\label{eq:semiC1}
		\mathcal{C}_{R,\ell} =& \quad\,  g_{2R+2,\ell}P_R - g_{2R+2,\ell+1}P_{R+1} + \xi_{\ell,R}\, g_{2R+2,\ell+1}P_R - \zeta_R\, g_{2R+2,\ell+1}P_{R-1}\nonumber \\ 
		&+ \zeta_{R+\ell+2}\, g_{2R+2,\ell+2}P_R - \zeta_R\, g_{2R+4,\ell-1}P_{R-1} + \zeta_R\, g_{2R+4,\ell}P_R\nonumber \\
		&-  \xi_{\ell+1,R-1}\zeta_R\, g_{2R+4,\ell}P_{R-1} + \zeta_R \zeta_{R-1}\, g_{2R+4,\ell}P_{R-2} \nonumber\\ 
		&- \zeta_R \zeta_{R+l+2}\, g_{2R+4,\ell+1}P_{R-1}\;.
	\end{align}
Here, we have defined
	\begin{equation}
		\zeta_R = \frac{(4R^2-p_{21}^2)(4R^2-p_{34}^2)}{64 R^2(4R^2-1)}\;, \quad \xi_{\ell,R} = \frac{p_{21}p_{34}(\ell+1)(2R+\ell+2)}{8R(R+1)(R+\ell+1)(R+\ell+2)}\;.
	\end{equation}
Note that $P_i$ with $i<\max(|p_{21}|,|p_{34}|)/2$ should be dropped from the expression above.

\section{About the protected part $\mathcal{G}_{\text{protected}}$}\label{appx:Gprotected}
	
In this appendix we discuss in detail the protected part $\mathcal{G}_{\text{protected}}(z,\zb;\alpha,\beta)$ in the decomposition \eqref{eq:GHdecp}
	\begin{equation}\label{eq:GHdecp2}
		\mathcal{G}(z, \zb ; \alpha,\beta)=\mathcal{G}_{\text{protected}}(z, \zb ; \alpha,\beta) + \frac{(z - \alpha)(\zb - \alpha)}{z \zb \alpha^2} \mathcal{H}(z, \zb;\alpha,\beta)\;.
	\end{equation}
We have already mentioned that this part is Bose symmetric and encodes information of the protected sector. However, we have not given the explicit form of $\mathcal{G}_{\text{protected}}$. In fact, the decomposition \eqref{eq:GHdecp2} is somewhat arbitrary. The requirement of Bose symmetry alone is not sufficient to determine $\mathcal{G}_{\text{protected}}$. If we shift $\mathcal{H}$ by an appropriately Bose symmetric function $\mathcal{A}$ which does not contain poles at $z=\alpha$, $\bar{z}=\alpha$ but is otherwise arbitrary, we obtain a new $\mathcal{G}_{\text{protected}}$
\begin{equation}\label{eq:Gpshift}
		\mathcal{G}_{\text{protected}} \rightarrow \mathcal{G}_{\text{protected}} + \frac{(z - \alpha)(\zb - \alpha)}{z \zb \alpha^2}\mathcal{A},\quad \mathcal{H}\rightarrow \mathcal{H} - \mathcal{A}\;.
	\end{equation}
 Note that the shift in $\mathcal{G}_{\text{protected}}$ vanishes under the twist $z=\alpha$ or $\bar{z}=\alpha$. Therefore, the new $\mathcal{G}_{\text{protected}}$ encodes the same protected information as the old one. This fact will be used later when we construct $\mathcal{G}_{\text{protected}}$ at tree level in AdS. 

 Note that generically, these $\mathcal{N}=2$ SCFTs do not admit  marginal couplings. However, in the case when they are constructed from $N$ D3-branes with $N_F$ probe D7-branes, i.e., 4d $\mathcal{N}=4$ SYM coupled to $\mathcal{N}=2$ matter, we do have a marginal coupling. This is because $N_F/N\ll 1$ and the beta function of the coupling is approximately zero. In this case, the theory has a free limit and the natural candidate for $\mathcal{G}_{\text{protected}}$ is the correlator at zero coupling. We will focus on this case and will see that this expectation is almost correct up to a small modification. More precisely, in this theory we are adding to $\mathcal{N}=4$ SYM $N_F$ $\mathcal{N}=2$ hypermultiplets. This gives rise to a $SU(N_F)$ flavor group which is identified with the color group in AdS. The scalars of the hypermultiplets are denoted by $q$ and $\bar{q}$. They transform in the fundamental and anti-fundamental representations of the flavor group $SU(N_F)$ and the gauge group $SU(N)$ respectively. Together with the scalars $\phi$ in $\mathcal{N}=4$ SYM we can construct two classes of gauge invariant operators of the schematic form
 \begin{equation}
		\bar{q}\, \phi^{p-2} q,\qquad \tr\phi^p\;,
	\end{equation}
 for $p=2,3,\ldots$. Roughly speaking, they are respectively dual to super gluons and super gravitons in AdS. However, to appropriately define single-particle states we need to further take into account mixing among these operators. This will be discussed in detail below. The free correlator $\mathcal{G}_{\text{free}}$ can then be calculated by performing Wick contractions for these single-particle operators.

\subsection{Single-particle operators}\label{appx:singleparticle}

Let us begin by making the notations more precise. The scalar fields of the $\mathcal{N}=2$ hypermultiplets, with all indices written explictly displayed, are
\begin{equation}
		q(x)=(q^{a,n})^i(x)\;,\qquad \bar{q}(x)=(\bar{q}^{\, a}_{\,m})_i(x)\;.
	\end{equation} 
Here $i$ is the $SU(N)$ index, $m$, $n$ are $SU(N_F)$ indices, and $a$ is the $SU(2)_R$ index. On the other hand, the field $\phi$ is in the adjoint representation of $SU(N)$ and is a singlet with respect to $SU(N_F)$. With respect to the orginal R-symmetry group $SO(6)_R$ of $\mathcal{N}=4$, $\phi$ transforms as a vector, labelled by the vector index $A=1,\ldots,6$. However, the addition of the $\mathcal{N}=2$ matter, $SO(6)_R$ breaks into $SO(4)\times SO(2)$, where $SO(4) \simeq SU(2)_R\times SU(2)_L$. Moreover, since the symmetry breaking corresponds to $S^3\subset S^5$, we should restrict to the $\phi$ components with $A=1,\ldots,4$
\begin{equation}
		\phi(x) = (\phi^{a\bar{a}})^i_{\ j}(x) = (\sigma^A)^{a\bar{a}} (\phi^A)^i_{\ j}(x)\;,\quad A=1,2,3,4\;.
	\end{equation}
Here $(\sigma^A)^{a\bar{a}}$ are the Pauli matrices. In addition to the $SU(2)_R$ index $a$, there is also an $SU(2)_L$ index  $\bar{a}$. The ``single-trace''\footnote{By that we mean the $SU(N)$ indices are contracted with the pattern of a single line.}  super gluon KK weight $p$ is defined to be
	\begin{equation}
		\widetilde{\mathcal{O}}^{I;a_0,\cdots,a_{p-1};\bar a_1,\dots,\bar a_{p-2}}_p (x) = (T^I)^m_{\ \ n}\, \bar{q}^{\,(a_0}_{\,\ m}\, \phi^{a_1 (\bar{a}_1} \cdots \phi^{a_{p-2}\bar{a}_{p-2})} q^{a_{p-1})n}\;,
	\end{equation}
where we have implicitly contracted all the gauge indices, and symmetrized all the $a$ and $\bar{a}$ indices respectively. $(T^I)^m_{\ \ n}$ is the generator of the flavor group $SU(N_F)$. It is easy to see that this operator has dimension $p$,  $SU(2)_R$ spin $\frac{p}{2}$, $SU(2)_L$ spin $\frac{p-2}{2}$, and transforms in the adjoint representation of $SU(N_F)$. As in \eqref{eq:vvb}, we can also get rid of the $a$ and $\bar{a}$ indices by contracting the operator with two-component polarization spinors $v$ and $\bar{v}$
	\begin{equation}
		\widetilde{\mathcal{O}}^I_p(x;v,\bar v)=\widetilde{\mathcal{O}}^{I;a_1,\cdots,a_p;\bar a_1,\dots,\bar a_{p-2}}_p (x) v_{a_1}\cdots v_{a_p} \bar v_{\bar a_1}\cdots \bar v_{\bar a_{p-2}}\;.
	\end{equation}
Note that these single-trace gluon operators are automatically orthogonal to any ``double-trace'' gluon operators $(\bar{q}\phi^{p_1-2}q)(\bar{q}\phi^{p_2-2}q)$, since they have different numbers of $q$ and $\bar{q}$. But they have nonzero inner products with the product of a single-trace super gluon and a single-trace super graviton of the form $(\bar{q}\phi^{p_1-2}q) (\tr\phi^{p_2})$ if $p=p_1+p_2$. Similarly, the single-trace super gluon operators can have nonzero overlap with higher product of operators which contains one single-trace super gluon and multiple single-trace super gravitons. To have a well-defined notion of single-particle states, we require that the single-particle super gluons $\mathcal{O}^I_p(x;v,\bar{v})$ are orthogonal to all multi-particle states. This can be achieved by adding to $\widetilde{\mathcal{O}}^I_p(x;v,\bar v)$ multi-trace operators with $1/N$ suppressed coefficients. The orthogonality condition then reads
\begin{equation}
		\left\langle \mathcal{O}^{I}_p(x_1;v_1,\bar{v}_1)\; \left((\bar{q}\phi^{p_1-2}q) \tr\phi^{p_2}\cdots \tr\phi^{p_n}\right)(x_2;v_2,\bar{v}_2) \right\rangle = 0\;,
	\end{equation} 
for any partition of $p$ into $p_1,\ldots,p_n$. We will normalize such operators as 
\begin{equation}
		\left\langle \mathcal{O}^{I_1}_p(x_1;v_1,\bar{v}_1)\, \mathcal{O}^{I_2}_p(x_2;v_2,\bar{v}_2) \right\rangle = \frac{\delta^{I_1 I_2}}{\left(x_{12}^2\right)^p} \left(v_{12}\right)^p\left(\bar{v}_{12}\right)^{p-2}\;.
	\end{equation}
In the free theory, correlation functions can be computed by Wick contractions
\begin{align}
		\parbox{4.8cm}{\tikz{\path (0,-0.4) -- (0,0.4);\node [anchor=center] at (0,0) {$\displaystyle(\bar{q}_{\,m})_{i_1}(x_1;v_1)\;\;(q^{n})^{i_2}(x_2;v_2)$};\draw [thick] (-2.06,0.2) -- ++(0,0.15) -- ++(2.49,0) -- ++(0,-0.15);}} =&\; \delta_{i_1}^{i_2}\,\delta_{m}^{n}\frac{v_{12}}{x_{12}^2}\;,\\
		\parbox{5.69cm}{\tikz{\path (0,-0.4) -- (0,0.4);\node [anchor=center] at (0,0) {$\displaystyle(\phi)^{i_1}_{\ j_1}(x_1;v_1,\bar{v}_1)\;\; (\phi)^{i_2}_{\ j_2}(x_2;v_2,\bar{v}_2)$};\draw [thick] (-2.5,0.2) -- ++(0,0.15) -- ++(2.9,0) -- ++(0,-0.15);}} =& \left(\delta^{i_1}_{j_2}\delta^{i_2}_{j_1}-\frac{1}{N}\delta^{i_1}_{j_1}\delta^{i_2}_{j_2}\right)\frac{v_{12}\bar{v}_{12}}{x_{12}^2}\;.
	\end{align}
We will also use such conventions for flavor group generators $(T^I)^m_{\ \ n}$
	\begin{equation}
		\tr(T^I T^J) = \frac{1}{2}\delta^{IJ}\;, \qquad f^{IJK} = -2i\left[\tr(T^I T^J T^K) - \tr(T^I T^K T^J)\right]\;.
	\end{equation}
From the orthogonality condition, it is straightforward to work out the first a few single-particle super gluon operators 
 	\begin{align}
		\mathcal{O}^I_2(x;v,\bar{v}) =& -i\,\sqrt{\frac{2}{N}}\; \bar{q}q\;,\\
		\mathcal{O}^I_3(x;v,\bar{v}) =& -i\,\sqrt{\frac{2}{N^2-1}}\; \bar{q}\phi q\;,\\
		\mathcal{O}^I_4(x;v,\bar{v}) =& -i\,\sqrt{\frac{2N}{(N^2-1)(N^2-4)}}\left( \bar{q}\phi^2 q - \frac{1}{N} (\bar{q}q)\tr\phi^2\right)\;,\\
		\mathcal{O}^I_5(x;v,\bar{v}) =& -i\,\sqrt{\frac{2(N^2+1)}{(N^2-1)(N^2-4)(N^2-9)}}\nonumber\\
		&\quad\times\left( \bar{q}\phi^3 q - \frac{1}{N} (\bar{q} q)\tr\phi^3 -\frac{2N^2-3}{N(N^2+1)}(\bar{q}\phi q)\tr\phi^2 \right)\;.
	\end{align}

We can also compute the three-point functions $\langle \mathcal{O}^{I_1}_{p_1}\, \mathcal{O}^{I_2}_{p_2}\, \mathcal{O}^{I_3}_{p_3}\rangle$ of these single-particle operators. The orthogonality condition ensures that they are non-vanishing only when $|p_1-p_2|+2\leq p_3 \leq p_1+p_2-2$. We find that the three-point coefficients are the same for all $p_1$, $p_2$ and $p_3$ at the leading order in $1/N$
	\begin{align}
		&\left\langle \mathcal{O}^{I_1}_{p_1}(x_1;v_1,\bar{v}_1)\, \mathcal{O}^{I_2}_{p_2}(x_2;v_2,\bar{v}_2)\, \mathcal{O}^{I_3}_{p_3}(x_3;v_3,\bar{v}_3) \right\rangle \nonumber\\
		=&  \text{(kinematic factor)}\times f^{I_1I_2I_3} \sqrt{\frac{2}{N}} \left[1+\order{\frac{1}{N^2}}\right]\;,
	\end{align}
where the kinematic factor is given by
	\begin{equation}
		\bar{v}_{12}^{-2}\bar{v}_{23}^{-2}\bar{v}_{31}^{-2}\left(\frac{v_{12}\bar{v}_{12}}{x^2_{12}}\right)^\frac{p_1+p_2-p_3}{2}\left(\frac{v_{23}\bar{v}_{23}}{x^2_{23}}\right)^\frac{p_2+p_3-p_1}{2}\left(\frac{v_{31}\bar{v}_{31}}{x^2_{31}}\right)^\frac{p_3+p_1-p_2}{2}\;.
	\end{equation}
This precisely matches with the result in \cite{Alday:2021odx} at strong coupling. The matching is expceted because the three-point functions of half-BPS operators are protected.  The computation of the three-point functions also indicates that we should identify $a_F = 2/N$ in this theory.

\subsection{Free correlator and the protected part}
	
The four-point functions of single-particle operators in the free theory can also be computed by Wick contractions. Let us first focus on the simplest case with $p_i=2$ 
\begin{align}
    \nonumber	\mathcal{G}_{2222,\text{free}} =&\ \delta^{I_1 I_2}\,\delta^{I_3 I_4} + \frac{(1-\alpha)^2}{\alpha^2} \frac{U^2}{V^2} \delta^{I_1 I_4}\,\delta^{I_2 I_3} + \frac{U^2}{\alpha^2} \delta^{I_1 I_3}\,\delta^{I_2 I_4} \\
    \nonumber		&+ 2\, a_F \left(-\frac{1-\alpha}{\alpha} \frac{U}{V}(T[1234]+ T[1432]) + \frac{U}{\alpha}(T[1342]+T[1243]) \right. \\
    &\left. \qquad\qquad + \frac{1-\alpha}{\alpha^2}\frac{U^2}{V}(T[1423]+T[1324])\right),
\end{align}
where we have used the abbreviation $T[12 \cdots n]={\rm tr}(T^{I_1} T^{I_2}\cdots T^{I_n})$. The leading term $\mathcal{G}^{(0)}_{2222,\text{free}}$ reproduces the disconnected  correlator $\mathcal{G}^{(0)}_{2222}$ on the AdS side. However, at tree level we find that $\mathcal{G}^{(1)}_{2222,\text{free}}$ contains color structures in trace basis which cannot be rewritten in terms of the structures $\cs$, $\ct$, $\cu$ as expected from the bulk computation.\footnote{At first sight, this seems to contradict  the fact that the chiral correlator \eqref{eq:chiralG} is independent of the coupling and should contain the same color structure in both limits. But one can check that after the chiral algebra twist  $\mathcal{G}_{2222,\text{free}}$ has the correct color structures
		\begin{equation*}
			\mathcal{G}_{2222,\text{free}}(z,\zb;\zb,\beta) = \bs + \frac{z^2}{(1-z)^2} \bt + z^2 \bu - a_F \left( \frac{z}{1-z}\ct + z \cu \right)\;.
	\end{equation*}
} To resolve this mismatch, we note the freedom in \eqref{eq:Gpshift} and add the following term to $\mathcal{G}_{2222,\text{free}}$
	\begin{align}
		-\frac{2}{3}a_F\frac{(z-\alpha)(\zb-\alpha)}{z\zb\alpha^2} \frac{U^2}{V} (T[1234]\!+\!T[1243]\!+\!T[1324]\!+\!T[1342]\!+\!T[1423]\!+\!T[1432])\;.
	\end{align}
The new $\mathcal{G}_{2222,\text{protected}}$ now has the correct color structures
	\begin{align}
		\nonumber \mathcal{G}_{2222,\text{protected}} =&\ \bs + \frac{(1-\alpha)^2}{\alpha^2} \frac{U^2}{V^2}\, \bt +  \frac{U^2}{\alpha^2}\, \bu + \frac{a_F}{3} \left(-\frac{1-\alpha}{\alpha} \frac{U}{V}(\ct-\cs)  \right. \\
		&\left. \ + \frac{U}{\alpha}(\cs-\cu) + \frac{1-\alpha}{\alpha^2}\frac{U^2}{V}(\cu-\ct)\right)\;.
	\end{align}
In general, to define $\mathcal{G}_{\text{protected}}$ for the free correlator of arbitrary KK modes with only $\mathtt{c}_s$, $\mathtt{c}_t$, $\mathtt{c}_u$ as color structures, we also need to add to it terms from the reduced correlator. More conveniently, this is equivalent to using the following replacement prescription in the free correlator
	\begin{align}
		T[1234]+ T[1432]\ \rightarrow&\ \frac{\ct-\cs}{6}\;, \\
		T[1342]+ T[1243]\ \rightarrow&\ \frac{\cs-\cu}{6}\;, \\
		T[1423]+ T[1324]\ \rightarrow&\ \frac{\cu-\ct}{6}\;.
	\end{align}
This gives the protected part of the correlator with the correct color structures while keeping the chiral correlator unchanged. Note that at strong coupling, $\mathcal{G}_{\text{protected}}$ is also defined by requiring that the tree-level reduced correlator is a sum of $D$-functions \cite{Alday:2021ajh} and explicit expressions are worked out for $\langle 22pp\rangle$ correlators. The protected part defined by shifting the free correlator is in agreement with  \cite{Alday:2021ajh} for these cases. 
	
The free correlator of $\langle 2222 \rangle$ vanishes automatically for one-loop and higher levels. But for generic cases, the free correlators are nonzero to all orders. For example, for $p_i=3$ Wick contraction gives 
	\begin{align}
		\mathcal{G}_{3333,\text{free}} =&\ \left(1+\frac{\frac{U}{\sigma} + \frac{U}{\sigma}\frac{\tau}{V}}{N^2-1}\right) \delta^{I_1 I_2}\,\delta^{I_3 I_4}+ \frac{U^2}{\alpha^2}\left(\frac{U}{\sigma}+\frac{1+\frac{U}{\sigma}\frac{\tau}{V}}{N^2-1}\right) \delta^{I_1 I_3}\,\delta^{I_2 I_4} \nonumber\\
		& + \frac{(1-\alpha)^2}{\alpha^2}\frac{U^2}{V^2}\left(\frac{U}{\sigma}\frac{\tau}{V}+\frac{1+\frac{U}{\sigma}}{N^2-1}\right) \delta^{I_1 I_4}\,\delta^{I_2 I_3} \nonumber\\
		&+ \frac{4}{N}\left(-\frac{1-\alpha}{\alpha} \frac{U}{V}\left(1+\frac{U}{\sigma}\frac{\tau}{V} - \frac{U/\sigma}{N^2-1} \right)(T[1234]+ T[1432]) \right.\nonumber\\
		& \quad\qquad + \frac{U}{\alpha}\left(1+\frac{U}{\sigma} - \frac{\frac{U}{\sigma}\frac{\tau}{V}}{N^2-1} \right)(T[1342]+T[1243])\nonumber\\
		&\left. \quad\qquad + \frac{1-\alpha}{\alpha^2}\frac{U^2}{V}\left(\frac{U}{\sigma} + \frac{U}{\sigma}\frac{\tau}{V} - \frac{1}{N^2-1} \right)(T[1423]+T[1324])\right).
	\end{align}
Expanding it in powers of $a_F$, we get the one-loop correction
	\begin{align}
		\mathcal{G}^{(2)}_{3333,\text{free}} =& \frac{1}{4}\frac{U}{\sigma}\left( 1+\frac{\tau}{V} \right)\delta^{I_1 I_2}\,\delta^{I_3 I_4} +\frac{1}{4} \frac{(1-\alpha)^2}{\alpha^2}\frac{U^2}{V^2}\left(1 + \frac{U}{\sigma} \right) \delta^{I_1 I_4}\,\delta^{I_2 I_3}\nonumber\\
		& + \frac{1}{4} \frac{U^2}{\alpha^2}\left(1 + \frac{U}{\sigma}\frac{\tau}{V} \right) \delta^{I_1 I_3}\,\delta^{I_2 I_4}\;.
	\end{align}
The color structures $\delta^{I_1 I_2}\,\delta^{I_3 I_4}$, $\delta^{I_1 I_4}\,\delta^{I_2 I_3}$, $\delta^{I_1 I_3}\,\delta^{I_2 I_4}$ are not expected from the one-loop gluon interactions in AdS. Instead, they are the structures of tree-level graviton exchanges which have the same order in $a_F$. Therefore, if we are focusing on gluon interactions we can conclude that this free correlator at one loop has no contribution from the gluon sector
	\begin{equation}
		\left. \mathcal{G}^{(2)}_{3333,\text{free}}\right|_\text{gluon sector} = 0\;.
	\end{equation}
Similarly by explicit checks one finds that the same conclusion holds for arbitrary KK weights
\begin{equation}
	\left.\mathcal{G}^{(2)}_{\{p_i\},\text{free}}\right|_\text{gluon sector} = 0\;.
	\end{equation}
We can also see this in a different way. Since the color structures only depend on how we contract $q$ and $\bar{q}$, the possible structures are linear combinations of  
\begin{align}
\nonumber		&\qquad\delta^{I_1 I_2}\,\delta^{I_3 I_4}\;,\qquad\qquad \delta^{I_1 I_4}\,\delta^{I_2 I_3}\;,\qquad\qquad \delta^{I_1 I_3}\,\delta^{I_2 I_4}\;,\\
		&T[1234]+ T[1432]\;,\quad T[1342]+ T[1243]\;,\quad T[1423]+ T[1324]\;,
	\end{align}
with constant coefficients independent of $N_F$. However, the one-loop box $\dst$ in the trace basis reads
	\begin{equation}
		\dst = - N_F(T[1234]+T[1432]) - \frac{1}{2}\left( \delta^{I_1 I_2}\,\delta^{I_3 I_4}+\delta^{I_1 I_4}\,\delta^{I_2 I_3}+\delta^{I_1 I_3}\,\delta^{I_2 I_4} \right),
	\end{equation}
where the factor $N_F$ can never be generated from Wick contractions. Therefore, we conclude that the one-loop gluon interactions in AdS do not generate a contribution to $\mathcal{G}_{\rm free}$. This statement directly generalizes to higher loops in the gluon sector of the theory, and so we always have
	\begin{equation}
		\left.\mathcal{G}^{(n)}_{\{p_i\},\text{free}}\right|_\text{gluon sector} = 0\;,\qquad n\geq 2\;.
	\end{equation}
As an important consequence of this discussion, at one and higher loops even though we always have the freedom in shifting the definition for the protected part $\mathcal{G}_{\{p_i\},\mathrm{protected}}^{(n)}$ in \eqref{eq:GHdecp2} by \eqref{eq:Gpshift}, it is possible and is in fact most convenient to simply set it to zero
\begin{equation}
		\mathcal{G}^{(n)}_{\{p_i\},\text{protected}}= 0\;,\qquad n\geq 2\;.
\end{equation}
This is the convention that we follow throughout our computations at loop level.

Finally, let us explain how the relation \eqref{eq:identityc} follows from the vanishing of the protected part at one loop. Since the chiral  correlator only comes from the protected part, it also vanishes at the one-loop level
	\begin{equation}
		f^{(2)}(z;\beta) = \mathcal{G}^{(2)}(z,\zb;\zb,\beta) = \mathcal{G}^{(2)}_{\text{protected}}(z,\zb;\zb,\beta) = 0\;.
	\end{equation}
In terms of the decomposition into superconformal blocks, the contribution from all protected operators should add up to be zero
	\begin{equation}
		f^{(2)}(z;\beta) = \sum_{R=R_0}^{R_\text{max}} \langle b^{(2)}(\beta) \rangle_R\, k_R(z) - \sum_{\ell=0}^\infty\sum_{R=R_0}^{R_\text{max}-1} \langle c^{(2)}(\beta) \rangle_{R,\ell}\, k_{R+\ell+2}(z) = 0\;,
	\end{equation}
where $R_0 = \max(|p_{21}|,|p_{34}|)/2$ and $R_\text{max} = \min(\Sigma_{12},\Sigma_{34})/2$. Taking the coefficient of $k_{R_\text{max}+\ell+1}(z)$ we obtain the sum rule \eqref{eq:identityc}
	\begin{equation}
		\langle c^{(2)} \rangle_{R_{\rm max}-1,\ell} + \langle c^{(2)} \rangle_{R_{\rm max}-2,\ell+1} + \dots +  \langle c^{(2)} \rangle_{R_0,\ell+R_{\rm max}-R_0-1} = 0\;.
	\end{equation}

\section{OPE data for unitarity recursion}\label{appx:data}
	
The unitarity recursion requires the input of disconnected and tree-level OPE data from different correlators. A standard method to compute these OPE data is to use the Lorentzian inversion formula \cite{Caron-Huot:2017vep,Simmons-Duffin:2017nub}. It is a Froissart-Gribov-type dispersion relation for conformal correlators that recovers the OPE data from the ``imaginary part'' of the correlator. The Lorentzian inversion formula generates a meromorphic function of $\tau$ and $\ell$, which has poles in $\tau$, and the OPE data is given by the residues at these poles. In 4d the Lorentzian inversion formula can be written as
	\begin{align}
		c_{\{p_i\},R,L}(\tau,\ell) =& N^{p_{21},p_{34}}_{\bar{h}}\int_0^1 \frac{\dd z}{z^2} \frac{\dd \zb}{\zb^2} k^{-p_{21},-p_{34}}_{1-h}(z)k^{-p_{21},-p_{34}}_{\bar{h}}(\zb) \frac{\zb-z}{z \zb} \text{dDisc} \left[\mathcal{H}^{I_1I_2I_3I_4}_{\{p_i\},R,L}(z,\zb) \right] \nonumber\\
		& + (-1)^{\ell+R-L}(1 \leftrightarrow 2)\;,
	\end{align}
where $h=\frac{\tau}{2}$, $\bar{h} = 1+\frac{\tau}{2}+\ell$ and
	\begin{equation}
		N^{p_{21},p_{34}}_{\bar{h}} = \frac{\Gamma(\bar{h}+\frac{p_{21}}{2})\Gamma(\bar{h}-\frac{p_{21}}{2})\Gamma(\bar{h}+\frac{p_{34}}{2})\Gamma(\bar{h}-\frac{p_{34}}{2})}{8\pi^2\Gamma(2\bar{h}-1)\Gamma(2\bar{h})}\;.
	\end{equation}
The ``imaginary part'' is given by the double discontinuity of the correlator
	\begin{equation}
		\text{dDisc}\left[ \mathcal{H}(z,\zb) \right] = \cos(\pi a) \mathcal{H}(z,\zb) - \frac{e^{i\pi a}}{2}\mathcal{H}^{\circlearrowleft}(z,\bar{z})-\frac{e^{-i\pi a}}{2}\mathcal{H}^{\circlearrowright}(z,\bar{z})\;,
	\end{equation}
where $a=\frac{p_{21}+p_{34}}{2}$, and $\circlearrowleft$ and $\circlearrowright$ means to analytically continue $\zb$ around $\zb=1$ counter-clockwisely and clockwisely respectively. The double discontinuity of the correlator kills all double-trace contributions in t-channel at the disconnected and tree levels, making it rather convenient for us to compute the OPE data by inversion formula. When we substitute $\text{dDisc}[\mathcal{H}]$ into the inversion formula, the integral can be factorized and worked out in a closed form. See, e.g., appendix A of \cite{Caron-Huot:2018kta} for details.
	
At the disconnected level, the general expression of the OPE data is found to be \cite{Drummond:2022dxd}
	\begin{align}
		\langle a^{(0)}\rangle_{pqpq,\tau,R,L,\ell}=&\quad \left(\delta _{p,q} (-1)^{R-L}\ \bt + (-1)^{\ell}\ \bu\right) \frac{r^{q-p,q-p}_{h}r^{q-p,q-p}_{\bar{h}}}{r^{q-p,q-p}_{R+1}r^{q-p,q-p}_{{L+1}}}\nonumber \\
		&\times \left(h-\frac{p+q}{2}+1\right)_{p+q-2} \left(\bar{h}-\frac{p+q}{2}+1\right)_{p+q-2}\nonumber \\
		&\times \frac{(\bar{h}-h) (h+\bar{h}-1)}{(p-1) (q-1) (h+R) (h-R-1) (\bar{h}+R) (\bar{h}-R-1)}\nonumber \\
		&\times \frac{(2 L+1) (2 R+1)}{\Gamma \left(\frac{p+q}{2}+L\right) \Gamma \left(\frac{p+q}{2}-L-1\right) \Gamma \left(\frac{p+q}{2}+R\right) \Gamma \left(\frac{p+q}{2}-R-1\right)}\;,
	\end{align}
where $r^{a,b}_h=\frac{\Gamma(h-a/2)\Gamma(h+b/2)}{\Gamma(2h-1)}$. The semi-short data is only nonzero when its twist sits at the double-trace threshold $\tau=p+q$
	\begin{equation}
		\left.\langle c^{(0)}\rangle_{pqpq,R,L,\ell}\right|_{R=\frac{p+q-2}{2}} = -\langle a^{(0)}\rangle_{pqpq,p+q-2,\frac{p+q-4}{2},L,\ell+1}\;.
	\end{equation}
	
The general expression of tree-level data is lacking, but we can compute it case by case using inversion formula. Since $\text{dDisc}$ kills all the double-trace contributions at tree level, we can replace $\text{dDisc}[\mathcal{H}]$ by its single-trace contributions (in t-channel)
	\begin{align}
		\left.\text{dDisc}\left[ -\ct\,\left(\frac{U}{\sigma}\right)^\frac{\Sigma_{12}}{2}\left(\frac{\tau}{V}\right)^\frac{\Sigma_{23}}{2}\left(\frac{\beta}{\beta'}\right)^2 \sum_{R=R_0}^{R_\text{max}} \mathcal{B}^{p_{23},p_{14}}_R(z',\zb';\alpha')P^{p_{23},p_{14}}_{R-1}(\beta') \right]\right|_\text{reduced part}\;,
	\end{align}
where $R_0 = \max(|p_{23}|,|p_{14}|)/2+1$, $R_\text{max}=\min(\Sigma_{23},\Sigma_{14})/2-1$ and $x'=1-x$. Note that we take the reduced part of the expression because we are computing $\text{dDisc}[\mathcal{H}]$ instead of $\text{dDisc}[\mathcal{G}]$.

With this method we find general expressions for both $\langle a^{(0)}\gamma^{(1)}\rangle_{p_1p_2p_3p_4} $ and $\langle a^{(1)}\rangle_{p_1p_2p_3p_4} $ when extremality $\mathcal{E}=2$. Without loss of generality, we set $p_4=p_1+p_2+p_3-4$. The expression with the unique representation $Y_{s_- ,s_- }(\alpha,\beta) $ in this case is
	\begin{equation}
		\langle a^{(0)}\gamma^{(1)}\rangle_{p_1p_2p_3p_4} = \langle a^{(0)}\gamma^{(1)}\rangle ^{dyn.} \, \ct-\langle a^{(0)}\gamma^{(1)}\rangle ^{dyn.}_{p_1 \leftrightarrow p_2} (-1)^{\ell}\, \cu\;,
	\end{equation}
at a specific $\tau$ and the dynamic part is
	\begin{equation}
		\langle a^{(0)}\gamma^{(1)}\rangle^{dyn.} =\frac{(-1)^{\frac{1}{2} \left(\Sigma_{12}-\Sigma_{34}\right)+1}  \Gamma _f}{(2\ell+\tau)!}{  \prod _{t=-\frac{1}{4} (\kappa _t+\kappa _u )-\ell}^{\frac{1}{2} \Sigma_{34}-1} \left(\frac{\tau }{2}-t\right)  \prod _{t=-\frac{1}{4} (\kappa _t-\kappa _u)-\ell}^{\frac{1}{2}\Sigma_{12}-1} \left(\frac{\tau }{2}-t\right)} \;,
	\end{equation}
where
	\begin{equation}
		\Gamma_f=\frac{2 \Gamma \left(\Sigma_{12}-2\right) \Gamma \left(\frac{1}{2} \left(\tau -\Sigma_{12}+2\right)\right) \Gamma \left(\frac{1}{2} \left(\tau -p_{12}\right)\right) \Gamma \left(\frac{1}{2} \left(\tau -p_{34}\right)\right) \Gamma \left(\frac{1}{2} \left(\tau +\Sigma_{34}-2\right)\right)}{\Gamma \left(p_1-1\right) \Gamma \left(p_2-1\right) \Gamma \left(p_3-1\right) \Gamma (\tau -1) \Gamma \left(\frac{1}{2} (s_-+  \Sigma_{12}-4)\right) \Gamma \left(\frac{1}{2} (s_-+ \Sigma_{34}-4)\right)}\;.
	\end{equation}
Moreover, we find the derivative relation \cite{Heemskerk:2009pn,Fitzpatrick:2011dm} also holds and  $\langle a^{(1)} \rangle$ can be written as 
	\begin{equation}
		\langle a^{(1)} \rangle_{p_1p_2p_3p_4} = \frac{\partial}{\partial \tau }\langle a^{(0)}\gamma^{(1)}\rangle_{p_1p_2p_3p_4}\;.
	\end{equation}
These formulas generate the data needed in the computation of $\langle22pp\rangle$ in section \ref{sec:o2o2opop}.
	
At higher extremalities it becomes more difficult to find general formulas. Therefore, we will just obtain the data required in the computations on a case by case basis using the inversion formula as described above. For example, in the computation of $\langle 3355\rangle$ in section \ref{sec:o3o3o5o5} (with extremality $\mathcal{E}=3$), the window region data are obtained by solving the operator mixing problem following the procedure presented in section \ref{sec:recursion}. At $\tau_0=6$ we have
	\begin{align}\label{eq:a1g1t6compute}
		\langle a^{(1)}\gamma^{(1)}\rangle_{3355}^{\tau_{0}=6}=&
		\begin{bmatrix}
			\langle a^{(0)}\gamma^{(1)}\rangle_{3322}\, & \langle a^{(0)}\gamma^{(1)}\rangle_{3333}  \, & \langle a^{(0)}\gamma^{(1)}\rangle_{3324}  
		\end{bmatrix}\nonumber\\
		&\times 
		\begin{bmatrix}
			\langle a^{(0)}\rangle_{2222}^{-1} & 0 & 0\\
			0 &\langle a^{(0)}\rangle_{3333}^{-1} & 0\\
			0&0& \langle a^{(0)}\rangle_{2424}^{-1} 
		\end{bmatrix}\times 
		\begin{bmatrix}
			\langle a^{(1)}\rangle_{2255}\, \\ \langle a^{(1)}\rangle_{3355}  \, \\  \langle a^{(1)}\rangle_{2455}
		\end{bmatrix}\;,
	\end{align}
and at $\tau_8=8$ we have
	\begin{align}\label{eq:a1g1t8compute}
		\langle a^{(1)}\gamma^{(1)}\rangle_{3355}^{\tau_{0}=8}=&
		\begin{bmatrix}
			\langle a^{(0)}\gamma^{(1)}\rangle_{3322}\, & \langle a^{(0)}\gamma^{(1)}\rangle_{3333}  \, & \langle a^{(0)}\gamma^{(1)}\rangle_{3324}  \, & \langle a^{(0)}\gamma^{(1)}\rangle_{3344}  \, & \langle a^{(0)}\gamma^{(1)}\rangle_{3335}  
		\end{bmatrix}\nonumber\\
		&\times 
		\begin{bmatrix}
			\langle a^{(0)}\rangle_{2222}^{-1} & 0 & 0 & 0 & 0 \\
			0 &\langle a^{(0)}\rangle_{3333}^{-1} & 0 & 0 & 0 \\
			0&0& \langle a^{(0)}\rangle_{2424}^{-1} & 0 & 0 \\
			0 & 0 & 0 & \langle a^{(0)}\rangle_{4444}^{-1} & 0 \\
			0 & 0 & 0 & 0 & \langle a^{(0)}\rangle_{3535}^{-1} 
		\end{bmatrix}\times 
		\begin{bmatrix}
			\langle a^{(1)}\rangle_{2255}\, \\ \langle a^{(1)}\rangle_{3355}  \, \\  \langle a^{(1)}\rangle_{2455}  \, \\  \langle a^{(1)}\rangle_{4455}  \, \\  \langle a^{(1)}\rangle_{3555} 
		\end{bmatrix}\;.
	\end{align}
Note that in the latter case it might seem that we should also include the exchange of the double-particle operator $[26]_{8,R,\ell}$. But the corresponding data $\langle a^{(0)}\gamma^{(1)}\rangle_{3326}$ comes from a correlator with extremality $1$ which vanishes in the single-particle basis. Therefore, effectively we only need to consider the mixing among other five operators as above. In these computations the tree-level data beyond $\mathcal{E}=2$ are $\langle a^{(0)}\gamma^{(1)}\rangle_{3333}$, $\langle a^{(1)}\rangle_{3355}$, $\langle a^{(0)}\gamma^{(1)}\rangle_{3344}$, $\langle a^{(0)}\gamma^{(1)}\rangle_{3335} $, $\langle a^{(1)}\rangle_{4455} $ and $\langle a^{(1)}\rangle_{3555} $, which are determined from the inversion formula separately. The detailed expressions are lengthy and we only include them in the ancillary file. The results of \eqref{eq:a1g1t6compute} and \eqref{eq:a1g1t8compute} have already been presented in section \ref{sec:o3o3o5o5}.

\bibliography{refs} 
\bibliographystyle{utphys}

\end{document}